\documentclass[superscriptaddress,groupedaddress,nofootnoteinbib,11pt,notoc]{article}
\pdfoutput=1
\usepackage{jcappub}

\usepackage{graphicx,color}
\usepackage{appendix}
\usepackage{latexsym,amsmath,amssymb,graphicx,booktabs}
\usepackage{epsfig,latexsym}
\usepackage{hyperref}
\numberwithin{equation}{section}% numera le equazioni seconde le sezioni , e.g. 1.15 invece che consecutivamente; anche le appendici, eq.~(A.1) etc. Richiede amsmath
\usepackage{url}

\definecolor{MyBlue}{rgb}{0.15,0.15,0.70}

\hypersetup{
colorlinks=true,
citecolor=MyBlue,
linkcolor=MyBlue,
urlcolor=MyBlue
}

\definecolor{lightgray}{gray}{0.9}

\def\deg{$^{\circ}$}
\newcommand{\lum}{$ \mathrm{erg}\, \mathrm{sec}^{-1}$}
\newcommand{\pflux}{$ \mathrm{ph}\, \mathrm{sec}^{-1}\, \mathrm{cm}^{-2}$}
\newcommand{\nn}{\nonumber}

\renewcommand\({\left(}
\renewcommand\){\right)}

\newcommand{\ra}{\rightarrow}

\def\lsim{\raise 0.4ex\hbox{$<$}\kern -0.8em\lower 0.62
ex\hbox{$\sim$}}

\def\gsim{\raise 0.4ex\hbox{$>$}\kern -0.7em\lower 0.62
ex\hbox{$\sim$}}

\def\lbar{{\hbox{$\lambda$}\kern -0.7em\raise 0.6ex
\hbox{$-$}}}

\newcommand\eq[1]{eq.~(\ref{#1})}

\newcommand\eqst[2]{eqs.~(\ref{#1})--(\ref{#2})}

\newcommand\p{\partial}

\newcommand\ee{\end{equation}}
\newcommand\be{\begin{equation}}
\def\bea{\begin{array}}
\def\eea{\end{array}}\def\ea{\end{array}}
\newcommand\ees{\end{eqnarray}}
\newcommand\bees{\begin{eqnarray}}
\def\nn{\nonumber}

\def\dslash{\hspace{-1mm}\not{\hbox{\kern-2pt $\partial$}}}
\def\Dslash{\not{\hbox{\kern-2pt $D$}}}
\def\pslash{\not{\hbox{\kern-2.1pt $p$}}}
\def\kslash{\not{\hbox{\kern-2.3pt $k$}}}
\def\qslash{\not{\hbox{\kern-2.3pt $q$}}}

\newcommand{\vk}{{\bf k}}

\def\p1{{\bf p}_1}
\def\p2{{\bf p}_2}
\def\k1{{\bf k}_1}
\def\k2{{\bf k}_2}

\newcommand{\dddM}{\kern 0.2em \raise 1.9ex\hbox{$...$}\kern -1.0em \hbox{$M$}}
\newcommand{\dddQ}{\kern 0.2em \raise 1.9ex\hbox{$...$}\kern -1.0em \hbox{$Q$}}
\newcommand{\dddI}{\kern 0.2em \raise 1.9ex\hbox{$...$}\kern -1.0em\hbox{$I$}}
\newcommand{\dddJ}{\kern 0.2em \raise 1.9ex\hbox{$...$}\kern-1.0em
\hbox{$J$}}
\newcommand{\dddcalJ}{\kern 0.2em \raise 1.9ex\hbox{$...$}\kern-1.0em
\hbox{${\cal J}$}}

\newcommand{\dddO}{\kern 0.2em \raise 1.9ex\hbox{$...$}\kern -1.0em
\hbox{${\cal O}$}}
\def\dddz{\raise 1.5ex\hbox{$...$}\kern -0.8em \hbox{$z$}}
\def\dddd{\raise 1.8ex\hbox{$...$}\kern -0.8em \hbox{$d$}}
\def\dddbd{\raise 1.8ex\hbox{$...$}\kern -0.8em \hbox{${\bf d}$}}
\def\ddbd{\raise 1.8ex\hbox{$..$}\kern -0.8em \hbox{${\bf d}$}}
\def\dddx{\raise 1.6ex\hbox{$...$}\kern -0.8em \hbox{$x$}}
\newcommand{\ode}{\Omega_{\rm DE}}
\newcommand{\oma}{\Omega_{M}}

\newcommand{\ola}{\Omega_{\Lambda}}

\newcommand{\rde}{\rho_{\rm DE}}
\newcommand{\wde}{w_{\rm DE}}

\graphicspath{ {./Graphs/} }

\title{Cosmology and dark energy  from joint gravitational wave-GRB observations}
\author[a]{Enis Belgacem,}
\author[b]{Yves Dirian,}
\author[a]{Stefano Foffa,}
\author[c]{Eric J. Howell,}
\author[a]{Michele Maggiore,}
\author[d]{Tania Regimbau}

\affiliation[a]{D\'epartement de Physique Th\'eorique and Center for Astroparticle Physics,\\
Universit\'e de Gen\`eve, 24 quai Ansermet, CH--1211 Gen\`eve 4, Switzerland}
\affiliation[b]{Center for Theoretical Astrophysics and Cosmology, Institute for Computational Science,
University of Z\"urich, CH-8057 Z\"urich, Switzerland}
\affiliation[c] {OzGrav-UWA, School of Physics and Astrophysics, University of Western Australia, Crawley WA 6009, Australia}
\affiliation[d]{Laboratoire d'Annecy-le-Vieux de Physique des Particules (LAPP),
Universit\'e Savoie Mont Blanc, CNRS/IN2P3, F-74941 Annecy, France\\
Artemis, Universit\'e C\^ote d'Azur, Observatoire de la C\^ote d'Azur CNRS, CS 34229, F-06304 Nice Cedex 4, France}

\abstract{Gravitational-wave (GW) detectors can contribute to the measurement of 
cosmological parameters and to testing the dark-energy sector of alternatives to $\Lambda$CDM, by using standard sirens. In this paper we focus on binary neutron stars with a counterpart 
detected through a gamma-ray burst (GRB), 
both at a second-generation network made by advanced LIGO+advanced Virgo+LIGO India+Kagra, and at
third-generation  (3G) detectors, discussing in particular the cases of a  single Einstein Telescope (ET), 
and of a
network of ET plus two Cosmic Explorer (CE). We construct mock catalogs of standard sirens, using different scenarios for the local merger rate and for the  detection of the electromagnetic counterpart.  For 3G detectors we estimate  the coincidences with a  GRB detector with the characteristics of the proposed THESEUS mission. We discuss   how these standard sirens with a GRB counterpart can improve the  determination of  cosmological parameters (and particularly of $H_0$) in $\Lambda$CDM, and we then study how to extract information on   dark energy, considering both a non-trivial dark energy equation of state and modified GW propagation. We find that a 2G detector network can already reach, over several years of data taking, an interesting sensitivity to modified GW propagation, while a single ET detector would have a remarkable potential for discovery. We also find that, to fully exploit the potential of a  ET+CE+CE network, it is necessary a much stronger program of search for electromagnetic counterparts (or  else to resort to  statistical methods for standard sirens), and furthermore gravitational lensing can become a limiting factor.}

\begin{document}
\maketitle
\flushbottom

\section{Introduction}

The first observations of gravitational waves (GWs) from binary black-holes coalescences \cite{Abbott:2016blz,Abbott:2016nmj,Abbott:2017vtc,Abbott:2017gyy,Abbott:2017oio,LIGOScientific:2018mvr}, as well as the first
observation of a neutron star binary coalescence~\cite{TheLIGOScientific:2017qsa}, together with the associated $\gamma$-ray burst (GRB)
\cite{Goldstein:2017mmi,Savchenko:2017ffs,Monitor:2017mdv} and the follow-up studies of the electromagnetic counterpart
(see \cite{GBM:2017lvd} and references therein) have opened the era of GW astronomy. In the near future, with advanced LIGO and advanced Virgo reaching their target sensitivity, and other detectors such as KAGRA and LIGO-India   joining the search, it is expected that such detections will  take place routinely.
On a longer timescale  the space interferometer LISA~\cite{Audley:2017drz}, that is expected to fly in 2034, and third-generation (3G) ground-based interferometers currently under study, such as the Einstein Telescope (ET) in Europe~\cite{Punturo:2010zz,Sathyaprakash:2012jk} and  Cosmic Explorer (CE) in the US~\cite{Dwyer:2014fpa},  will have the potential of detecting a large number of coalescing compact binaries at cosmological  redshifts. The detection of the GWs from a coalescing binary allows  a direct measurement of its luminosity distance $d_L$~\cite{Schutz:1986gp}, so these sources are referred to as ``standard sirens", the GW analogue of standard candles.
Much work has been devoted to investigating the cosmological informations that could be obtained from such  measurements, either when the redshift of the source is obtained thanks to the detection of an  electromagnetic counterpart, or using statistical methods~\cite{Holz:2005df,Dalal:2006qt,MacLeod:2007jd,Nissanke:2009kt,Cutler:2009qv,Sathyaprakash:2009xt,Zhao:2010sz,DelPozzo:2011yh,Nishizawa:2011eq,Taylor:2012db,Camera:2013xfa,Tamanini:2016zlh,Caprini:2016qxs,Cai:2016sby,DelPozzo:2017kme,Belgacem:2017ihm,Belgacem:2018lbp,Mendonca:2019yfo}.

In this paper we contribute to the currently ongoing effort for exploring the scientific potential of 3G   interferometers (see e.g. \cite{Sathyaprakash:2019nnu,Sathyaprakash:2019rom,Sathyaprakash:2019yqt}), by
performing an updated study of  the cosmological information that can be obtained from the observation of standard sirens. We focus on binary neutron stars (BNS) with an electromagnetic counterpart. This  type of analysis is appropriate for ground-based detectors, since in this case BNS merge within the bandwidth of the detector, and thus can in principle be detected in coincidence with an electromagnetic signal that can allow us to identify the host galaxy and therefore obtain the redshift.\footnote{For the space interferometer LISA the best studied example of standard siren with an expected  electromagnetic counterpart is the coalescence of supermassive BH binaries at large redshift, $z\, \lsim \, 8$ (see~\cite{Tamanini:2016zlh} and references therein).  Thus,  the construction of the source catalog and the methodology are completely different from those used here. A study  of the perspective for  observing  dark energy and modified gravity with   LISA using  supermassive BH binaries has been recently presented in \cite{Belgacem:2019pkk}.}
Using the strategy presented in \cite{2012PhRvD..86l2001R,2014PhRvD..89h4046R,Regimbau:2014nxa, 2015PhRvD..92f3002M,2016PhRvD..93b4018M,2017PhRvL.118o1105R},
we begin by constructing  mock catalogs of  BNS detections at GW detectors. We consider first
a  network of second-generation (2G)  detectors composed by   advanced LIGO-Hanford, advanced LIGO-Livingston, advanced Virgo,  Kagra and LIGO India, assumed to be at their target sensitivity. We will refer to this as the HLVKI network.
We will then consider  third-generation (3G) detectors  studying  two different  configurations, namely  a single ET detector, and a network made of a single Einstein Telescope plus   two Cosmic Explorers.\footnote{The  ET+CE+CE network is the baseline configuration that is studied in the 3G Science Case document, that is currently being developed by GWIC 3G Committee \url{https://gwic.ligo.org/}, the GWIC 3G Science Case Team and the International 3G Science Team Consortium.}

We will then study the possibility of simultaneous detection of an electromagnetic counterpart, focusing on the case of a joint GW-GRB detection.
A single GW detector, even in a triangular configuration as  planned for ET, cannot provide the localization of a coalescing binary with a significant accuracy. However,  the detection of a temporally coincident GRB can still allow for the measurement of the redshift of the source; indeed, GRB satellites such as Neil Gehrels Swift Observatory (\emph{Swift}; \url{https://swift.gsfc.nasa.gov/}) regularly obtain redshifts of GRBs, without the need of a GW localization.\footnote{Swift attempts to obtain redshifts by using its UV/Optical Telescope, UVOT, or through arc-second accuracy positional information that is relayed to ground based telescopes.} In particular, for 3G detectors we will estimate  the expected number and the redshift distributions of coincidences between  GW events and the electromagnetic signal observed at a GRB detector with the characteristics of the proposed  THESEUS mission~\cite{Amati:2017npy,Stratta:2017bwq,Stratta:2018ldl}, that could be in operation at the same time as 3G  detectors.

A network of third-generation GW detectors, such as the ET+CE+CE configuration,  would instead localize the source, whose redshift could then be measured also by  optical/IR telescopes. For instance, the electromagnetic signal from a kilonova associated to a BNS coalescence could be detected up to $z\simeq 0.55$ by optical imaging at LSST (\url{https://www.lsst.org/}) and Subaru (\url{https://www.naoj.org/}),  up to $z\simeq 0.76$ with infrared imaging at WFIRST (\url{https://wfirst.gsfc.nasa.gov/}), and up to $z\simeq 0.37$ by optical spectroscopy at E-ELT (\url{https://www.eso.org/sci/facilities/eelt/}). However, it is currently difficult to estimate how much telescope time will be devoted by these facilities to the follow-up of GW events. In this paper we limit ourselves to the coincidence with GRB detectors, but we should keep in mind that, for a network with significant localization capabilities such as ET+CE+CE, at $z\,\lsim\, 0.5$ a significant number of  coincidences with optical/IR telescopes is in principle possible, to the extent that the number of such coincidences could be much larger than those obtained   from GRB detections, so for such a network our estimates will be conservative.
To construct our mock source catalogs we will examine different possibilities  for the local merger rate and for the probability of determining the redshift through the detection  of an associated GRB.
This will lead to different scenarios, more conservative or more optimistic, that will be presented in Section~\ref{sect:catalog}. In particular, in Section~\ref{sect:estimates} we will present detailed and ready-to-use results for the redshift distribution of the GW events at 3G detectors and of the coincidences with GRBs, as well as expressions  for the observational error $\Delta d_L/d_L$ as a function of $z$.

In Section~\ref{sect:LCDM}, using a Markov Chain Monte Carlo (MCMC) analysis, we  study how the detection of standard sirens  from  our mock  catalogs  would contribute to the knowledge of the cosmological parameters,  in the context of $\Lambda$CDM, considering the HLVKI network in Section~\ref{sect:LCDMwith2G} and 3G detectors  in Sections~\ref{sect:LCDMwithET} and~\ref{sect:LCDMwithET+CE+CE}. For both 2G and 3G detectors, we combine the contribution from standard sirens 
with  existing datasets  from cosmic microwave background (CMB), baryon acoustic oscillation (BAO) and type Ia supernovae (SNe), to remove degeneracies and improve
the accuracy of cosmological parameter reconstruction. For 3G detectors, we find that useful results can also  sometimes be obtained already just from standard sirens, without external datasets.

In Section~\ref{sect:DE} we  extend the analysis performed for $\Lambda$CDM, by considering a more general dark energy (DE) sector.
Beside the effect of the DE equation of state, we will study the effect induced by modified GW propagation.
Indeed, as we will recall in Section~\ref{sect:DEandmodGW}, it has been  realized in recent years that modified gravity models that predict a non-trivial DE equation of state  also  predict deviations from general relativity in the GW propagation  across cosmological distances, even in theories where the speed of gravity is equal to
$c$~\cite{Deffayet:2007kf,Yunes:2010yf,Saltas:2014dha,Gleyzes:2014rba,Lombriser:2015sxa,Nishizawa:2017nef,Arai:2017hxj,Belgacem:2017ihm,Amendola:2017ovw,Linder:2018jil,Pardo:2018ipy,Belgacem:2018lbp,Lagos:2019kds}.
As recently found in \cite{Belgacem:2017ihm,Belgacem:2018lbp}, in a generic modified gravity model the  effect on the luminosity distance induced by modified GW propagation dominates over that from the DE equation of state. Furthermore, it is specific to GW observations, making it a prime observable for the study of dark energy and modified gravity at 3G GW detectors. In
ref.~\cite{Belgacem:2018lbp} has been proposed  a simple parametrization of the effect on the luminosity distance induced by modified GW propagation,  in terms of two parameters $(\Xi_0,n)$, that complements the $(w_0,w_a)$ parametrization of the DE equation of state. Recently, in \cite{Belgacem:2019pkk} it has been found that this parametrization covers almost all viable modified gravity models that have been considered in the literature. An analysis of the possibility of observing modified GW propagation at ET was already presented in \cite{Belgacem:2018lbp}.
In that paper,  following the standard working hypothesis used to date in the literature, it was simply assumed  that, given the detection rate at ET, which is estimated as ${\cal O}(10^5)$ BNS/yr, in a few years one could collect
${\cal O}(10^3)$ events with an electromagnetic counterpart. To go beyond this simple estimate, one needs a model for how the counterpart is detected, to see first of all whether the figure of  ${\cal O}(10^3)$ events with counterpart is realistic. Furthermore, once one specifies how the counterpart is observed, one can study how
this  affects the redshift distribution of the standard sirens with observed electromagnetic counterpart which
in  \cite{Belgacem:2019pkk}, again following the standard assumption in the literature,
was just assumed to follow a simple distribution determined by the star formation rate.
In Section~\ref{sect:DE} of the present paper we will improve on these results by using our more realistic scenarios for both the catalog of GW sources and the detection of the electromagnetic counterparts (restricted to the case of a GRB observation), and we will extend the study done for ET  also to the 2G HLVKI network, and to the ET+CE+CE network.
Section~\ref{sect:conl} contains our conclusions.

\section{Construction of mock source catalogs}\label{sect:catalog}

\subsection{GW events}\label{sect:GWevents}

In order to simulate a catalog of binary neutron stars coalescences, we first produce an extra-galactic population of neutron star binaries using the Monte Carlo algorithm developed in~\cite{2012PhRvD..86l2001R,2014PhRvD..89h4046R,Regimbau:2014nxa, 2015PhRvD..92f3002M,2016PhRvD..93b4018M,2017PhRvL.118o1105R}. We use the fiducial model of \cite{2018PhRvL.120i1101A} for the distribution of the parameters and we proceed as follows for each source:
the location in the sky $\hat{\Omega}$, the cosine of the orientation
$\iota$, the polarization $\psi$ and the phase of the signal at coalescence $\phi_0,$ are drawn from uniform distributions.
The redshift is drawn from a (normalized) probability distribution $p(z)$,
\begin{equation}\label{red1}
p(z)=\frac{R_z(z)}{\int_0^{10} R_z(z) dz}\, ,
\end{equation}
where $R_z(z)$ is the merger rate density per unit redshift, in the observer frame. It can be
expressed as\footnote{We correct here a typo in eq.~(2) of \cite{2017PhRvL.118o1105R}, where a spurious $\int dz$ appears on the right-hand side.}
\begin{equation}\label{red2}
R_z(z) =  \frac{R_m(z)}{1+z} \frac{dV(z)}{dz} \, ,
\end{equation}
where $dV/dz$ is the comoving volume element and $R_m$ is the rate per volume in the source
frame. The latter is given by
\begin{equation}\label{RnzRf}
R_m(z) = \int_{t_{\min}}^{t_{\max}} R_f[t(z)-t_d]P(t_d) dt_d\, ,
\end{equation}
where $R_f(t)$ is the formation rate of massive binaries, $P(t_d)$ is the distribution
of the time delay $t_d$  between the formation of the massive progenitors and
their merger,  and $t(z)$
is the age of the Universe at the time of merger. We assume that $R_f(t) $ in \eq{RnzRf} follows the cosmic star formation rate, for which   we use the
recent model of \cite{2015MNRAS.447.2575V}. We further assume that the time delay distribution follows $P(t_d) \propto t_d^\alpha$,
with $\alpha=-1$ for $t_d>t_{\min}$, where $t_{\min}=20$ Myr is the minimum delay time for a massive binary to evolve until coalescence, and $t_{\max}$ is a maximum time delay, set  equal to the Hubble time. The overall normalization  is  fixed by requiring that
the value of $R_m$ at $z=0$ agrees with the local rate estimated from the O1 LIGO observation run and the O2 LIGO/Virgo observation run~\cite{LIGOScientific:2018mvr}, using  the median rates obtained from the GstLAL pipeline. The result  depends on the  assumption for the mass distribution of the neutron stars. For  a flat mass distribution
\be\label{rateRmflat}
R_m(z=0)=662 \, {\rm Gpc}^{-3}\, {\rm yr}^{-1}\, ,
\ee
while for a Gaussian mass distribution
\be\label{rateRmGau}
R_m(z=0)=920 \, {\rm Gpc}^{-3}\, {\rm yr}^{-1}\, .
\ee
In the following we will refer to them  just as the ``O2 rates'', and we will 
give our results both for the flat distribution and for the Gaussian distribution.
To  have a quantitative measure of how the results depend on our astrophysical assumptions, we will 
also generate alternative catalogs of GW events by assuming  a Madau-Dickinson star formation rate~\cite{Madau:2014bja} and an exponential time delay between formation and merger with an e-fold time of 100~Myr~\cite{Vitale:2018yhm}. This will also allow us to compare with  
the results presented in Table~1 of ref.~\cite{Sathyaprakash:2019rom}, which also computes the number of BNS detected  per year at the HLVKI network  and at the ET+CE+CE network under these astrophysical assumptions [and using a local comoving BNS merger rate of $ 1000\, {\rm Gpc}^{-3} {\rm yr}^{-1}$, that is close  to our rate for a Gaussian mass distribution, \eq{rateRmGau}]. 

\begin{figure}[t]
 \centering
 \includegraphics[width=0.65\textwidth]{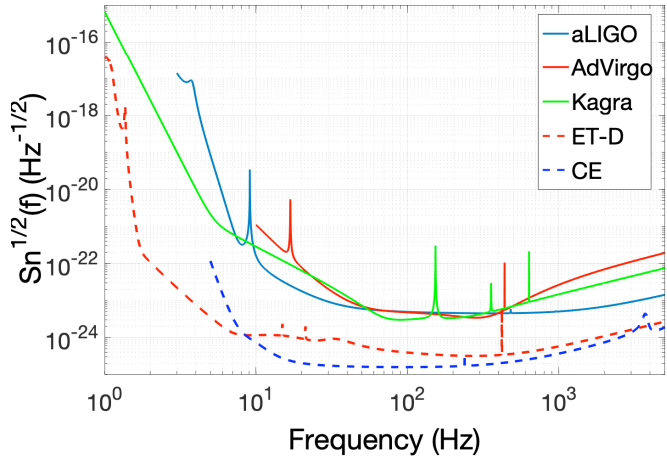}
 \caption{The strain sensitivities of advanced and 3G GW detectors. For ET we use the ET-D sensitivity curve.}
  \label{fig:noise}
\end{figure}

Next, for each BNS generated by this procedure, we determine if its resultant GW emission is detectable with a given GW detector network. We consider three cases: (1) a 2G network composed by   advanced LIGO-Hanford+advanced LIGO-Livingston+advanced Virgo+Kagra+LIGO India (HLVKI). (2) A single 3G detector, chosen according to current estimates for the sensitivity of the  Einstein Telescope. (3) A three-detector network made by ET and two CE. The sensitivity curves that we use are shown in Fig.~\ref{fig:noise}.\footnote{The ET and CE sensitivity curves, as well as the assumed locations of ET (in Europe) and two CE (in the US) correspond to the choices currently used to develop the Science Case for 3G detectors.}
The
signal-to-noise ratio (SNR), $\rho_a$, detected by matched filtering with an optimum filter
in the ideal case of Gaussian noise, in a detector labeled $a$, is
\begin{equation}
\rho_{a}^2 = 4 \int_0^\infty df\, \frac{|F_{+,a}\tilde{h}_{+}
 +F_{\times,a}\tilde{h}_{\times}  |^2}{S_{n,a}}\, ,
\end{equation}
where $f$ is the GW frequency in the observer frame,
$\tilde{h}_{+}$ and $\tilde{h}_{\times}$ the Fourier transforms
of the GW strain amplitudes of $+$ and $\times$ polarizations, $F_{+,a}$ and $F_{\times,a}$ are the
antenna response functions to the GW $+$ and $\times $ polarizations,
and $S_{n,a}(f)$ is the one-sided noise power spectral density (PSD)
of detector $a$. The coherent SNR, assuming uncorrelated noises among the detectors, is simply given by the
quadrature sum of the individual SNRs, $\rho_{\rm tot}^2 = \sum_a \rho_a^2$. The triangular configuration of ET provides three independent differential signals between the arms, equivalent to three detectors, and again the  coherent SNR is given by the
quadrature sum of the individual SNRs
for these three equivalent detectors.

For low-mass systems such as BNS the SNR in one detector is dominated by the inspiral part of the signal and is then given
\begin{equation}
\rho_a^2 = \frac{5}{6} \frac{[G\mathcal{M}(1+z)]^{5/3} {\cal F}_a^2}{c^3 \pi^{4/3} d_L^2(z)} \int_{f_{\mathrm{min}}}^{f_{\rm insp}(z)} df \,  \frac{f^{-7/3}}{S_{n,a}(f)}\, .
\label{eq:SNR}
\end{equation}
Here $\mathcal{M}$ is the intrinsic chirp mass, a combination of the two component masses, $d_L(z)$ is the luminosity distance, $G$ is the gravitational constant, $c$ is the speed of light, $f_{\min}$ is the low frequency limit of the detector and $f_\mathrm{insp}(z)=f_\mathrm{insp}/(1+z)$ is the observed (redshifted) gravitational-wave frequency at the end of the inspiral phase.
The factor
\begin{equation}
{\cal F}_a^2 = \frac{(1+\cos^2 \iota)^2}{4}  F_{+,a}^2 + \cos^2 \iota \, F_{\times,a}^2\, ,
\end{equation}
characterizes the detector response.  In order to decide which detectors contribute to the combined SNR, we assume that each detector has a duty cycle of
80\%. We then classify  the event as detectable
if the combined SNR among the detectors in the network, $\rho_{\rm  tot}$, is larger than a SNR threshold level, that we take to  be $\rho_{\rm  threshold}=12$.\footnote{Notice that we do not ask for the signal to be seen at least in two detectors, but we only use the SNR cutoff. In any case, for 5 detectors, the probability to have just one detector online is less than $1\%$ [more precisely, $(0.2)^4\times 0.8\times 5=0.64\%$] , so our HLVKI catalogs are basically not affected by the inclusion of events seen in just one detector.}    Both for the HLVKI network and for 3G detectors we assume for definiteness  10 years of running (which, given the 80\% duty cycle for each detector, corresponds to a shorter stream of actual coincident data). This assumption should be taken as a limiting case (which,  for the 2G case, is also necessary to have a sufficient sample of events to obtain the convergence of our MCMC); however, the results for a shorter time span $T$ can be obtained basically by rescaling our results by a factor $\sqrt{T/{\rm 10\, yr}}$, corresponding to the fact that, with a large number of events $N$, the error scales roughly as $1/\sqrt{N}$.

In order to generate our mock catalogs of measured luminosity distances of standard sirens, we assume a fiducial $\Lambda$CDM model with $\oma=0.3087$ and $H_0=67.64 \, {\rm km}\, {\rm s}^{-1}\, {\rm Mpc}^{-1}$, corresponding to the mean values obtained from the  CMB+BAO+SNe dataset that we will use,  which is presented in  detail in Section~\ref{sect:LCDM}.
Extracting randomly the redshift of the source from the theoretical distribution obtained from
\eqst{red1}{RnzRf} and using our fiducial cosmological model, we obtain a value of $d_L(z)$ for each source. To take into account the observational error in the reconstruction of the luminosity distance from the GW data, we scatter randomly the values of $d_L(z)$ according to a Gaussian distribution with a width $\Delta d_L(z)$ equal to the expected error in the reconstruction. For each generated event, this is estimated from
$\Delta d_L/d_L=1/{\rm SNR}$, following e.g. ref.~\cite{Dalal:2006qt}.
 Note  that, comparing with the result of an actual mock parameter reconstruction,  one finds that,  because of the degeneracy with the inclination angle,
this can result in an
underestimate of the actual value of  $\Delta d_L/d_L$,  by a factor which has a significant scatter from event to event, but is generically $\sim 2$~\cite{Nissanke:2009kt}. However, for GW signals detected in coincidence with a GRB (which are the signals that  we consider in this paper), assuming that the GRB is beamed within an angle of about $25^{\circ}$ one finds that the correlation between distance and inclination is substantially broken, and the above estimate becomes more accurate~\cite{Nissanke:2009kt}.

Beside the instrumental error, we must consider the error due to lensing. Following \cite{Sathyaprakash:2009xt,Zhao:2010sz}, we model it as
\be\label{errorlensing}
\(\frac{\Delta d_L(z)}{d_L(z)}\)_{\rm lensing}\simeq 0.05 z\, ,
\ee 
and we add it in quadrature to the instrumental error.
However, we will see below that, for the sources at $z<1.5$, that will largely dominate our results,  the lensing error is subdominant with respect to the instrumental error.

If the source is at very low redshift, once determined the measured redshift as discussed in Section~\ref{sect:counter}, to obtain the cosmological redshift
we must correct for the peculiar Hubble flow. This is estimated adding an error on $z$ corresponding to  a recessional velocity of the host galaxy of $200$~km/s, as in ref.~\cite{Chen:2017rfc}.

\subsection{Electromagnetic counterpart}\label{sect:counter}

In general, to identify the counterpart, one can consider two possible strategies. The first, that has been implemented successfully with GW170817, consists in having a network of GW detectors,  that allows us to localize  the source relatively well. Then, the follow-up with telescopes working, e.g., in the optical or IR can identify the host galaxy and determine its redshift. The second possibility, that can be applied even when no GW localization is available (as, for instance, with a single ET detector) is to use   the temporal coincidence of the GW event with a short GRB; for many short GRBs, the redshift has indeed been determined from the X-ray afterglow, that can be accurately localized by {\em Chandra} or \emph{Swift}/XRT. For instance, in the sample of 67 \emph{Swift} short GRBs discussed in~\cite{Berger:2013jza}, 53 events were rapidly followed up with the on-board X-ray Telescope, leading to 47 detections of the source.

The estimate of the number of coincidences between GW events and electromagnetic observations depends crucially, of course, on the rate of expected GW events, as well as on the network of GRB satellites and telescopes available at the time. We therefore discuss  the 2G and 3G cases separately.

\subsubsection{GRB coincidences  with the HLVKI network}

We begin by investigating the possibility of detecting in coincidence a GW signal at the HLVKI network and a GRB with the current generation of GRB satellites. We assume here that the Fermi-GBM can make a coincident detection and that \emph{Swift} can slew to the combined GW/GRB error box and identify an X-ray counterpart. We note here that for 170817A \emph{Swift} was occulted by Earth at time of Fermi trigger, so imaging by the X-ray telescope (XRT) took place around 1\,hr post trigger. At that time it was able to cover 90\% of the GW skymap to rule out any bright sources \citep{evans_swift_2017}.

For a GRB detected in coincidence with a GW signal we
require that the peak flux is above the flux limit of the satellite. Based on the modeling of \citep{Howell_rates_2018} we assume a Gaussian structured jet profile model of GRB170817A given by
\begin{equation}
L (\theta_{\mathrm{V}}) = L_{\mathrm{c}} \exp \left( -\frac{\theta_{\mathrm{V}}^{2}}{2\theta^{2}_{\mathrm{c}}} \right)
 \,, \label{eq_structure_gaussian}
\end{equation}
\noindent with $L (\theta)$ the luminosity per unit solid angle, $\theta_{\mathrm{V}}$ the viewing angle and  $L_{\mathrm{c}}$ and $\theta_{\mathrm{c}}$ structure parameters that define  the angular profile. The structured jet parameter is given by $\theta_{\mathrm{c}} = 4.7$\deg. The value of $ L_{\mathrm{c}}$ is given by $L_{\mathrm{c}} = L_{\mathrm{p}}/4\pi$ erg s$^{-1}$ sr$^{-1}$, where $L_{\mathrm{p}}$ is the peak luminosity of  each burst, which is obtained by sampling $\Phi (L_p)dL_p$.
We assume the standard broken power-law distribution of the form
\begin{equation}
\Phi(L_{\mathrm{p}}) \propto
\biggl \lbrace{
\begin{array}{ll}
\hspace{1.0mm}\left(L_{\mathrm{p}}/L_{*}\right) ^{\alpha}\, , \hspace{10.0mm} L_{\mathrm{p}} < L_{*}\\
\hspace{1.0mm}\left(L_{\mathrm{p}}/L_{*}\right) ^{\beta}\, ,  \hspace{10.0mm} L_{\mathrm{p}} \geq L_{*}
\end{array} }
\end{equation}

\noindent where $L_{\mathrm{p}}$ is the peak luminosity assuming isotropic emission   in the rest frame in the 1-10000\,keV energy range,  $L_{*}$ is a characteristic value separating the two regimes, and the slopes describing these regimes are given by $\alpha$ and $\beta$  respectively. Following \citep{WandermanPiran2015MNRAS} we use the values $\alpha = -1.95$, $\beta = -3$ and $L_{*} = 2\times 10^{52}$\,\lum. Given a source at luminosity distance $d_{L}$ one can convert $4\pi L(\theta_{\mathrm{V}})$ to an observed peak flux as a function of viewing angle, $F_{P}(\theta_{\mathrm{V}})$, obtained from the value of the GW inclination angle. A Fermi-GBM detection is recorded if the value of $F_{P}(\theta_{\mathrm{V}})$ is greater than the flux limit of 1.1\,\pflux\, in the 50-300 keV band for Fermi-GBM \citep{Howell_rates_2018}, noting that 95\% of the bursts detected in the 64\,ms timescale are within this limit.\footnote{See the Fermi-GBM burst catalogue \url{https://heasarc.gsfc.nasa.gov/W3Browse/fermi/fermigbrst.html}} We further assume the total time-averaged observable sky fraction of the Fermi-GBM, which is 0.60 \citep{Burns2016ApJ}. Using this procedure, among the events in the GW BNS catalog generated as discussed in sect.~\ref{sect:GWevents}, we select those that have an observed  GRB counterpart.

\begin{figure}[h]
 \centering
 \includegraphics[width=\textwidth]{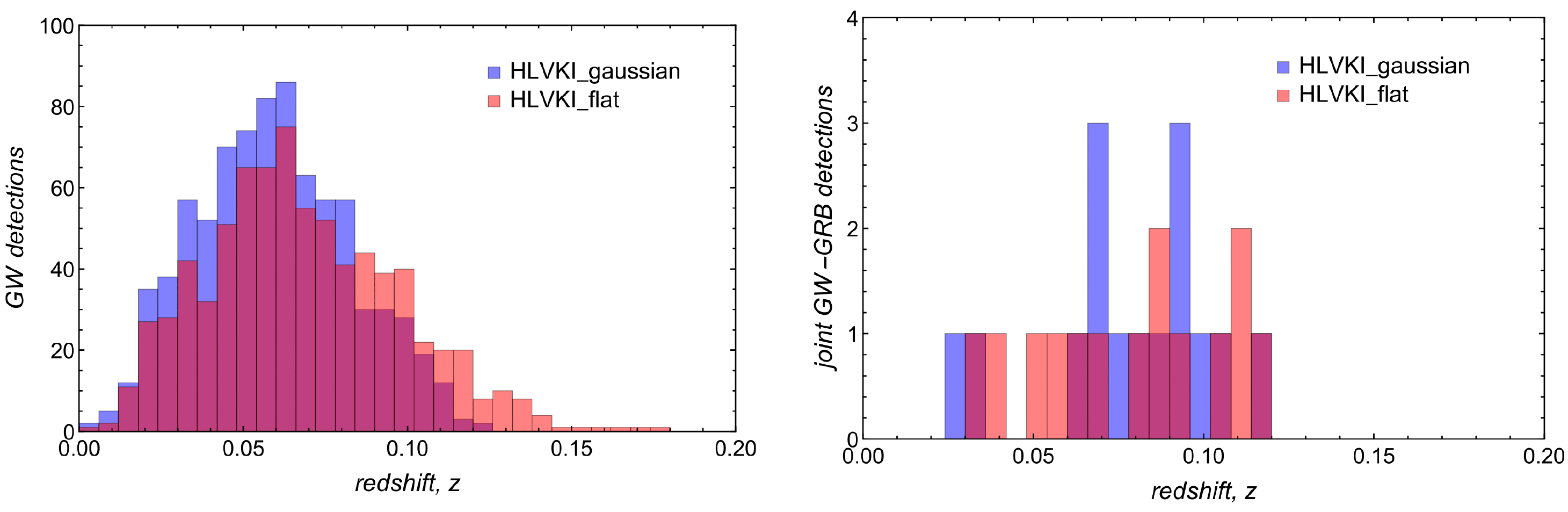}
 \caption{Left panel: the redshift distribution of 10 yr of GW events  from a  realization of the mock catalog  at the  HLVKI network,  for the flat and the Gaussian mass distribution. Right panel:
the redshift distribution of 10-years of GW-GRB coincidences between the  HLVKI network and  the current generation of GRB satellites.}
  \label{fig:HLVKI_histogram}
\end{figure}

Table~\ref{tab:2G} shows the number of BNS sources along with the number of coincident GRB detections determined using the procedure above, for the HLVKI network.  We see that 10 years of observation would yield of order 14-15 joint detections. In   Fig.~\ref{fig:HLVKI_histogram} we show the redshift distribution of the GW events (left panel) and of the  GW-GRB coincidences (right panel), for a realization of our catalog. In app.~\ref{app:catalog} we give, in Table~\ref{tab:cat2G}, the explicit values of $z, d_L$ and $\Delta d_L$ for  the events in the specific catalog corresponding to the right panel of Fig.~\ref{fig:HLVKI_histogram}, where $d_L$ is the `measured' luminosity distance reconstructed from $z$ using our fiducial $\Lambda$CDM model and scattering randomly those fiducial values of $d_L(z)$ according to a Gaussian distribution with a width equal to the error $\Delta d_L(z)$ (as explained in Section~\ref{sect:GWevents}).

To test the impact of changing our astrophysical assumptions, we have also generated a  catalog of GW events assuming a Madau-Dickinson star formation rate  and an exponential time delay between formation and merger with an e-fold time of 100~Myr, as in
ref.~\cite{Sathyaprakash:2019rom}. In this case, assuming again a duty cycle of 80\% and a network SNR threshold level $\rho_{\rm  threshold}=12$, we find 
that the number of BNS detected  at the HLVKI network, for the Gaussian mass distribution,  is 64/yr, to be compared with the value
48/yr reported in ref.~\cite{Sathyaprakash:2019rom}. 
%\red{To be understood with what duty cycle and threshold; Samaya, Sathya?}

\vspace*{5mm}

\begin{table}[h]
\begin{center}
\begin{tabular}{l|cc|cc}
\hline
\multicolumn{1}{c}{Network} & \multicolumn{2}{|c}   {GW events}                       & \multicolumn{2}{|c}{Joint GW-GRB events}  \\ \hline
\multicolumn{1}{c|}{} &  \multicolumn{1}{c}{Flat} & \multicolumn{1}{c}{Gaussian} & \multicolumn{1}{|c}{Flat} & \multicolumn{1}{c}{Gaussian}\\ \hline \hline
          HLVKI          &       768                &       814               &    14                  &       15               \\ \hline
\end{tabular}
\caption{Number of GW events detected by second generation (2G) networks in 10 years, and
the expected GW-GRB coincidences obtained by assuming a GRB detector with the characteristics of  Fermi-GBM. We show detection rates for BNS populations generated using O2 rates corresponding to both flat and Gaussian mass distributions.
\label{tab:2G}}
\end{center}
\end{table}

%\clearpage

\subsubsection{GRB coincidences with  3G detectors}\label{sect:count3G}

For  3G detectors the estimates are of course more uncertain. Indeed, the identification of the counterpart depends
on the network of GRB satellites and of telescopes  at the time when 3G detectors  will operate, as well as on issues that are presently difficult to predict, such as the prioritization that will be given by various telescopes to the follow-up of GW signals.

The  proposed THESEUS mission~\cite{Amati:2017npy,Stratta:2017bwq,Stratta:2018ldl} could be particularly useful for performing coincidences between GW events and GRBs, even in the absence of localization from the GW signal.
A crucial difference with the 2G case is that, at the sensitivity level of 3G detectors, there will be many more GW events compared to what GRB satellites could detect. The main reason for this is that the GRB instruments are limited by their flux sensitivity for the more distant GRB emissions at wider viewing angles.
For instance, it was estimated in \cite{Sathyaprakash:2009xt} that ET will be able to detect  ${\cal O}(10^5-10^6)$ BNS mergers  per year.  As we will see below our estimate, given in Table~\ref{tab:3G}, is
that ET will detect about $(6-7)\times 10^5$ events in 8~yr of actual data taking, corresponding to a rate
of order $(0.8-0.9)\times 10^5/ {\rm yr}$, consistent with previous estimates, although somewhat smaller.
In any case, according to the estimate in \cite{Stratta:2017bwq}, only about $15-35$ coincident short GRB (sGRB) per year will be detected by THESEUS with its X-Gamma ray Imaging Spectrometer (XGIS); we will see below that our results, using somewhat different assumptions for the GRB luminosity and BNS rate, gives a slightly higher number of coincidences, but still of this order of magnitude.
Beside the collimated prompt GRB emission, more isotropic soft X emission is also expected from the afterglow. This could be detected by the Soft X-ray Imager (SXI) on board THESEUS, leading  possibly to a few hundred more detections per year~ \cite{Stratta:2017bwq}. In any case, the number of joint GW-GRB detections will be a very small fraction of the number of GW events.

With a network of at least three GW detectors, accurate localization of the GW signal becomes possible, allowing for electromagnetic follow-up observations, that could determine the  redshift of the source.
 If the source localization is already available through GWs, LSST could detect the counterpart up to $z\simeq 0.55$ and WFIRST up to $z\simeq 0.76$, and many more telescopes in the UV, optical, IR, radio
could detect the counterpart at smaller redshifts, say $z\sim 0.1-0.3$.
However, the follow-up of $O(10^3)$ well localized GW events at $z\sim 0.5$ would require the equivalent of 1~yr of  dedicated LSST time, which is not realistic. Currently, a more realistic estimate is that LSST might use of order of $1\%$  of its time for GW follow up, so
it will be challenging for  LSST to  observe more than ${\cal O}(10)$ counterparts per year, at $z\sim 0.5$. The localization cost is much smaller at $z\sim 0.1$, where ${\cal O}(100)$  events per year could be a more realistic expectation, but this will depend on the science prioritization in the 2030s, when 3G detectors will hopefully operate.\footnote{We thank Matthew Bailes, Samaya Nissanke and Bangalore Sathyaprakash for discussions on these issues.}
Given these large uncertainties, in this paper we will limit ourselves to the coincidences with GRB detectors. 

We repeat our simulations for the coincidence with a single ET detector and with a ET+CE+CE network,  assuming that a THESEUS type satellite will be used for coincidence searches. For the GRB detection we assume a duty cycle of 80\% due to a reduction of 20\% as the satellite passes through the Southern Atlantic Anomaly, a flux limit of 0.2\,\pflux\, in the 50--300 keV band and a sky coverage fraction of 0.5 \citep{Stratta:2018ldl}. We note that the XGIS will be able to localise sources to around 5 arcmin only within the central 2~sr of its field of view (FOV); outside this central region localisation will be coarse at best.\footnote{We thank Giulia Stratta for valuable discussions on the localisation capabilities of THESEUS-XGIS.} We therefore consider two scenarios: one, that we will denote as `optimistic', in which all the events detected by XGIS have a measured redshift, and one, that we will denote as  `realistic',
where we assume that only around 1/3 of the sGRBs detected by XGIS could provide redshift estimates.

\subsection{Events rates, redshift distributions and analytic estimates for $\Delta d_L(z)/d_L(z)$}
\label{sect:estimates}

In our MCMCs we will use a given realization of the catalog of events obtained with the procedure discussed above. It is however useful to describe the qualitative features of these catalogs, such as the redshift distributions of the events and the average value of $\Delta d_L(z)/d_L(z)$ as a function of redshift. This will provide a physical insight into which sources contribute most, to compare with previous works, and to provide ready-to-use formulas that can be  applied to future studies.

\subsubsection{Events at a single ET detector}

\begin{table}[]
\begin{center}
\begin{tabular}{l|cc|cc}
\hline
\multicolumn{1}{c}{Network}  & \multicolumn{2}{|c}{GW events}                          & \multicolumn{2}{|c}{Joint GW-GRB events }   \\ \hline
\multicolumn{1}{c|}{} &  \multicolumn{1}{c}{Flat}  & \multicolumn{1}{c}{Gaussian}  & \multicolumn{1}{|c}{Flat} & \multicolumn{1}{c}{Gaussian} \\ \hline \hline
          ET            &     621,700                  &    688,426                   &      389  (128)              &          511   (169)          \\
          ET+CE+CE          &      5,420,656                &   7,077,131                   &    644  (213)               &       907     (299)          \\ \hline
\end{tabular}
\caption{Number of GW BNS events detected by third generation (3G) networks in 10 years of data taking (assuming a $80\%$ duty cycle for each detector) and the corresponding GW-GRB coincidences obtained by assuming a GRB detector with the characteristics of THESEUS-XGIS; numbers in parenthesis show the number of sources with arcmin localisation. BNS populations are generated using the O2 rates corresponding to `flat' and `Gaussian' mass distributions.
\label{tab:3G}}
\end{center}
\end{table}

\begin{figure}
 \centering
 \includegraphics[width=1.0\textwidth]{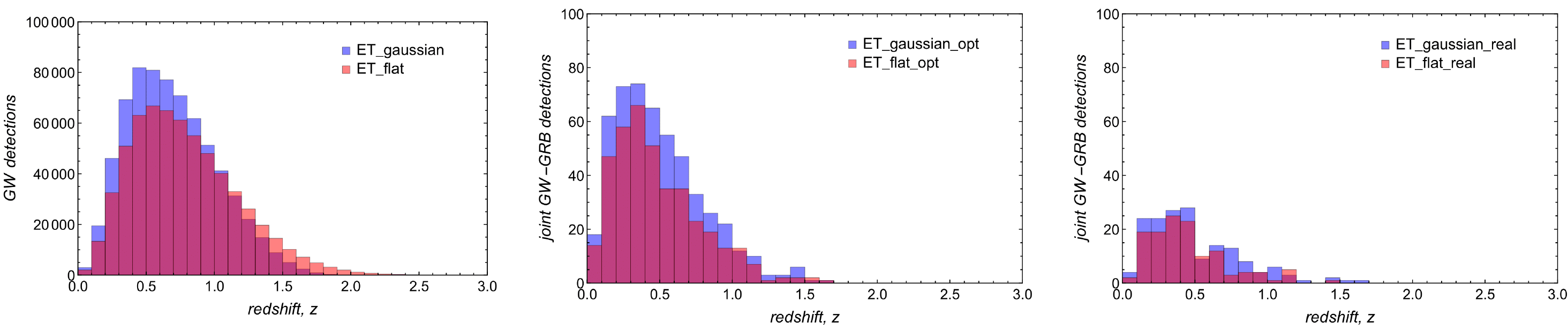}
 \caption{Left panel: the redshift distributions of 10-years of BNS detections  by a ET detector.  Middle panel: the coincident detections made by THESEUS in the `optimistic' scenario for the FOV. Right panel: the coincident detections  in the `realistic' scenario. Notice that the vertical scale for the left panel is very different from that in the middle and right panels.
 }
  \label{fig:redshift_dist}
\end{figure}

\begin{figure}
 \centering
 \includegraphics[width=0.45\textwidth]{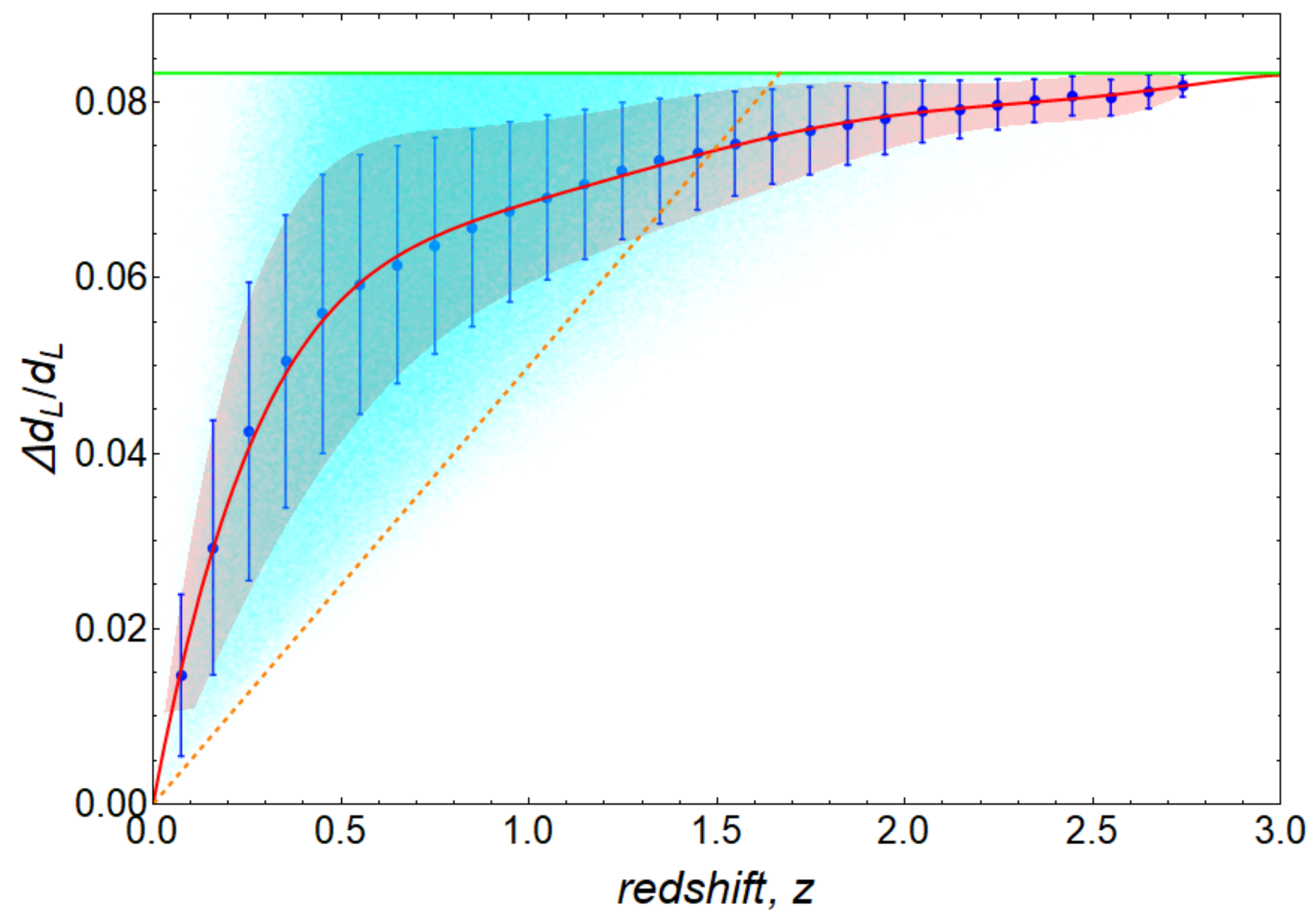}
 \includegraphics[width=0.45\textwidth]{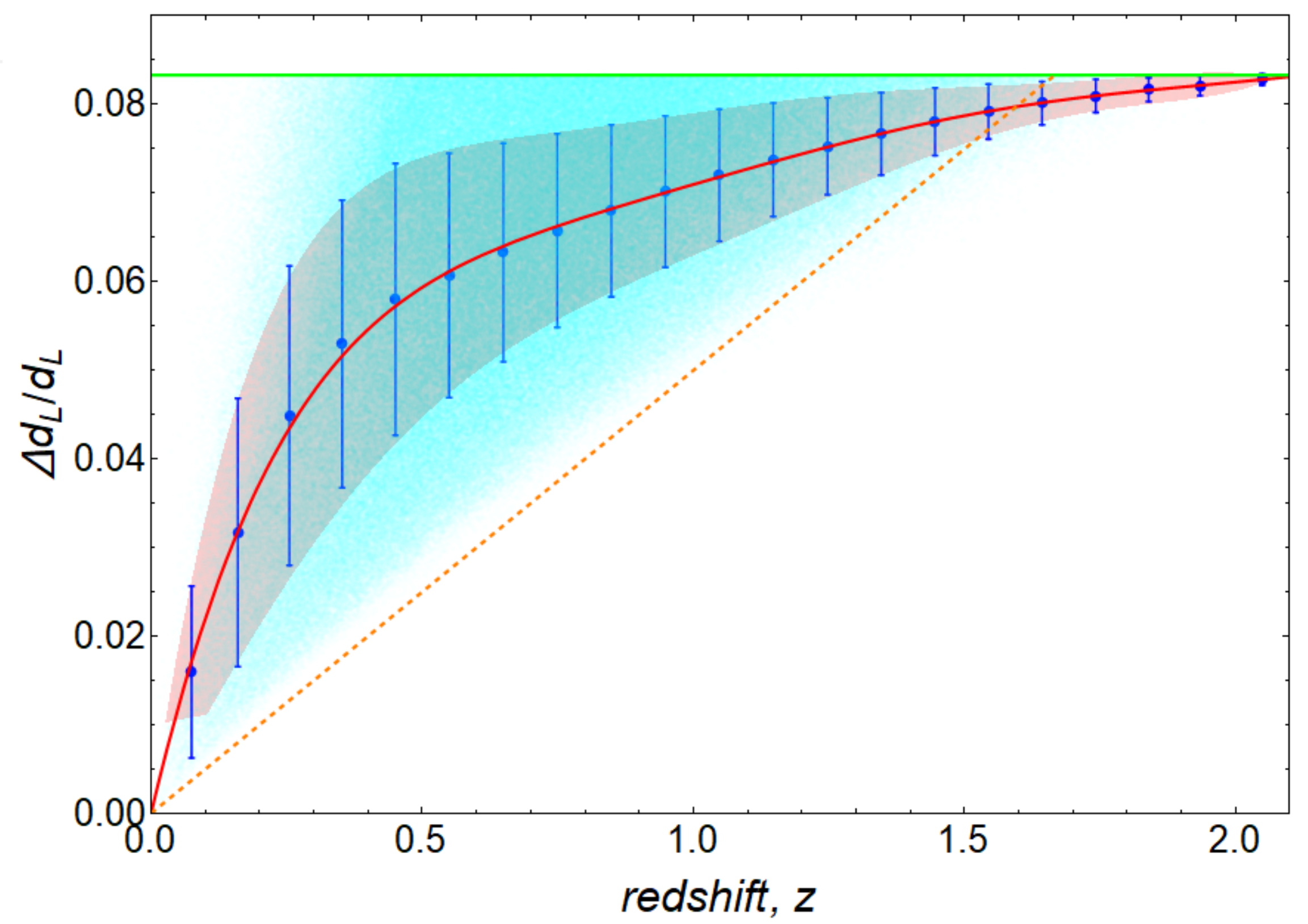}
 \caption{The ET instrumental contribution to the relative error $\Delta d_L/d_L$ for the  specific realization of the catalog of BNS detections shown in
Fig.~\ref{fig:redshift_dist}. All the GW events, with or without a detected EM counterpart, are taken into account. Left panel: for the flat distribution of neutron star masses.  Right panel: for the Gaussian distribution of neutron star masses. In each panel, the  cyan shaded area corresponds to all the BNS events, while the coordinates of the blue points are given by the mean values of the redshift and of  $\Delta d_L/d_L$ in each redshift bin, with the bins chosen as  in Fig.~\ref{fig:redshift_dist}. The blue error bars are the standard deviations of 
$\Delta d_L/d_L$ in each redshift bin. The horizontal green line at  $\Delta d_L/d_L=1/12$ is fixed by the SNR threshold of 12. The red line and the light red region are the fits to the blue points and error bars, given explicitly in the text.  The orange dotted line is the error on $\Delta d_L/d_L$  induced by lensing.
The redshift ranges shown in the two panels differ as a result of the different maximum values of redshift reached in the two corresponding catalogs. }
  \label{fig:ET_rel_err}
\end{figure}

Table \ref{tab:3G} shows the results of our simulations for the 3G era in terms of the number of GW signals from BNS, along with the number of  joint GW/sGRB detections; the number of events with arcmin localisation are shown in parenthesis. 

For a  single ET detector  our estimate of the rate of BNS detection is between $6.2\times 10^5$ and $6.9\times 10^5$ events in 10 yr, having assumed a duty cycle of 80\%, which, in the case of a single detector, corresponds to 8~yr of actual data. This corresponds to a rate (normalized to the actual time of data taking)
\be
R\simeq (0.8-0.9)\times 10^5\, {\rm BNS/yr}\, ,
\ee
consistent with previous estimates, although somewhat smaller. This can be traced to the fact that 
we have used a threshold for the network SNR, obtained by combining the three arms of ET, given by $\rho_{\rm  threshold}=12$, while previous work, e.g. 
ref.~\cite{Zhao:2010sz}, used $\rho_{\rm  threshold}=8$.
We also see from Table~\ref{tab:3G}
that,  with a  single ET detector, we should expect around $39-51$ coincident sGRB/GW events in one year of observation using the XGIS and the SXI detectors. These numbers differ from the $15-35$ events quoted  in \citep{Stratta:2017bwq}
 for two main reasons. Firstly, the assumed luminosity function and BNS rate differs from that assumed in \citep{Stratta:2017bwq}. Secondly, our calculations assume a structured jet profile based on GRB170817A. From our sample of detections one could expect around $13-17$ events yr$^{-1}$ to have arcmin localisations.

Figure \ref{fig:redshift_dist} shows the redshift distributions of 10-years of BNS detections using ET, along with the joint GW-GRB distributions with optimistic and realistic scenarios for the FOV of THESEUS.
We see that the vast majority of the joint GW-GRB detections is at $z \, \lsim \, 1$. In this realization of the catalog, the ET event with the highest redshift is at $z\simeq 2.91$ for the flat mass distribution and at $z\simeq 2.10$ for the Gaussian mass distribution, while
the  joint detection with the highest redshift is at $z\simeq 1.63$ (found both in optimistic and realistic scenario for the THESEUS FOV, and for the Gaussian mass distribution).
We find that the higher-$z$ sGRB detections have viewing angles close to the jet axis corresponding with almost face-on BNSs; this selection effect was highlighted in \citep{Regimbau:2014nxa}.

Figure \ref{fig:ET_rel_err} shows the ET instrumental contribution to the relative error on luminosity distance, considering all the events in the  specific realization of the catalog of BNS detections presented in the left panel of Fig.~\ref{fig:redshift_dist}. The events are organized in the same redshift bins as Fig.~\ref{fig:redshift_dist} and, for the events belonging to each bin, the mean value of $\Delta d_L/d_L$ and its standard deviation  $\sigma(\Delta d_L/d_L)$  are evaluated. Of course, the mean value of $\Delta d_L/d_L$ increases with distance, until we reach the threshold at SNR=12, and therefore the value $\Delta d_L/d_L=1/12$, beyond which we no longer record the triggers as detections. In contrast, its variance eventually decreases; this is due to the fact that, in a given bin at some intermediate redshift, we have  events with different possible orientations with respect to the detector, and therefore different SNR. As the redshift increases toward the horizon of the detector, only the events with optimal orientation can go above the threshold.

It is useful to  provide a fit for the mean value and for the standard deviation of $\Delta d_L/d_L$, as a function of the redshift. In order to increase the significance of the fit, in the case of flat mass distribution the two highest redshift bins containing events of the catalog shown in Fig.~\ref{fig:redshift_dist} have been excluded (they only contain a very low number of events and we do not show them in the figure). Using sixth degree polynomials intercepting zero, for the flat distribution (f) of neutron star masses, the fit to the mean value of $\Delta d_L/d_L$ and to its standard deviation
are given by 
\be\label{eq:mean_ET_flat}
\overline{\left(\frac{\Delta d_L}{d_L}\right)}_{\rm f}=0.230896 z -0.345265 z^2 +0.282128 z^3-0.124715 z^4+0.028027 z^5-0.002508 z^6 \, ,
\ee
\be\label{eq:SD_ET_flat}
\mathrm{\sigma}{\left[\left(\frac{\Delta d_L}{d_L}\right)_{\rm f}\right]}=0.129548 z-0.334210 z^2+0.369258 z^3-0.205939 z^4+0.056642 z^5-0.006100 z^6 \, .
\ee
In the case of Gaussian distribution (G) the corresponding fits  are given by
\be\label{eq:mean_ET_gaussian}
\overline{\left(\frac{\Delta d_L}{d_L}\right)}_{\rm G}=0.261577 z - 0.456248 z^2 + 0.442636 z^3-0.232171 z^4+0.061693 z^5 -0.006502 z^6 \, ,
\ee
\be\label{eq:SD_ET_gaussian}
\mathrm{\sigma}{\left[\left(\frac{\Delta d_L}{d_L}\right)_{\rm G}\right]}=0.154942 z - 0.490698 z^2+0.683342 z^3-0.490253 z^4+0.175418 z^5 - 0.024721 z^6 \, .
\ee
Tables~\ref{tab:relerr_ET_flat_opt}, \ref{tab:relerr_ET_gaussian_opt}, \ref{tab:relerr_ET_flat_real} and \ref{tab:relerr_ET_gaussian_real}, in App.~\ref{app:catalog}, list the  ET instrumental contribution to the relative error 
$\Delta d_L/d_L$ in each redshift bin, for the realizations of the catalogs of joint GW-GRB detections displayed in the middle and right panels of Fig.~\ref{fig:redshift_dist}.

\begin{figure}[t]
 \centering
 \includegraphics[width=0.55\textwidth]{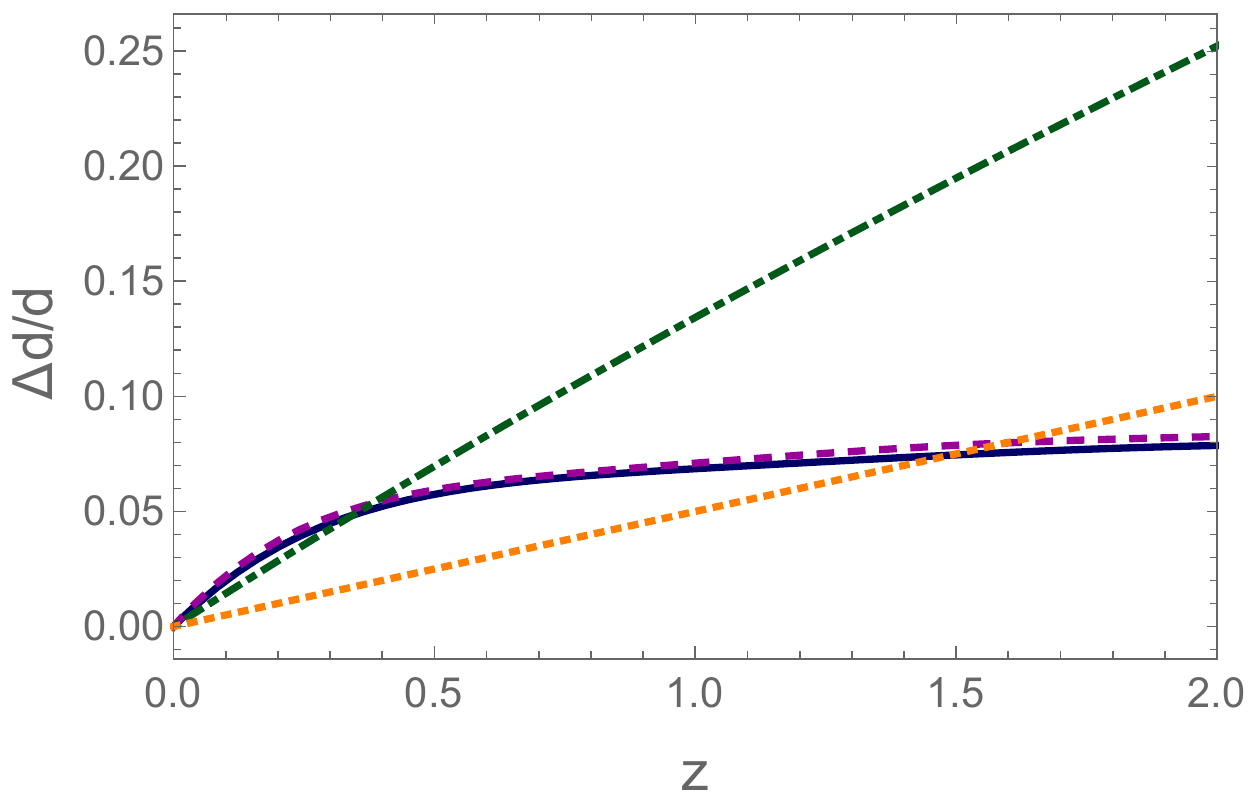}
 \caption{Our fits to $\Delta d_L/d_L$ given in \eq{eq:mean_ET_flat} (blue solid line)  
 and in \eq{eq:mean_ET_gaussian} (magenta dashed line) compared to the error due to lensing
 given in  \eq{errorlensing} (orange dotted line) and to the fit (\ref{eq:fitZhao}) to the instrumental error given in 
 ref.~\cite{Zhao:2010sz} (green dot-dashed line).
 }
  \label{fig:compareFits}
\end{figure}
 
It is instructive to compare our fits to the instrumental error on $\Delta d_L/d_L$, with the fit  to the instrumental error found in \cite{Zhao:2010sz}, which is 
\be
\overline{\left(\frac{\Delta d_L}{d_L}\right)}=0.1449 z - 0.0118  z^2+0.0012 z^3\, ,
\label{eq:fitZhao}
\ee
as well as  with the lensing error (\ref{errorlensing}). The comparison is shown in Fig.~\ref{fig:compareFits}. We see first of all that our fits to the instrumental error are consistent with that of
ref.~\cite{Zhao:2010sz} at low redshifts, say $z\, \lsim \, 0.3$, but are smaller at large redshift. The sensitivity curve for ET used in ref.~\cite{Zhao:2010sz} [see their eq.~(19)] is relatively close to the one that we use, shown in Fig.~\ref{fig:noise}. The reason for the difference is  rather that we use a SNR threshold  $\rho_{\rm  threshold}=12$, while  
ref.~\cite{Zhao:2010sz} uses $\rho_{\rm  threshold}=8$. On the one hand, this implies that we get less events, and up to smaller redshift, compared to ref.~\cite{Zhao:2010sz}. On the other hand, since $\Delta d_L/d_L$ is estimated as $1/{\rm SNR}$, the events that we retain are those with smaller value of $\Delta d_L/d_L$ and therefore also the average value of $\Delta d_L/d_L$ over the events in a redshift bin is smaller. This effect becomes more important as $z$ increases, as  there are many more events that are sub-threshold with respect to $\rho_{\rm  threshold}=12$, but above threshold with respect to $\rho_{\rm  threshold}=8$.

It is also interesting to compare the lensing contribution to $\Delta d_L/d_L$ to the observational error, in the light of this understanding of the  dependence of the average observational error on the threshold. Comparing the fit obtained in ref.~\cite{Zhao:2010sz} with $\rho_{\rm  threshold}=8$, given by the  green dot-dashed line in Fig.~\ref{fig:compareFits}, 
to the lensing error, given by the orange dotted line, one would be tempted to conclude that, at ET, the error induced by  lensing is negligible at all redshifts. In fact, this statement needs some qualification. The  green dot-dashed line   represents the average of $\Delta d_L/d_L$ over an ensemble of events, in the same redshift bin, selected by requiring ${\rm SNR}\, \geq\, \rho_{\rm  threshold}=8$. However, within each bin, the events that are most useful for cosmological studies  are those with the smallest errors on $\Delta d_L/d_L$, i.e. with the highest SNR. Already selecting only the events with 
${\rm SNR}\, \geq\, \rho_{\rm  threshold}=12$ reduces significantly the instrumental error  averaged over  such events, and  we see that for this ensemble of events lensing becomes larger than the mean value of 
 $\Delta d_L/d_L$  at $z\,\gsim\, 1.5$.

We have seen above that the vast majority of the GW-GRB coincidences are at $z<1$. Therefore, in our analysis, lensing will  indeed be subdominant. However, it must be kept in mind that, if one restricts the analysis to the `golden events', i.e. the loudest and best characterized events in each redshift bin (corresponding to the lower edge of the cyan shaded area in Fig.~\ref{fig:ET_rel_err}), the effect of lensing will become more and more important. The contribution to the error on $\Delta d_L/d_L$ from lensing, estimated as in \eq{errorlensing}, is shown as the orange dotted line in Fig.~\ref{fig:ET_rel_err}. We see that, for $z<1$, it is comparable to the lower edge of the distribution of events given by the cyan shaded area, and therefore its inclusion, in quadrature with the observational error, at most degrades by a factor $\sqrt{2}$ the error on $\Delta d_L/d_L$  on these very few events at the edge of the distribution. These events are interesting because they are those with the smallest error, but for $z<1$ they are extremely rare. 
For larger redshifts the situation is different and we see that, say at $z\simeq 1.5$, for a non-negligible fraction of the events   the error from lensing can be the limiting factor.

\subsubsection{Events at a ET+CE+CE network}

In the second line of Table~\ref{tab:3G} we show the results for  the ET+CE+CE network. This configuration, featuring three 3G detectors, can be considered an extreme case.
Other cases, such as one ET detector and two advanced 2G detectors, will be intermediate between this case and a single 3G detector.\footnote{Currently, are under study a series of improvements of the advanced LIGO detectors, that should eventually lead to the Voyager detectors. The Voyager phase in the US could happen partly in temporal overlap with  ET in Europe.} 
The result for the GW events in Table~\ref{tab:3G} in the Gaussian case corresponds to a detection rate of 
710k events/yr.  To understand the dependence of these results on the astrophysical assumptions, we  have also  generated a  catalog of GW events, for ET+CE+CE, assuming a Madau-Dickinson star formation rate  and an exponential time delay between formation and merger with an e-fold time of 100~Myr. In that case we find 840k events/yr, in  good agreement with the value 990k events/yr found, under similar  hypothesis, but a slightly larger rate $R_m(z=0)=1000 \, {\rm Gpc}^{-3}\, {\rm yr}^{-1}$,  in
ref.~\cite{Sathyaprakash:2019rom}.

For the coincidences with GRBs
we find that, for an ET+CE+CE network, we would get of order $64-90$ coincident events per year, and around $20-30$ events per year  will  have arcmin localisations.
Observe  that, even if the ET+CE+CE network  has a number of GW detections larger than a single ET by a factor ${\cal O}(10)$, the number of coincidences with GRBs is higher only by a factor less than 2. This already tells us that the bottleneck, for joint GW-GRB  detections, is on the GRB side, that cannot keep pace with the GW detection rate of 3G  detectors.

\begin{figure}[t]
 \centering
 \includegraphics[width=1.0\textwidth]{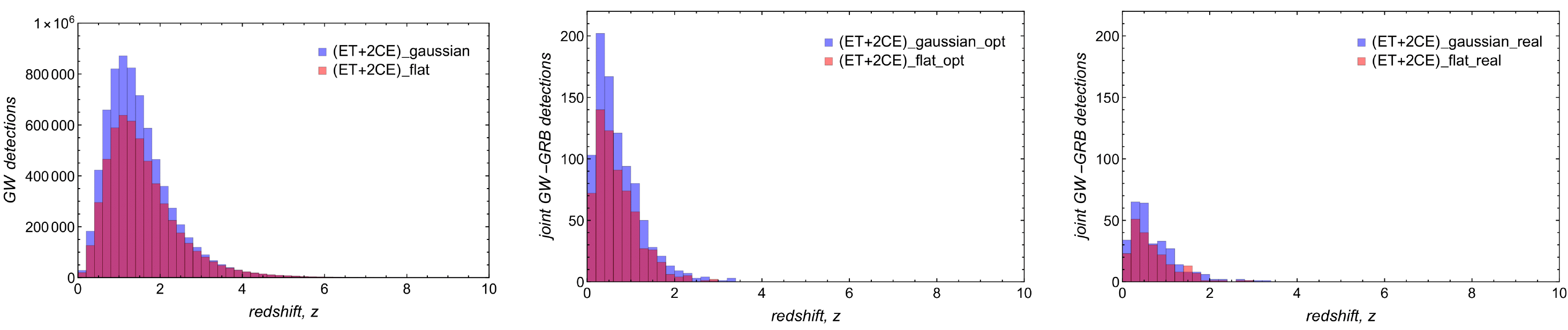}
 \caption{As in Fig.~\ref{fig:redshift_dist} for the ET+CE+CE network and its coincidences with THESEUS. For uniformity, we use the same redshift range in the three panels.
 }
  \label{fig:redshift_distET2CE}
\end{figure}

\begin{figure}[t]
 \centering
 \includegraphics[width=0.45\textwidth]{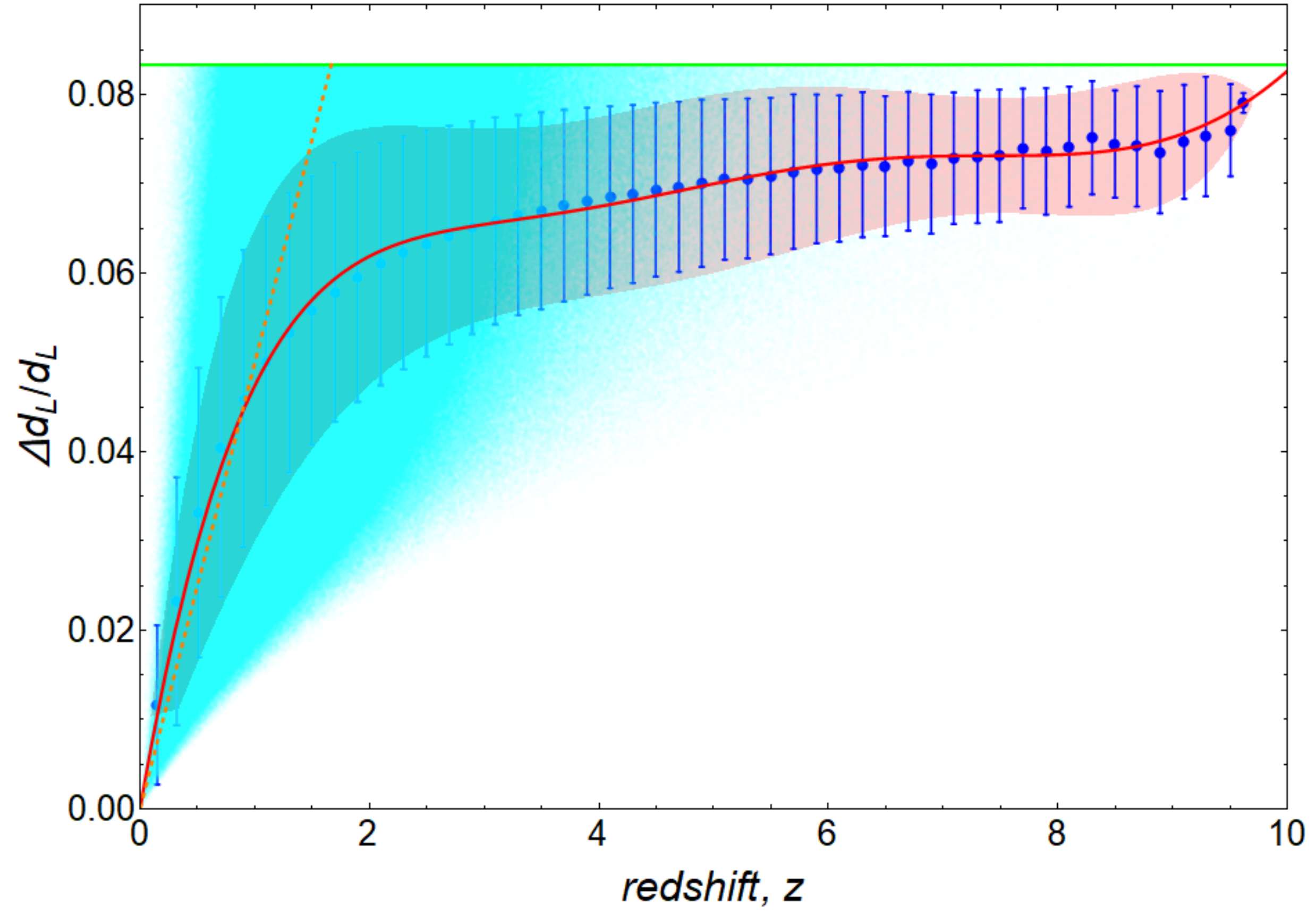}
 \includegraphics[width=0.45\textwidth]{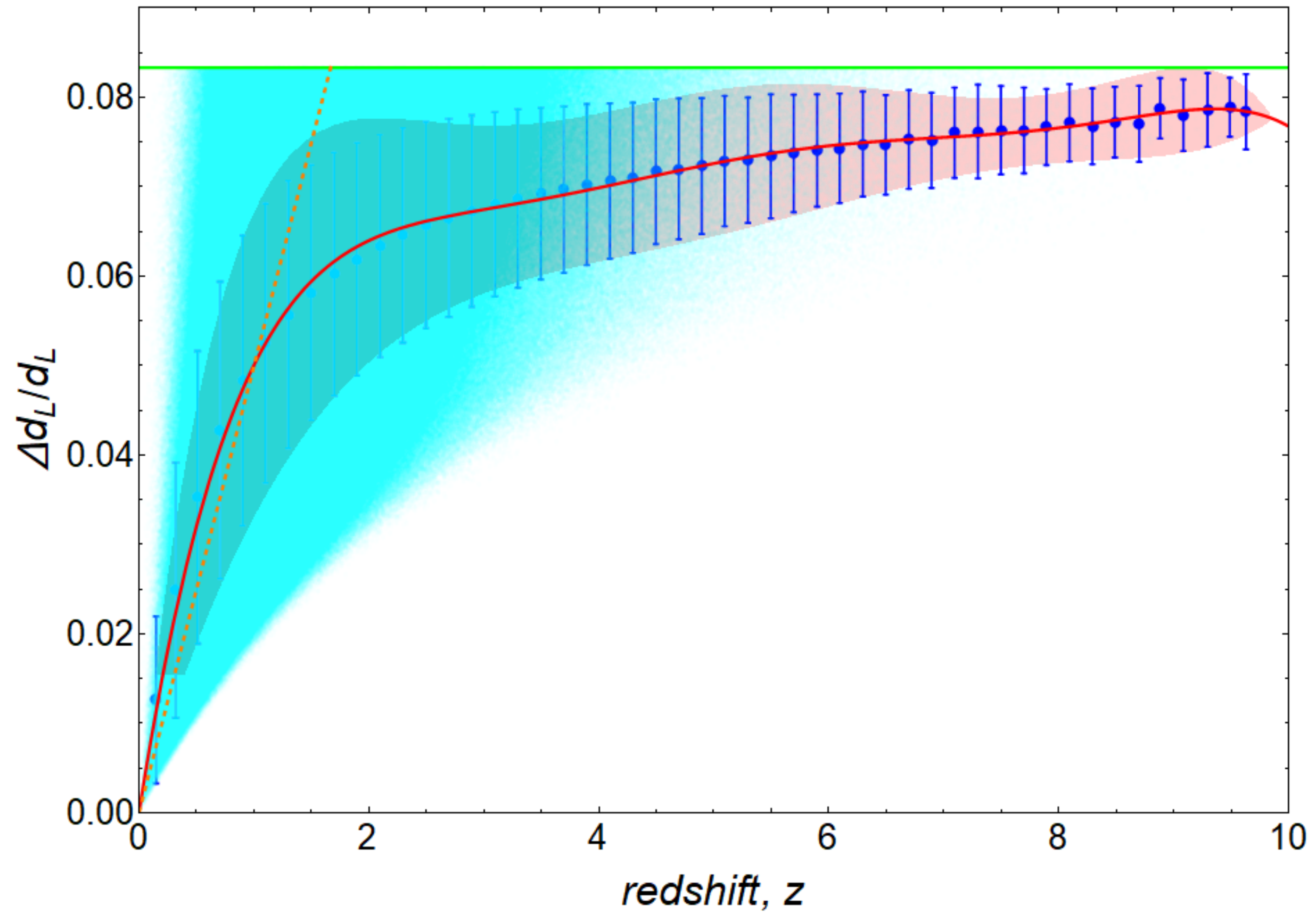}
 \caption{As in Fig.~\ref{fig:ET_rel_err}, for the ET+CE+CE network.}
  \label{fig:ET+2CE_rel_err}
\end{figure}

Figure \ref{fig:redshift_distET2CE} shows the redshift distributions of BNS detections using ET+CE+CE, 
over 10~yr,  along with the joint GW-GRB distributions with optimistic and realistic scenarios for the FOV of THESEUS.  In this realization of the catalog, the ET+CE+CE event with the highest redshift is at $z\simeq 9.63$ for the flat mass distribution and at $z\simeq 9.66$ for the Gaussian mass distribution; this is due to the fact that a single CE already has a reach to BNS of order $z\simeq 9$ (while we have seen above that, for BNS, ET alone reaches $z\simeq 2-3$). For the
joint GW-GRB detections we find that the one with the highest redshift is at $z\simeq 3.38$ (found both in optimistic and realistic scenario for the THESEUS FOV, and for the Gaussian mass distribution). Notice that, in the left panel,  events with $z\,\gsim \, 6$  are not visible on the vertical scale used in the figure, but are indeed present.

Fig.~\ref{fig:ET+2CE_rel_err} shows the instrumental error for the luminosity distance in our catalog of events for ET+CE+CE, with the same meaning of the lines and shaded areas as in Fig.~\ref{fig:ET_rel_err}. 
The distribution now extends to much higher redshifts,  because, as we have seen, CE (assuming the current design configuration, with 40~km arms) has a much larger horizon to BNS. 
For the flat distribution  of neutron star masses the fit for the mean value of the 
instrumental error and to its standard deviation 
is given by 

\begin{figure}[t]
 \centering
 \includegraphics[width=0.55\textwidth]{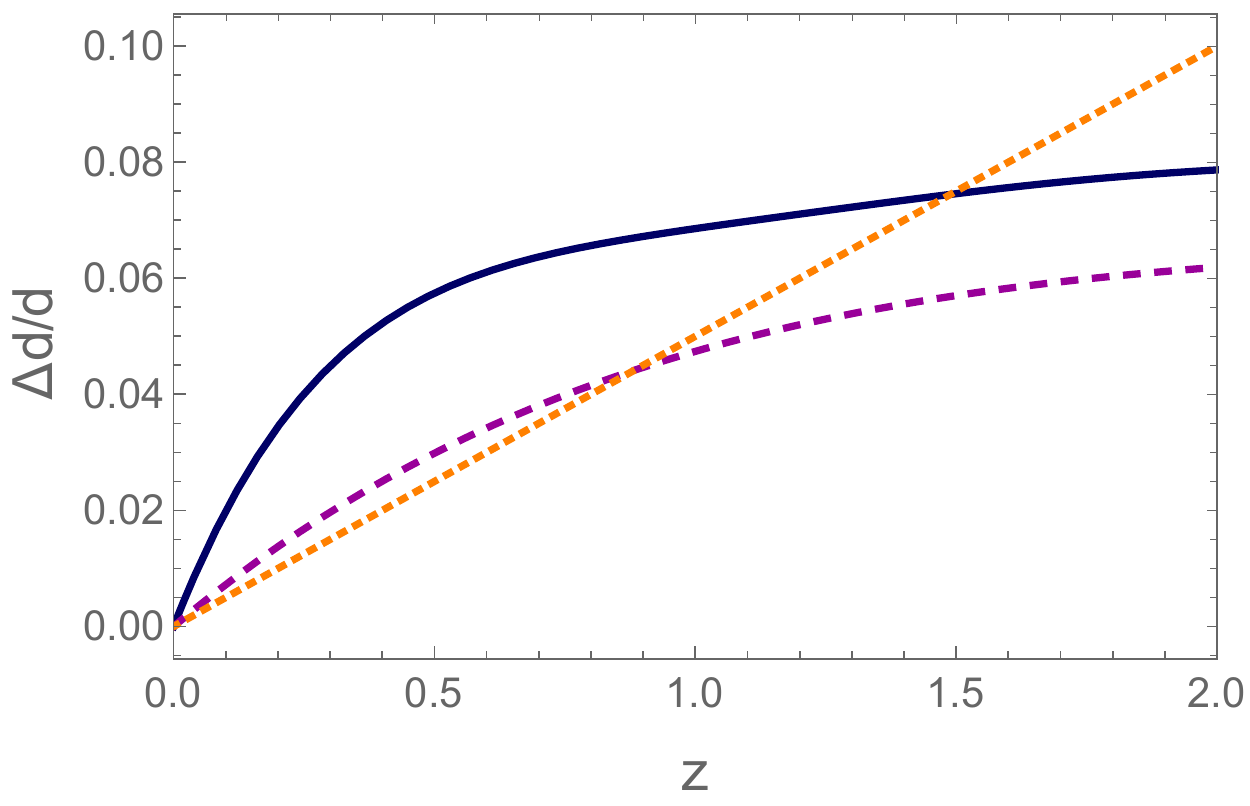}
 \caption{The fit to the average value of  the instrumental error on $\Delta d_L/d_L$ for ET
 (blue solid line) and for ET+CE+CE (magenta dashed line)  (both, for definiteness, for the flat mass distribution),
compared to the error due to lensing (orange dotted line).}
  \label{fig:compareETandET2CE}
\end{figure}

\bees
\overline{\left(\frac{\Delta d_L}{d_L}\right)}_{\rm f}&=& 
0.0756734 z -0.0360303 z^2+0.00882417 z^3-0.00113238 z^4\nn\\
&&+7.18676\times 10^{-5} z^5-1.75818\times 10^{-6} z^6\, ,\label{eq:mean_ET+2CE_flat}\\
\mathrm{\sigma}{\left[\left(\frac{\Delta d_L}{d_L}\right)_{\rm f}\right]}&=&0.0411611 z-0.0334084 z^2+0.0115623 z^3-0.00198248 z^4\nn\\
&&+1.65371\times 10^{-4} z^5-5.35185\times 10^{-6} z^6 \, ,
\label{eq:SD_ET+2CE_flat}
\ees
while for the Gaussian distribution
\bees
\overline{\left(\frac{\Delta d_L}{d_L}\right)}_{\rm G}&=& 
0.0828382 z-0.0427639 z^2+0.0115373 z^3-0.00166096 z^4\nn\\
&&+1.21048\times 10^{-4} z^5-3.50769\times 10^{-6} z^6\, ,\label{eq:mean_ET+2CE_gaus}\\
\mathrm{\sigma}{\left[\left(\frac{\Delta d_L}{d_L}\right)_{\rm G}\right]}&=&
0.0400958 z-0.0323344 z^2+0.0109567 z^3-0.00183302 z^4\nn\\
&&+1.48911\times 10^{-4} z^5-4.687178\times 10^{-6} z^6 \, ,
\label{eq:SD_ET+2CE_flat}
\ees
However, we also see from Fig.~\ref{fig:ET+2CE_rel_err} that the contribution to the error from lensing, given by the dotted orange line, becomes quickly dominant at $z\, \gsim \, 1.5$ [assuming that the linear extrapolation (\ref{errorlensing}) is still  correct at large redshifts]. For the study of joint GW-GRB detections that we have performed this has a limited impact, as for  ET+CE+CE the bulk of the joint GW-GRB detections is  at 
$z\, \lsim \, (1-1.5)$. Given the localization capability of the ET+CE+CE network, one could still hope to extract cosmological information from the very large number of standard sirens at larger redshifts, through statistical methods. However, at these redshifts  the dominant contribution to $\Delta d_L/d_L$ will come from lensing (similarly to what happens for LISA). This is clearly seen  from Fig.~\ref{fig:compareETandET2CE}, where we plot the instrumental error $\Delta d_L/d_L$ at ET and at ET+CE+CE, in the range $z<2$ where they can be compared, and the error due to lensing. We see that, even if the instrumental error from the ET+CE+CE network is obviously better than for a single ET, above $z\simeq 1$ the error in ET+CE+CE starts to be dominated by lensing, so that at $z\, \gsim \, 1.5$ 
both ET and ET+CE+CE are dominated by lensing and therefore eventually this becomes the limiting factor in both configurations.

This limitation could however be turned into a virtue. The situation is indeed similar to the one that was discussed a decade ago in ref.~\cite{Cutler:2009qv}, in the context of a study for the Big Bang Observer (BBO). It was found that also in BBO the error is almost entirely dominated by lensing. This means that, once one has determined the dependence of $d_L$ on $z$ from the full ensemble of sources (and possibly by combining standard sirens with CMB, BAO and SNe), the scatter around this mean value for each single BNS event is basically a measurement of the gravitational lensing magnification along that line of sight. Given the very large number of sources, this will produce a map of the lensing magnification across the sky; in particular, the corresponding two-point correlation function gives a measurement of the convergence power spectrum, and 
provides important information on cosmological structure formation.

\section{Constraints from standard sirens on $H_0$ and $\oma$  in $\Lambda$CDM}\label{sect:LCDM}

In this section we study how the addition of the standard sirens  with   GRB counterpart that could be observed with the HLVKI network or with 3G detectors, would contribute to the knowledge of cosmological parameters in $\Lambda$CDM.
In $\Lambda$CDM the luminosity distance to standard sirens is given by
\be\label{dLem}
d_L(z)=\frac{1+z}{H_0}\int_0^z\,
\frac{d\tilde{z}}{\sqrt{\oma (1+\tilde{z})^3+\ola }}\, ,
\ee
where, as usual,  $\oma$ and $\ola$ are the present  fraction of matter energy density and of vacuum energy density and, assuming flatness, $\ola\simeq1-\oma$ (we do not write explicitly the contribution of radiation, which is completely negligible at the redshifts relevant for standard sirens). The measurement of the luminosity distances from a set of coalescing binaries therefore gives constraints on $H_0$ and $\oma$.

The most accurate results are obtained by combining  the constraints from standard sirens with
other cosmological datasets such as CMB, BAO and SNe, to
remove the degeneracies between cosmological parameters.
A priori it would be interesting to study also
the constraints that emerge   using only standard sirens, that, even if less constraining,  are conceptually interesting because they have systematics completely different from those of  electromagnetic observations. However, for 2G detectors with a counterpart identified through a GRB,  we find that the number of sources is too small to obtain significant results from standard sirens alone, and we will present only the results obtained by combining standard sirens with CMB+BAO+SNe. For 3G detectors, we will also show the separate results from standard sirens.
When combining standard sirens with  CMB, BAO and SNe, we use the following datasets:

\begin{itemize}

\item{\em CMB.} We use the 2015 \textit{Planck}  \cite{Planck_2015_1} measurements of the angular (cross-)power spectra, including  full-mission lowTEB data for low multipoles ($\ell \leq 29$) and the high-$\ell$ Plik  TT,TE,EE (cross-half-mission) ones for the high multipoles ($\ell > 29$) of the temperature and polarization auto- and cross- power spectra \cite{Ade:2015rim}.
We also include the temperature$+$polarization (T$+$P) lensing data, using  only the conservative multipole range $\ell =40-400$  \cite{Planck_2015_Lkl,Planck_2015_lens}.

\item {\em Type Ia supernovae.} We use the JLA data for  SN~Ia provided by the SDSS-II/SNLS3 Joint Light-curve Analysis~\cite{Betoule:2014frx}.

\item {\em Baryon Acoustic Oscillations} (BAO). We use the isotropic constraints provided by 6dFGS at $z_{\rm eff}=0.106$ \cite{Beutler:2011hx}, SDSS-MGS DR7 at $z_{\rm eff}=0.15$ \cite{Ross_SDSS_2014} and BOSS LOWZ at $z_{\rm eff}=0.32$  \cite{Anderson:2013zyy}, as well as  the anisotropic constraints from CMASS at $z_{\rm eff}=0.57$ \cite{Anderson:2013zyy}.

\end{itemize}

We then run a MCMC, using the  CLASS Boltzmann code \cite{Class}. For the baseline $\Lambda$CDM model we use the standard set of six independent
cosmological parameters:  the Hubble parameter today
$H_0 = 100 h \, \rm{km} \, \rm{s}^{-1} \rm{Mpc}^{-1}$, the physical baryon and cold dark matter density fractions today $\omega_b = \Omega_b h^2$ and $\omega_c = \Omega_c h^2$, respectively, the amplitude  $A_s$ and tilt $n_s$ of the primordial scalar perturbations,   and  the reionization optical depth  $\tau_{\rm re}$. We keep the sum of neutrino masses fixed, at the value $\sum_{\nu}m_{\nu}=0.06$~eV, as in the {\em Planck} baseline analysis~\cite{Planck_2015_CP}.

Since standard sirens, within $\Lambda$CDM, are only sensitive to $H_0$ and $\oma$, we
focus  on the  two-dimensional likelihoods in the $(\oma, H_0)$ plane (although, of course, when we combine standard sirens with CMB+BAO+SNe, the fact that the addition of standard sirens allows a more accurate determination of $H_0$ and $\oma$ also has a beneficial effects on the determination of the other parameters, because it helps to reduce the degeneracies).

\subsection{Results for the  HLVKI network}\label{sect:LCDMwith2G}

\begin{figure}[t]
\centering
\includegraphics[width=0.4\textwidth]{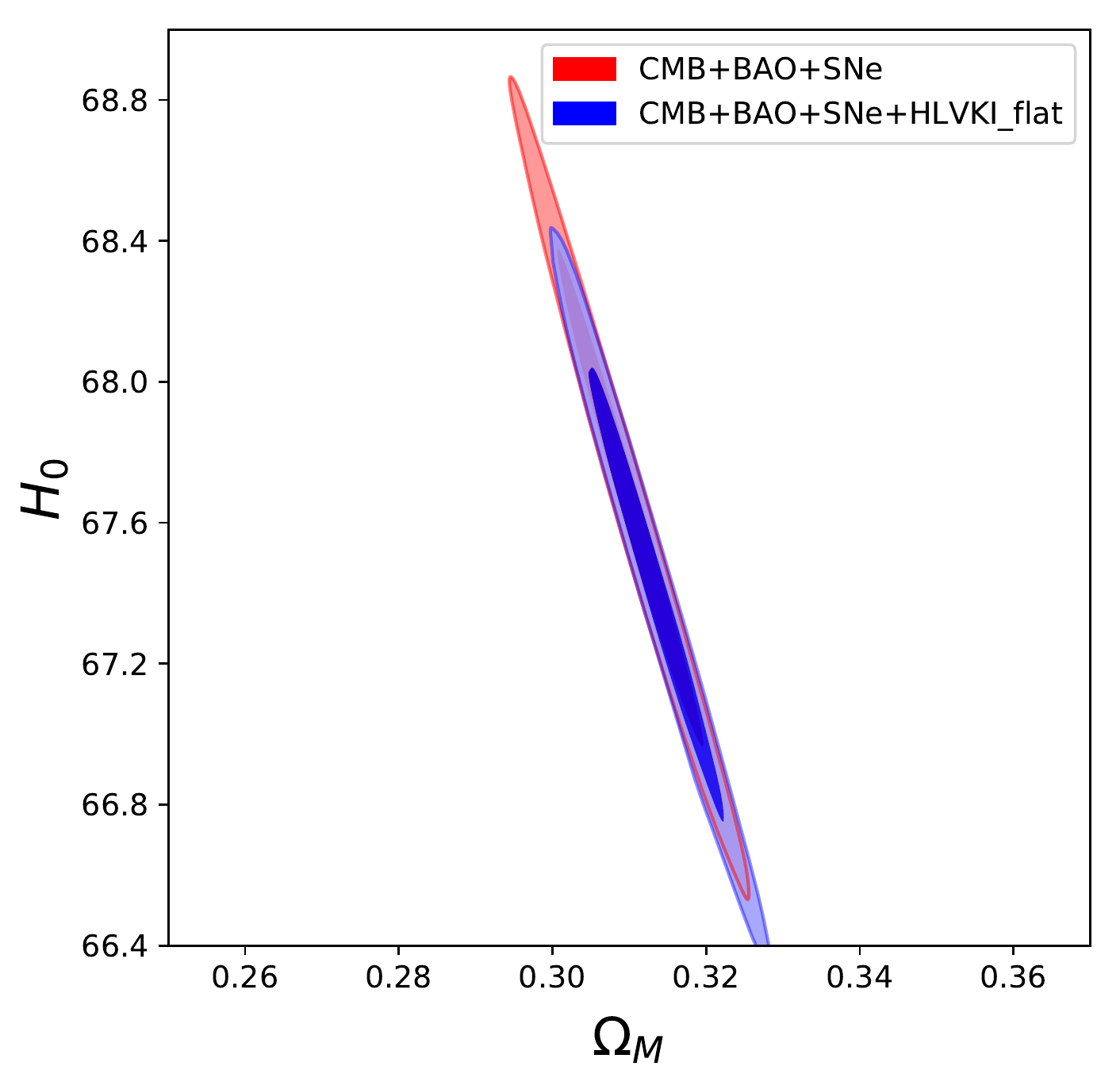}
\includegraphics[width=0.4\textwidth]{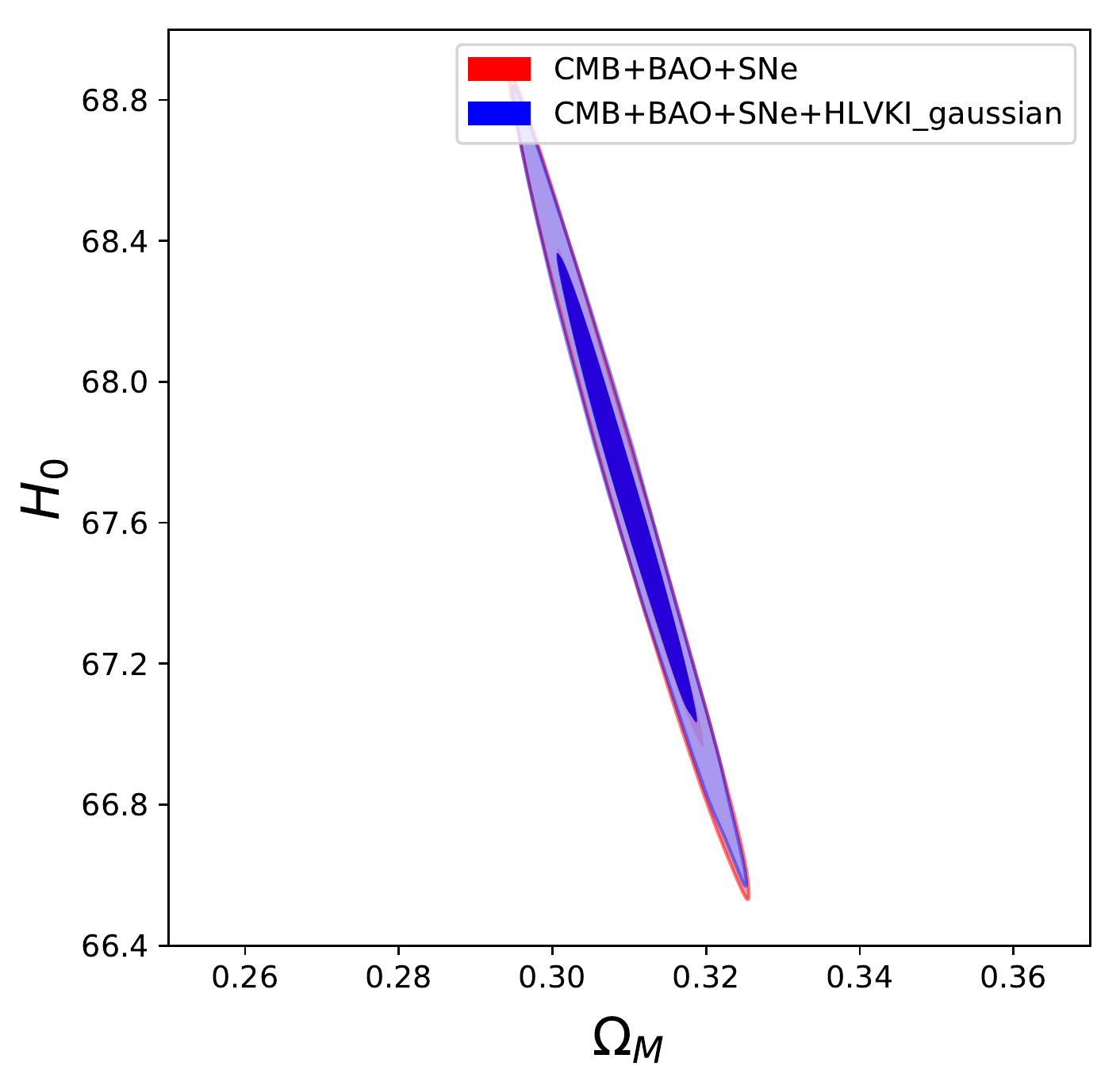}
\caption{The  $1\sigma$ and $2\sigma$
contours  of the two-dimensional likelihood in the $(\oma,H_0)$ plane, in $\Lambda$CDM, from CMB+BAO+SNe (red), and the result obtained by combining standard sirens at the HLVKI network with CMB+BAO+SNe (blue). Left: in the case of flat neutron star mass distribution. Right: in the case of gaussian neutron star mass distribution.}
\label{fig:HLVKI_LCDM}
\end{figure}

The result is shown in Fig.~\ref{fig:HLVKI_LCDM}, where we compare  the likelihood in the $(\oma,H_0)$ plane  obtained from  a MCMC using the above CMB+BAO+SNe dataset (red contours) with those obtained from  the combined datasets
CMB+BAO+SNe+standard sirens (blue contours), for the two distributions of neutron star masses.  The contours from standard sirens only are not shown since the MCMC fails to converge due to the small number of sources.

In table~\ref{tab:LCDMwith2G} we show the relative errors $\Delta H_0/H_0$ and
$\Delta\oma/\oma$ obtained from the
corresponding one-dimensional marginalized likelihood. We see that,
for the  HLVKI   network, the addition of  joint GW-GRB detections   to the current cosmological dataset does not improve substantially the accuracy on $H_0$ and $\oma$. This should be traced to the fact  that, as we see from Table~\ref{tab:2G}, if we  perform coincidences of the GW events with GRBs, we only have of order 15 joint detections, even over a 10-yr period. On the other hand, at the typical redshifts of the events seen by 2G detectors, the network of optical and infrared telescopes is expected to provide  many more electromagnetic counterparts. In the present paper, even for 2G detectors  we have restricted ourselves to a study of coincidences with GRBs only, also as a benchmark for the study of GW-GRB coincidences with 3G detectors. However, it is clear that for 2G detectors the contribution of optical/IR telescopes will be crucial and could raise substantially the number of standard sirens with observed counterpart.

Our results are broadly consistent with the analysis of \cite{Chen:2017rfc,Feeney:2018mkj}, which show that, to obtain a measurement of $H_0$ below $1\%$ at 2G detectors  with standard sirens only,
${\cal O}(50-100)$ standard sirens with counterpart are needed. This would allow  to address  the discrepancy between the local measurement of $H_0$ and the value inferred by  {\em Planck} and BAO observations assuming $\Lambda$CDM, which has now reached the $4.4\sigma$
level~\cite{Riess:2019cxk}.

\begin{table}[t]
\centering
\begin{tabular}{|c|c|c|c|}
 \hline
                               &  CMB+BAO+SNe& combined, flat & combined, gaussian     \\ \hline
$\Delta H_0/H_0$ &  0.72\%                   & 0.65\%             & 0.66\%         \\
$\Delta\oma/\oma$ &  2.11\%                  & 1.91\%             &  1.96\%             \\
\hline
\end{tabular}
\caption{Accuracy ($1\sigma$ level) in the reconstruction of $H_0$ and $\oma$ with  CMB+BAO+SNe only,  and  the combined result CMB+BAO+SNe+standard sirens, using the HLVKI detector network and the flat and gaussian mass distributions.
\label{tab:LCDMwith2G}}
\end{table}

\subsection{Results for ET}\label{sect:LCDMwithET}

\begin{figure}[t]
\centering
\includegraphics[width=0.4\textwidth]{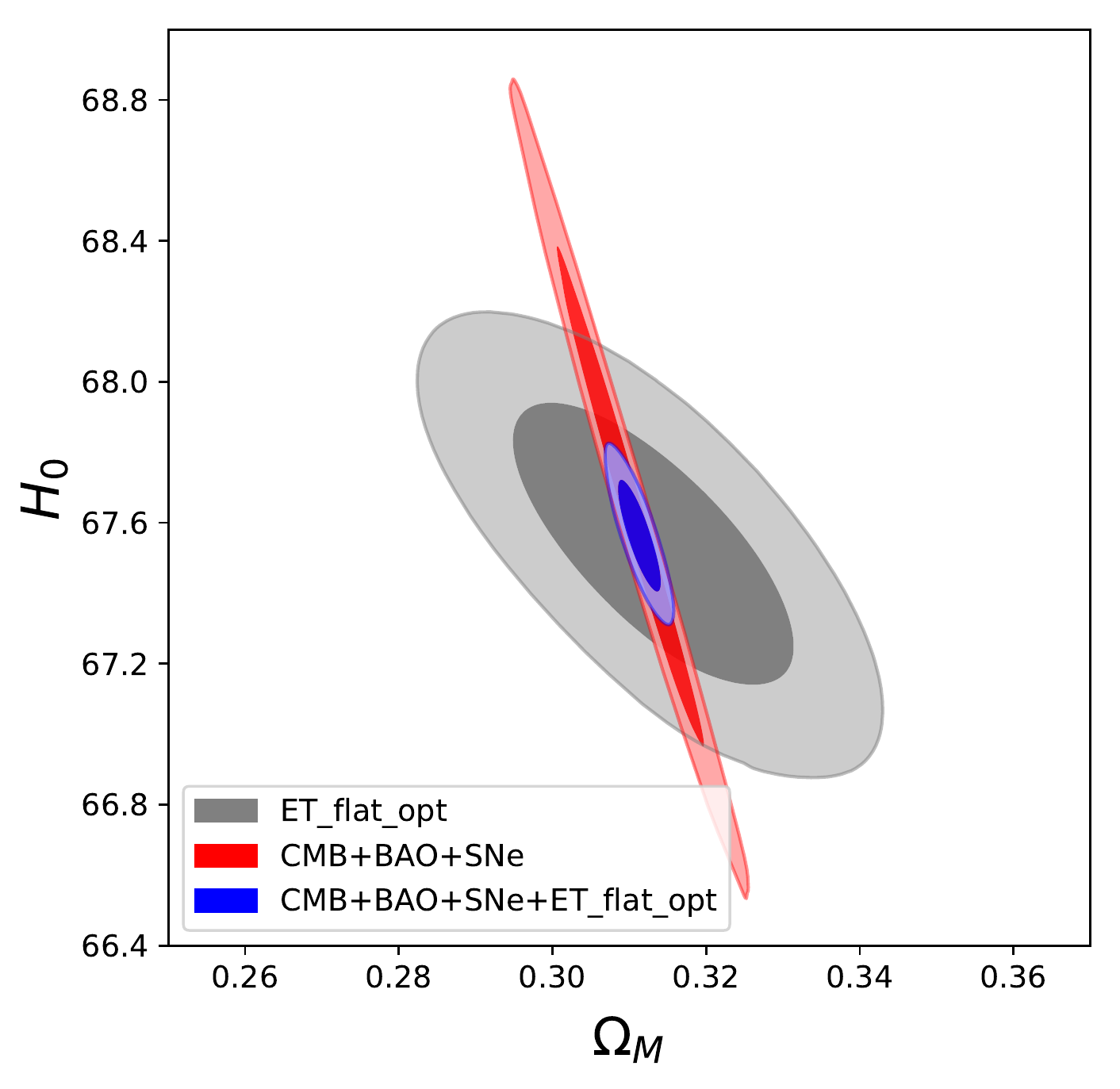}
\includegraphics[width=0.4\textwidth]{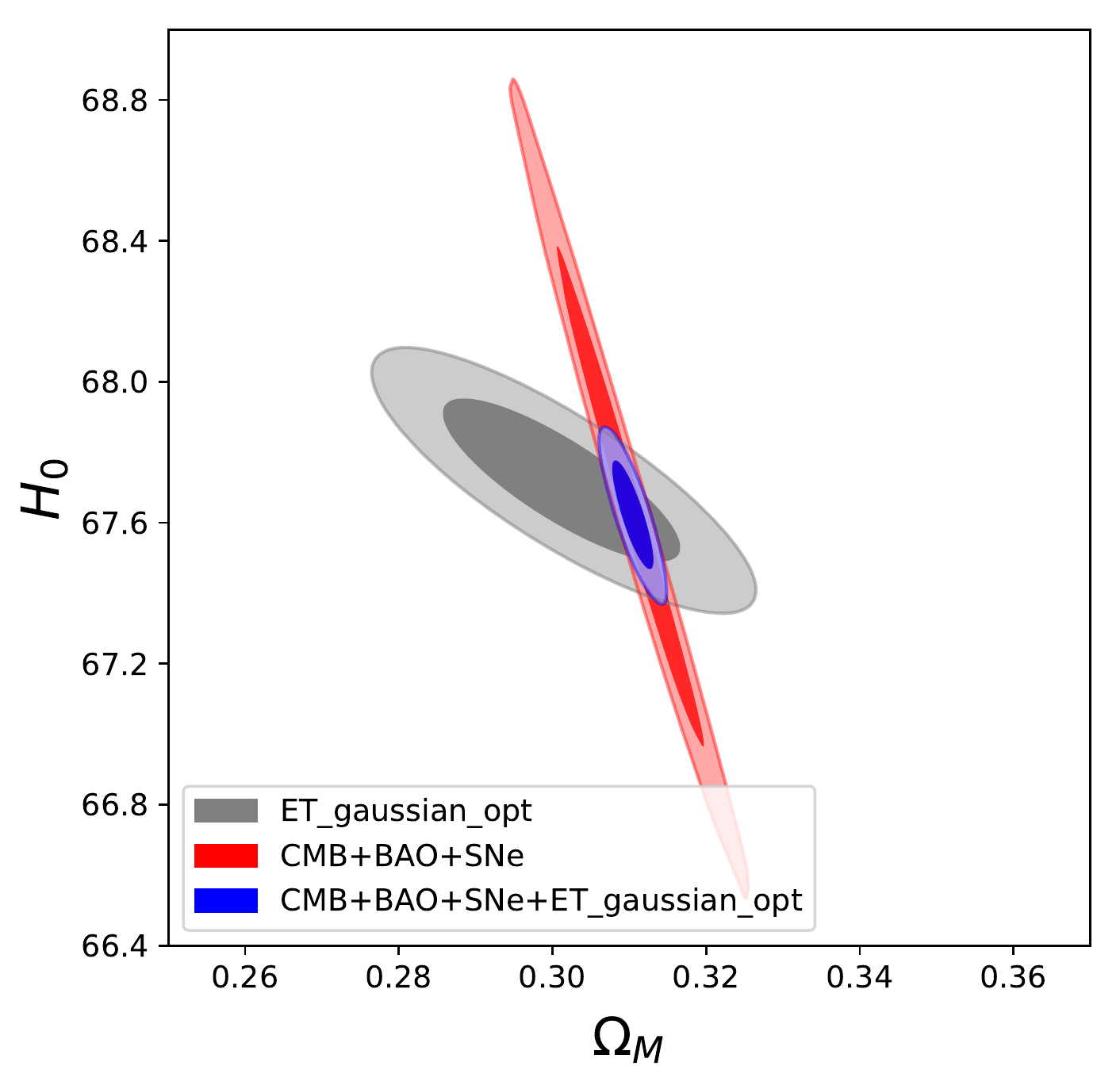}
\caption{The  $1\sigma$ and $2\sigma$
contours  of the two-dimensional likelihood in the $(\oma,H_0)$ plane, in $\Lambda$CDM, from CMB+BAO+SNe (red),  standard sirens at ET  with a GRB counterpart determined by THESEUS (gray), and the result obtained by combining standard sirens   with CMB+BAO+SNe (blue). Left: in the case of flat neutron star mass distribution. Right: in the case of gaussian neutron star mass distribution. We use the optimistic estimate for the FOV of THESEUS.}
\label{fig:ET_LCDM_opt}
\end{figure}

\begin{figure}[t]
\centering
\includegraphics[width=0.4\textwidth]{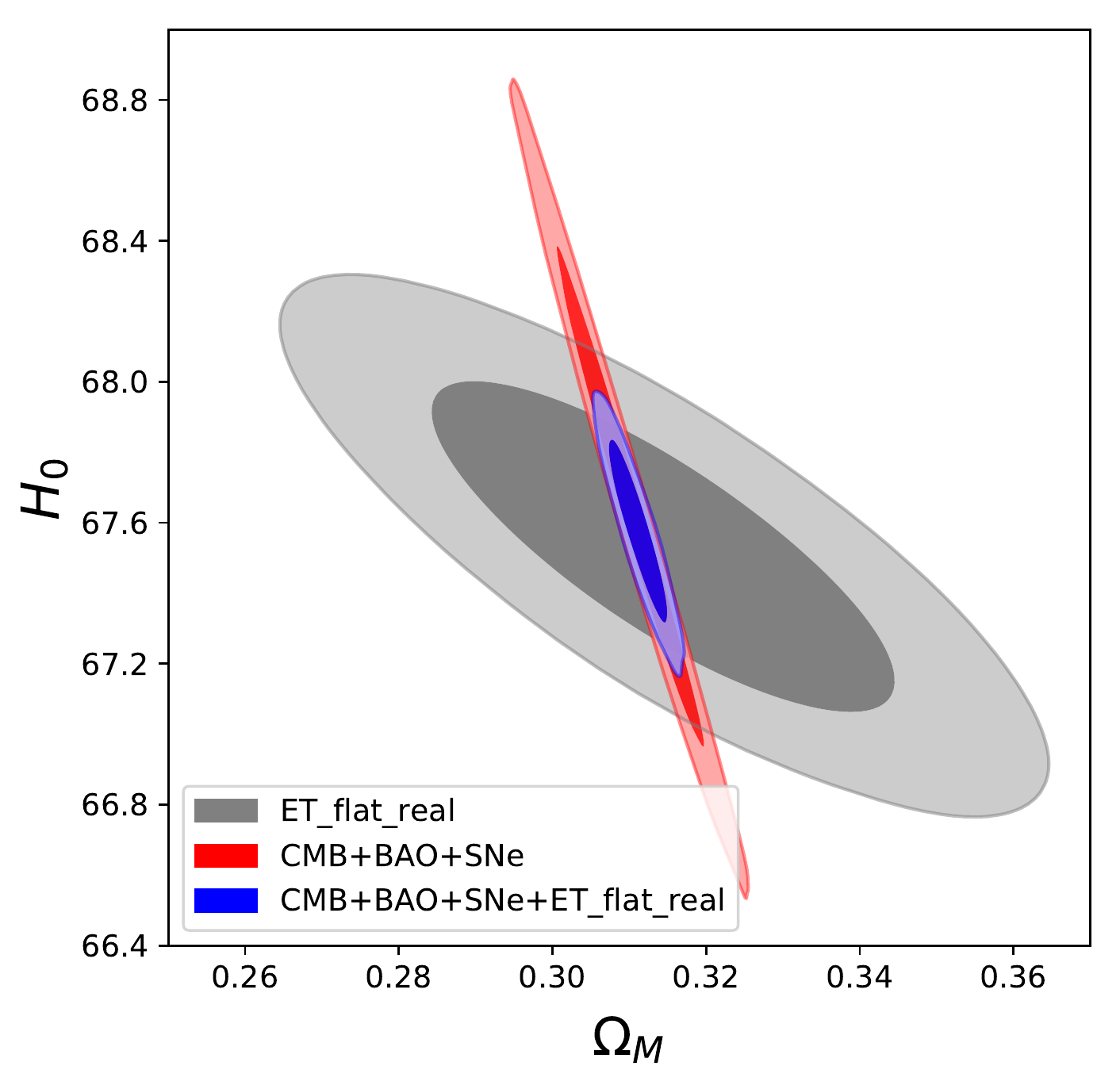}
\includegraphics[width=0.4\textwidth]{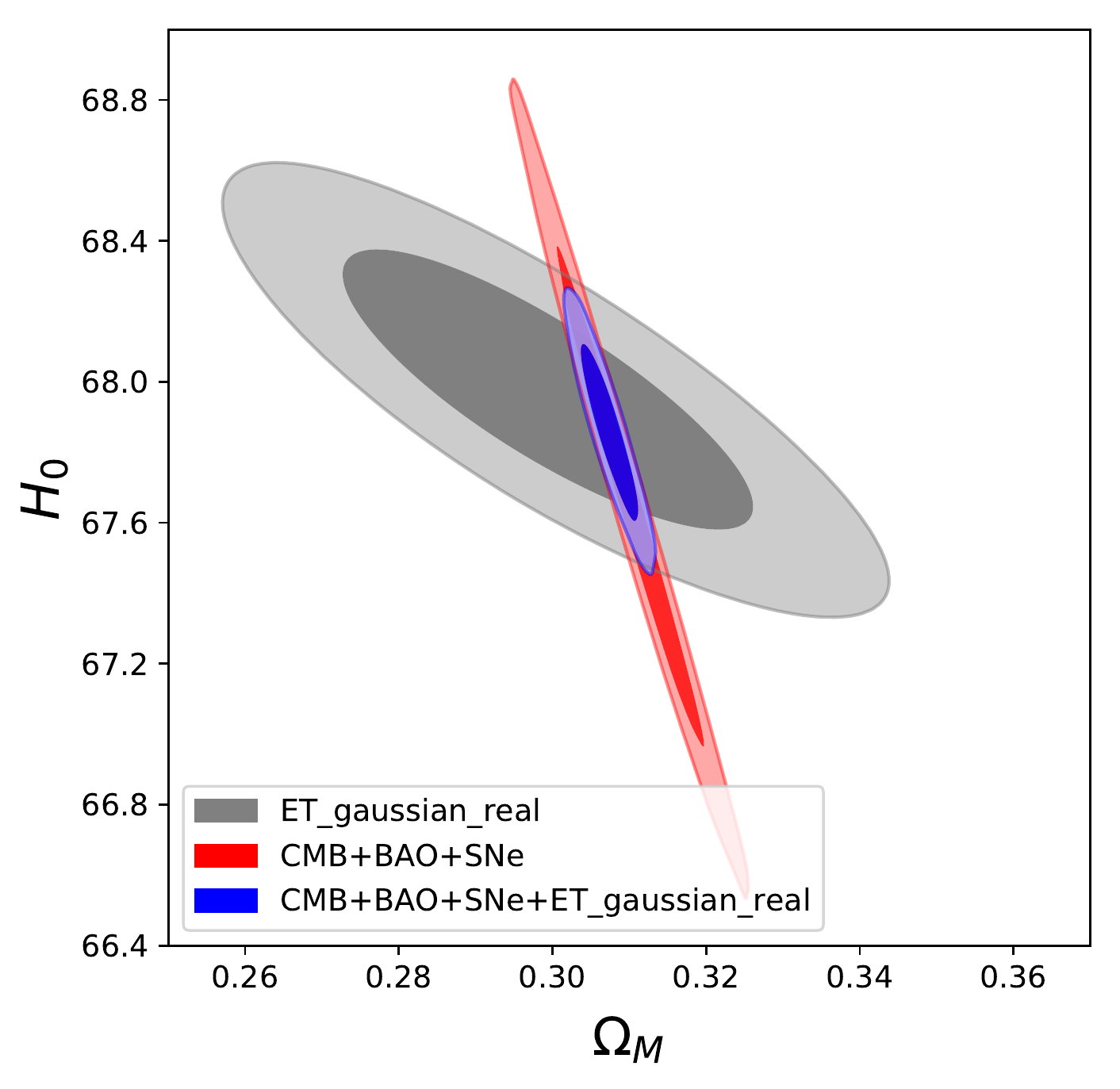}
\caption{As in Fig.~\ref{fig:ET_LCDM_opt}, with the realistic estimate for the FOV of THESEUS.}
\label{fig:ET_LCDM_real}
\end{figure}

We next consider the case of joint detections between a single ET detector and a  GRB detected by THESEUS, again restricting at first to $\Lambda$CDM.
The result is shown in Figs.~\ref{fig:ET_LCDM_opt} and  \ref{fig:ET_LCDM_real}, where we compare  the likelihood in the $(\oma,H_0)$ plane  obtained from  a MCMC using our CMB+BAO+SNe dataset (red contours) with those obtained from standard sirens only (gray contour) and
 the combined datasets
CMB+BAO+SNe+standard sirens (blue contours). In particular, in Fig.~\ref{fig:ET_LCDM_opt} we show the result
for the two distribution of neutron star masses in the optimistic scenario for the FOV of THESEUS, while in
Fig.~\ref{fig:ET_LCDM_real} we show the result
for the two distribution of neutron star masses in the realistic scenario for the FOV.

First of all observe that, despite the fact that the mock catalog of standard sirens has been generated by taking as fiducial cosmological model $\Lambda$CDM with the values of $H_0$ and $\oma$ obtained from these
CMB+BAO+SNe data,  the  contour obtained from standard sirens only is not always centered  on the mean values given by the CMB+BAO+SNe contour. This is an unavoidable consequence of the fact that, in order to simulate the result from actual observations, we have scattered the values of $d_L(z)$ with a variance $\Delta d_L(z)$ given by the expected observational error, see the discussion in Section~\ref{sect:catalog}. Of course, this is the situation that would take place in an actual observation (although the actual position of the gray contour will depend on the particular realization of the random scattering of the data around their mean value), and all that we should expect is that the contours are consistent at, say, the $(1-2) \sigma$ level, which is indeed the case. However, one should be  aware of the fact that, if we combine the standard sirens and the CMB+BAO+SNe dataset in a realization where the two contours do not overlap well, we get a larger error on the parameters with respect to what is obtained in a realization where the contours happens to overlap well. Once again, this is exactly the situation that will be faced in the actual experiment.

The  accuracy on $H_0$ and $\oma$ from the corresponding one-dimensional likelihood are given in
Table~\ref{tab:LCDMwithETopt} (for  the optimistic FOV of THESEUS) and
Table~\ref{tab:LCDMwithETreal} (for  the realistic FOV).
From these tables we see that standard sirens at ET, already before combining them with other cosmological datasets, give an accuracy on $H_0$ between $0.2\%$ and $0.4\%$, depending on the scenarios considered. This is
a very interesting accuracy, that would allow  to conclusively arbitrate the tension between the local $H_0$ measurement
and the {\em Planck}-$\Lambda$CDM value, with totally different systematic compared to SNe. In particular, this measurement would  have the potential of  falsifying $\Lambda$CDM.

\begin{table}[t]
\centering
\begin{tabular}{|c|c|c|c|c|c|}
 \hline
                               &  CMB+BAO+SNe&ET, &ET,           & combined,  & combined,      \\
                               &                               &flat &gaussian   &flat              & gaussian \\
\hline
$\Delta H_0/H_0$ &  0.72\%                   & 0.28\%             & 0.23\%    & 0.16\% &0.15\%     \\
$\Delta\oma/\oma$ &  2.11\%                  & 3.68\%             &  3.38\%      &0.59\%  & 0.57\%     \\
\hline
\end{tabular}
\caption{Accuracy ($1\sigma$ level) in the reconstruction of $H_0$ and $\oma$ with   only CMB+BAO+SNe,  with only standard sirens (with the flat and gaussian mass distributions, respectively) and  the combined results CMB+BAO+SNe+standard sirens, using ET  and THESEUS and assuming the optimistic FOV of THESEUS.
\label{tab:LCDMwithETopt}}
\end{table}

\begin{table}[t]
\centering
\begin{tabular}{|c|c|c|c|c|c|}
 \hline
                               &  CMB+BAO+SNe&ET, &ET,           & combined,  & combined,      \\
                               &                               &flat &gaussian   &flat              & gaussian \\
\hline
$\Delta H_0/H_0$ &  0.72\%                   & 0.42\%             & 0.39\%    & 0.26\% &0.25\%     \\
$\Delta\oma/\oma$ &  2.11\%                  & 6.17\%             &  5.88\%      &0.82\%  & 0.82\%     \\
\hline
\end{tabular}
\caption{As in Table~\ref{tab:LCDMwithETopt},
assuming the realistic FOV of THESEUS.
\label{tab:LCDMwithETreal}}
\end{table}

If instead the value of $H_0$ from standard sirens should agree with the CMB+BAO+SNe value obtained using $\Lambda$CDM, it would then make sense to combine these datasets. As we see from the tables, in that case the overall accuracy on $H_0$ could reach $(0.15-0.25)\%$.

%\newpage
%\clearpage

\subsection{Results for ET+CE+CE}\label{sect:LCDMwithET+CE+CE}

\begin{figure}[th]
\centering
\includegraphics[width=0.4\textwidth]{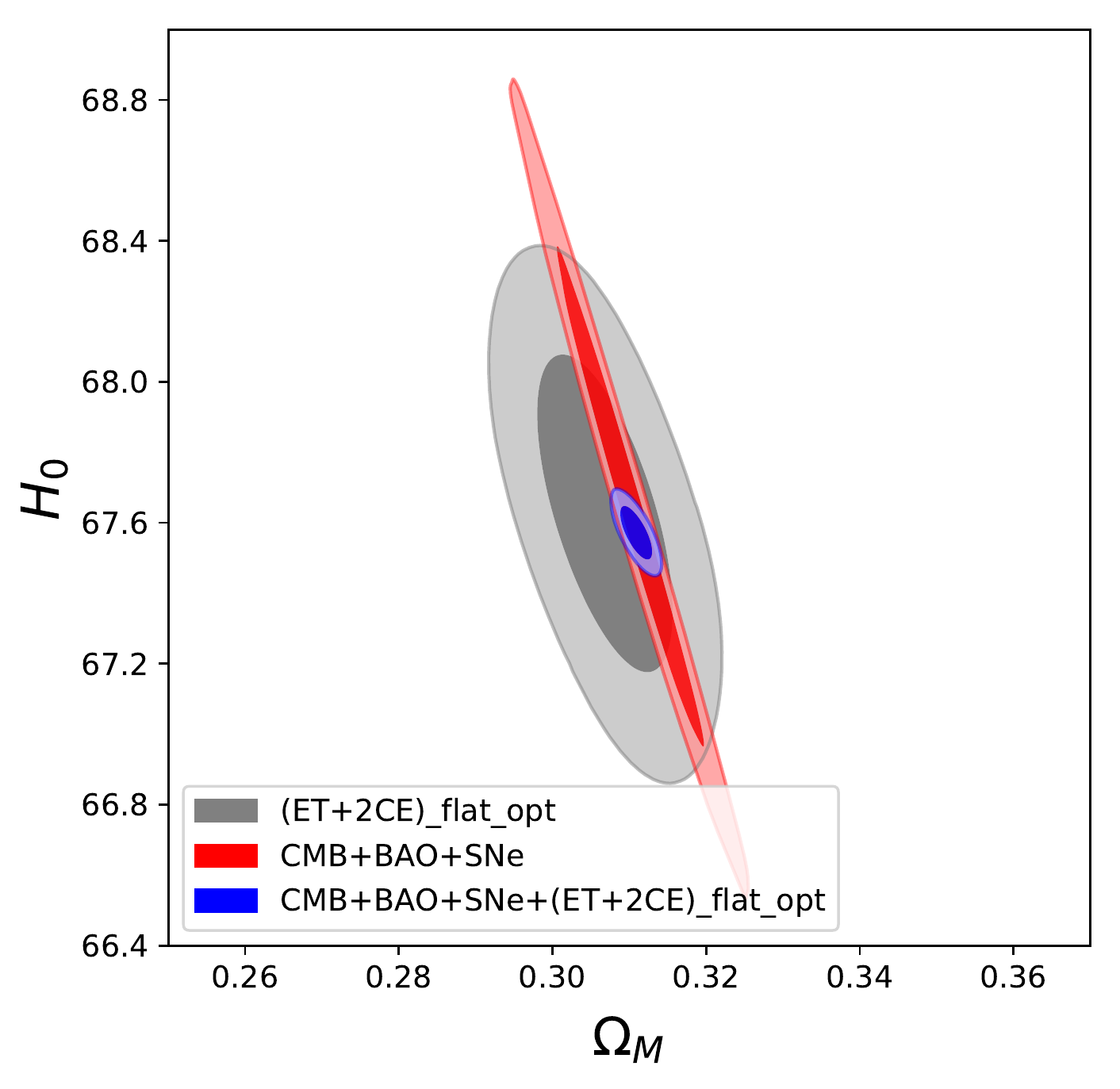}
\includegraphics[width=0.4\textwidth]{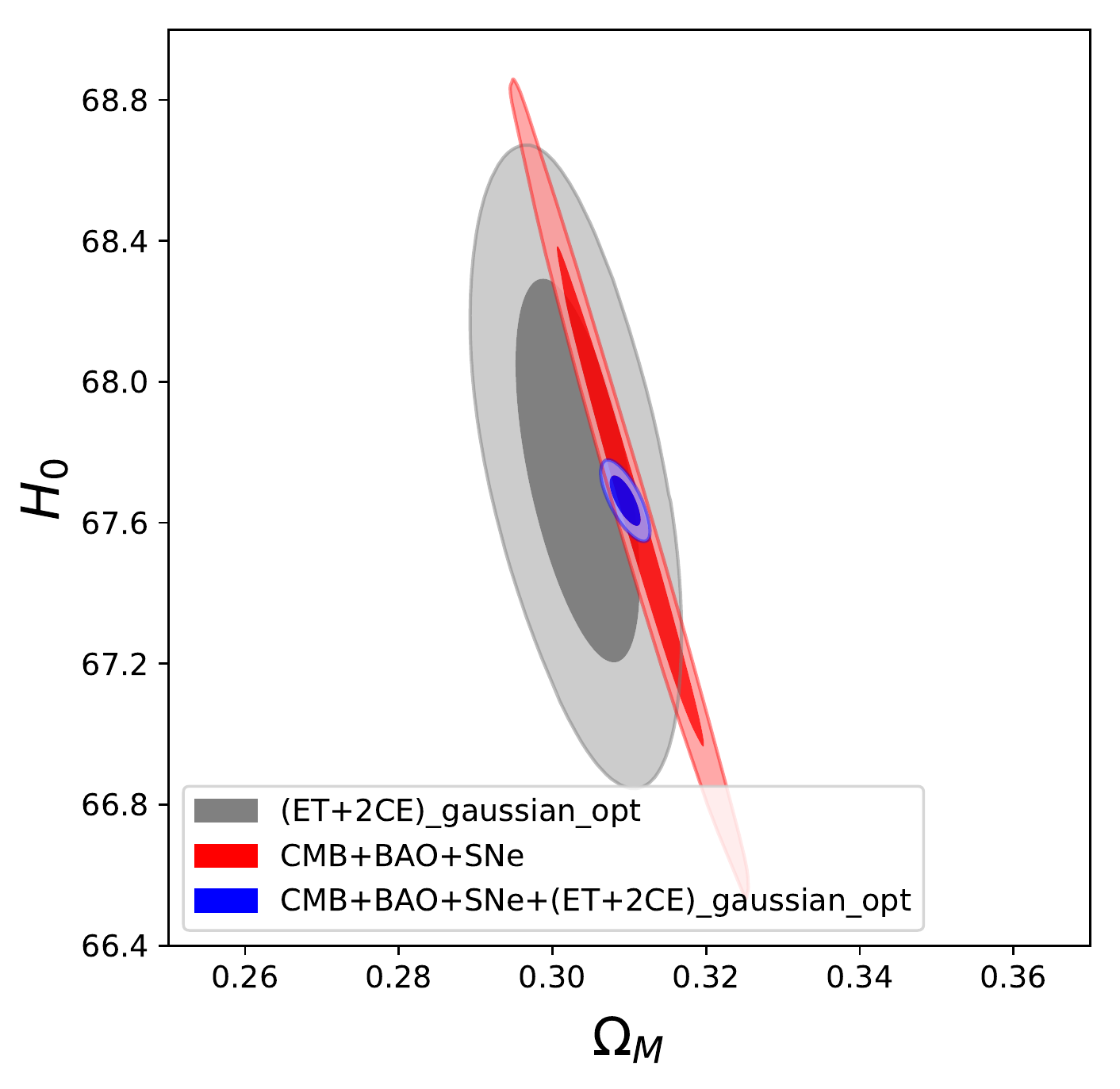}
\caption{As in Fig.~\ref{fig:ET_LCDM_opt}, for the ET+CE+CE network and optimistic FOV of THESEUS.}
\label{fig:ET+2CE_LCDM_opt}
\end{figure}

\begin{figure}[hb]
\centering
\includegraphics[width=0.4\textwidth]{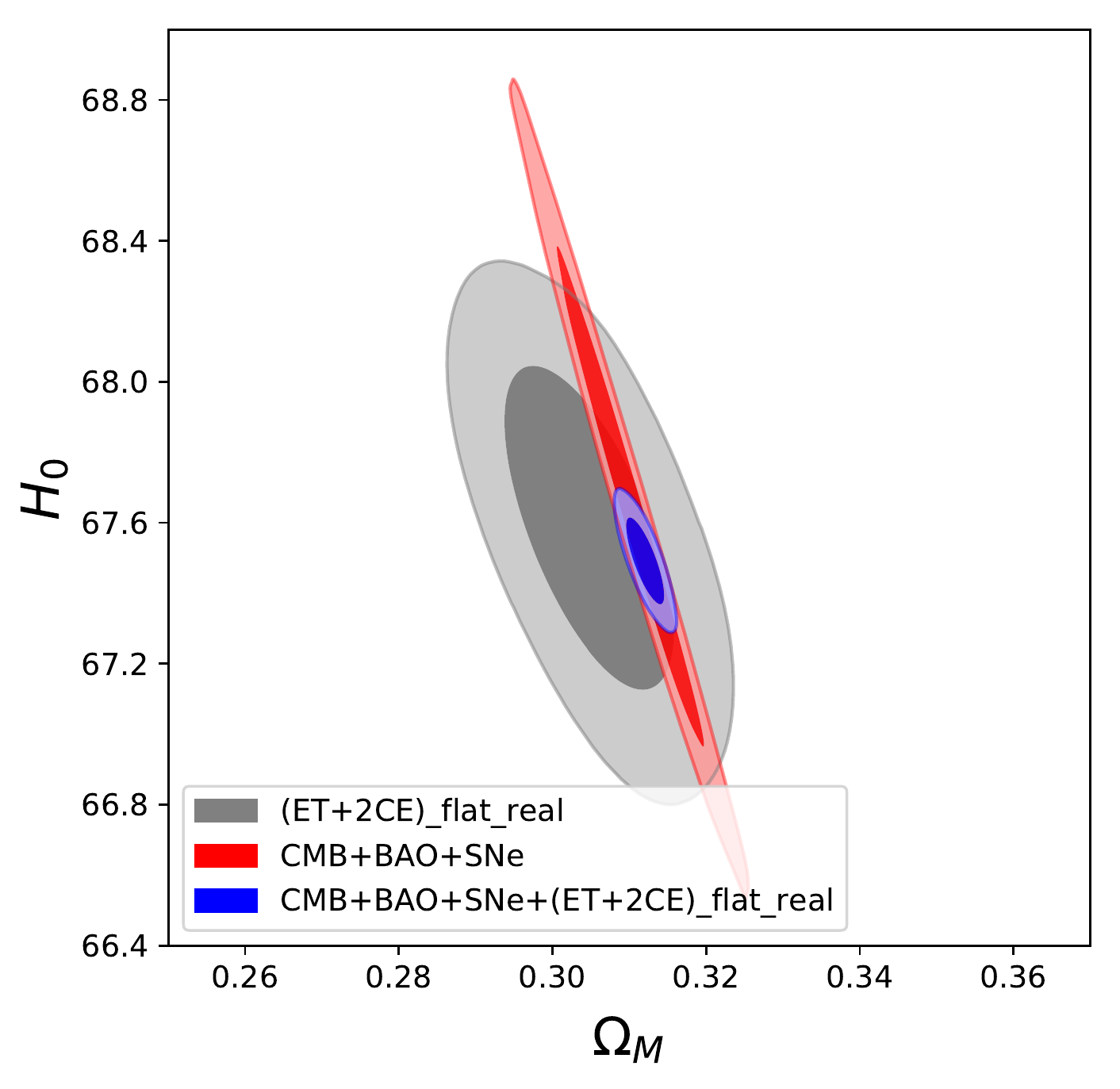}
\includegraphics[width=0.4\textwidth]{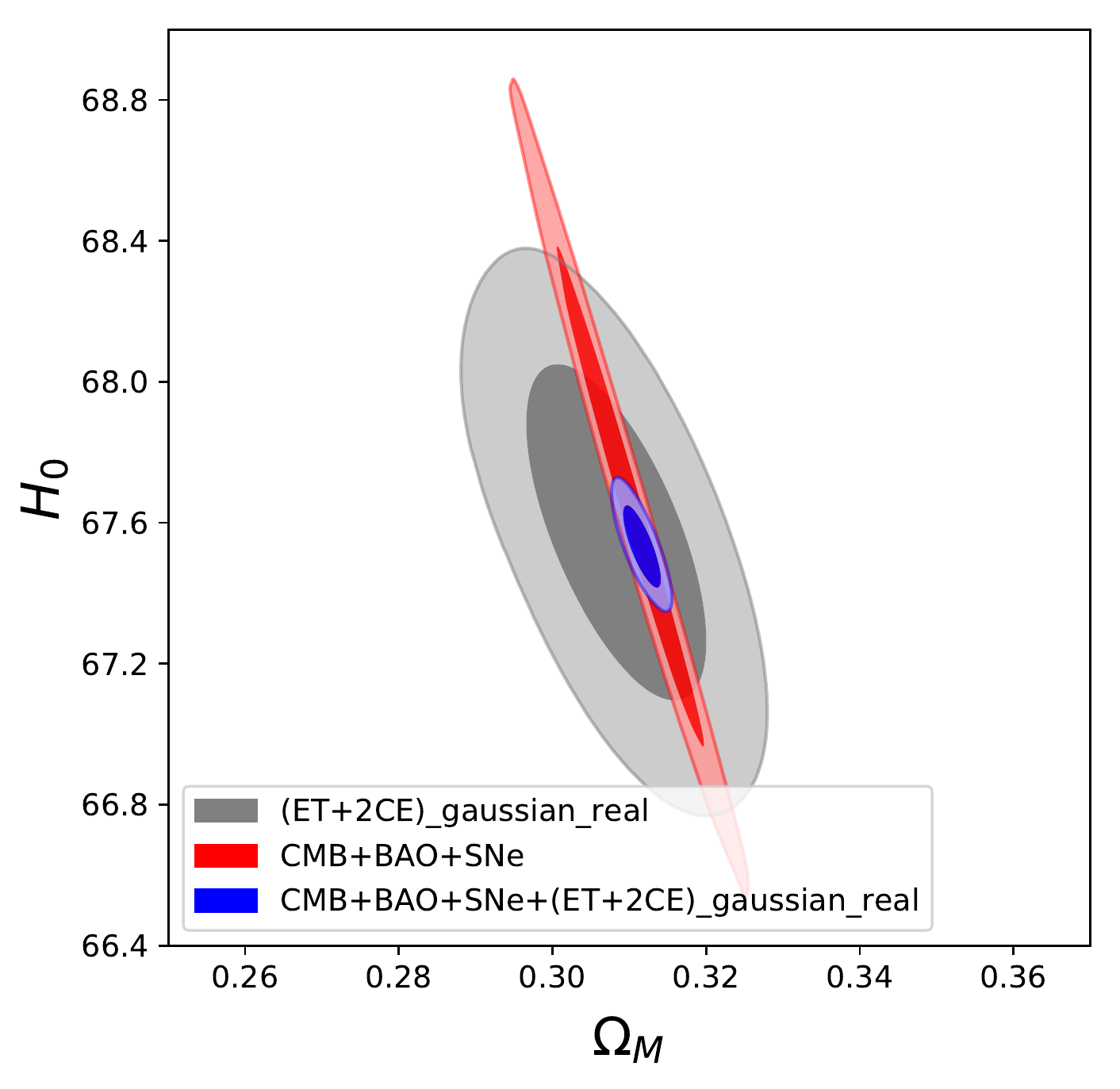}
\caption{As in Fig.~\ref{fig:ET_LCDM_real}, for the ET+CE+CE network and realistic estimate for the FOV of THESEUS.}
\label{fig:ET+2CE_LCDM_real}
\end{figure}

We finally consider the ET+CE+CE network. The two-dimensional likelihoods in the $(\oma,H_0)$ plane are shown in Figs.~\ref{fig:ET+2CE_LCDM_opt} and  \ref{fig:ET+2CE_LCDM_real}, and the corresponding $1\sigma$ accuracies from the one-dimensional likelihoods are shown in Tables~\ref{tab:LCDMwithET+2CEopt} (for  the optimistic FOV of THESEUS) and
Table~\ref{tab:LCDMwithET+2CEreal} (for  the realistic FOV).  We see that, in this case, with standard sirens only we get an accuracy of $H_0$ of  about $0.2\%$, while, combining with the other cosmological datasets we can reach an accuracies of order $(0.07-0.12)\%$. While in itself this would be a remarkable accuracy, still it is not significantly better than that reached with a single ET detector, as we see comparing with Tables~\ref{tab:LCDMwithETopt} and \ref{tab:LCDMwithETreal}. This result can be understood by looking at the number of events in our catalogs, shown in Table~\ref{tab:3G}. Despite the fact that the ET+CE+CE network has a number of GW detections higher by a factor ${\cal O}(10)$ compared to a single ET, when we look at joint GW-GRB detections the increase in the number of events is less than a factor of 2. In other worlds, the bottleneck here is on the GRB side. It is crucial to observe, however, that a three detector network such as ET+CE+CE will have excellent localization capabilities. Thus, at least for the events at $z\,\lsim\, 0.5$, the follow-up with optical and IR telescope will be possible, and will probably lead to a significant increase in the number of standard sirens with electromagnetic counterpart. As we discussed in Section~\ref{sect:count3G}, realistic estimates are currently difficult because they also depend on choices such as the amount of telescope time that will be devoted by the various  facilities to the follow-up of GW events, and in this paper we will not attempt such an estimate. However, it should be borne in mind that, for a ET+CE+CE network, the  joint GW-GRB detections that we are studying in this paper might provide just a fraction of the whole sample of GW signals with electromagnetic counterpart. Correspondingly, the accuracies that can be obtained on $H_0$ and $\oma$ at ET+CE+CE could be significantly better, compared to the figures that we find.

\vspace{30mm}

\begin{table}[h]
\centering
\begin{tabular}{|c|c|c|c|c|c|}
 \hline
                               &  CMB+BAO+SNe&ET+CE+CE, &ET+CE+CE,           & combined,  & combined,      \\
                               &                               &flat                &gaussian                  &flat              & gaussian \\
\hline
$\Delta H_0/H_0$ &  0.72\%                  & 0.20\%           & 0.22\%                  & 0.07\%      &0.07\%     \\
$\Delta\oma/\oma$ &  2.11\%                  & 1.43\%             &     1.31\%            &0.43\%       & 0.42\%     \\
\hline
\end{tabular}
\caption{Accuracy ($1\sigma$ level) in the reconstruction of $H_0$ and $\oma$ with   only CMB+BAO+SNe,  with only standard sirens (with the flat and gaussian mass distributions, respectively) and  the combined results CMB+BAO+SNe+standard sirens, using ET+CE+CE  and THESEUS, and assuming the optimistic FOV of THESEUS.
\label{tab:LCDMwithET+2CEopt}}
\end{table}

\vspace{10mm}

\begin{table}[h]
\centering
\begin{tabular}{|c|c|c|c|c|c|}
 \hline
                               &  CMB+BAO+SNe&ET+CE+CE, &ET+CE+CE,           & combined,  & combined,      \\
                               &                               &flat                &gaussian                  &flat              & gaussian \\
\hline
$\Delta H_0/H_0$ &  0.72\%                  & 0.24\%             & 0.23\%               & 0.12\%       &0.11\%     \\
$\Delta\oma/\oma$ &  2.11\%                  & 2.12\%             &  2.09\%               &0.51\%        & 0.52\%     \\
\hline
\end{tabular}
\caption{As in Table~\ref{tab:LCDMwithETopt},
assuming the realistic FOV of THESEUS.
\label{tab:LCDMwithET+2CEreal}}
\end{table}

\clearpage
\newpage

\vspace{30mm}

\clearpage
\newpage

\section{Constraints  from standard sirens on dark energy}\label{sect:DE}

\subsection{Testing the dark energy sector:  DE equation of state and  modified GW propagation}\label{sect:DEandmodGW}

We next extend $\Lambda$CDM by considering a non-trivial DE sector. A generic modified gravity theory, that generates a dynamical DE density $\rde(z)$,  will  produce a different evolution at the background level, so that \eq{dLem} is now replaced by
\be\label{dLemDE}
d_L(z)=\frac{1+z}{H_0}\int_0^z\,
\frac{d\tilde{z}}{\sqrt{\oma (1+\tilde{z})^3+\ode(\tilde{z}) }}\, ,
\ee
where
\be\label{4rdewdeproofs}
\ode(z)  =\Omega_{\rm DE}\exp\left\{ 3\int_{0}^z\, \frac{d\tilde{z}}{1+\tilde{z}}\, [1+\wde(\tilde{z})]\right\}\, ,
\ee
$\Omega_{\rm DE}\equiv \ode(z=0)\simeq 1-\oma$, and $\wde(z)$ is the DE equation of state. At the phenomenological level, the DE equation of state is usually parametrized as
\cite{Chevallier:2000qy,Linder:2002et}
\be\label{w0wa}
w_{\rm DE}(z)= w_0+ w_a \, \frac{z}{1+z}\, .
\ee
On top of this, any concrete modified gravity model will  produce deviations at the level of cosmological perturbations, both in the scalar sector and in the tensor sector. Modified perturbations in the scalar sectors affect the growth of cosmological structures or weak lensing, but are not directly relevant for standard sirens. In contrast, modifications in the tensor perturbations can affect the luminosity distance measured by standard sirens~\cite{Deffayet:2007kf,Yunes:2010yf,Saltas:2014dha,Gleyzes:2014rba,Lombriser:2015sxa,Nishizawa:2017nef,Arai:2017hxj,Belgacem:2017ihm,Amendola:2017ovw,Linder:2018jil,Pardo:2018ipy,Belgacem:2018lbp}. Recall that, in GR, the free propagation of tensor perturbations over a
Friedmann-Robertson-Walker (FRW) background is governed by the equation
\be\label{4eqtensorsect}
\tilde{h}''_A+2{\cal H}\tilde{h}'_A+k^2\tilde{h}_A=0\, ,
\ee
where $\tilde{h}_A(\eta, \vk)$ are  the Fourier modes of the GW amplitude, $A=+,\times$ labels the two polarizations, the prime denotes the derivative with respect to cosmic time $\eta$,  $a(\eta)$ is the scale factor,
${\cal H}=a'/a$, and we use units $c=1$. In a generic modified gravity model both the coefficient of the term
$k^2\tilde{h}_A$ and of the term $2{\cal H}\tilde{h}'_A$ can in principle be modified. A modification of the $k^2\tilde{h}_A$ term would affect the  speed of propagation of GWs and is by now excluded,
at the level  $|c_{\rm gw}-c|/c< O(10^{-15})$, by the observation of  GW170817/GRB~170817A \cite{Monitor:2017mdv}. However, one can find explicit examples of viable modified gravity models where the equation for the free propagation of tensor perturbation takes the form
\be\label{prophmodgrav}
\tilde{h}''_A  +2 {\cal H}[1-\delta(\eta)] \tilde{h}'_A+k^2\tilde{h}_A=0\, ,
\ee
where $\delta(\eta)$ is a function of conformal time or, equivalently, of redshift. In this case the speed of GWs, that is determined by the coefficient of the $k^2\tilde{h}_A$ term, is not affected. However, the coefficient of the `friction term', i.e. of the term   $2 {\cal H} \tilde{h}'_A$, changes with respect to its value in GR.
In fact, the recent analysis in
\cite{Belgacem:2019pkk} shows that all the best studied models of modified gravity, such as
Horndeski or the more general   degenerate higher order scalar-tensor (DHOST) theories, non-local infrared modifications of gravity, or  bigravity theories,  have a propagation equation with a non-trivial function $\delta(\eta)$, even when they predict $c_{\rm gw}=c$.

In that case, it is possible to show that standard sirens do not measure the usual ``electromagnetic" luminosity distance
$d_L^{\,\rm em}(z)$ given by \eq{dLemDE}, but rather a ``GW luminosity distance" $d_L^{\,\rm gw}(z)$, related to $d_L^{\,\rm em}(z)$ by~\cite{Belgacem:2017ihm,Belgacem:2017cqo} (see also Sect.~19.6.3 of \cite{Maggiore:2018zz})
\be\label{dLgwdLem}
d_L^{\,\rm gw}(z)=d_L^{\,\rm em}(z)\exp\left\{-\int_0^z \,\frac{dz'}{1+z'}\,\delta(z')\right\}\, .
\ee
Thus, standard sirens can in principle test the DE sector of a cosmological model through the functions $\wde(z)$ and $\delta(z)$. To perform phenomenological studies, it is convenient to have simple parametrization of these functions. The function $\wde(z)$ is usually parametrized as in \eq{w0wa}, in terms of the two parameters $(w_0,w_a)$. For modified GW propagation a convenient parametrization, in terms of two parameters $(\Xi_0,n)$,  has been proposed in \cite{Belgacem:2018lbp},
\be\label{eq:fit}
\frac{d_L^{\,\rm gw}(z)}{d_L^{\,\rm em}(z)}=\Xi_0 +\frac{1-\Xi_0}{(1+z)^n}\, .
\ee
This expression correctly reproduces the fact that, as $z\ra 0$,  $d_L^{\,\rm gw}/d_L^{\,\rm em}=1$, since as the redshift of the source goes to zero, there is no effect from modified gravity propagation. On the opposite limit of large redshifts, in \eq{eq:fit} $d_L^{\,\rm gw}/d_L^{\,\rm em}$ goes to a constant value $\Xi_0$. This is motivated by the fact that,
in  typical  models where  the modifications from GR only appear close to the recent cosmological epoch,
one expects that $\delta(z)$ would go to zero at large redshift so, from
\eq{dLgwdLem},
$d_L^{\,\rm gw}(z)/d_L^{\,\rm em}(z)$ should saturate to a constant value at large $z$.

An explicit example of  models with modified GW propagation of this type is given by the `RT' and `RR' nonlocal modifications of gravity that were  proposed in \cite{Maggiore:2013mea} and \cite{Maggiore:2014sia}, respectively, and whose cosmological consequences have been studied in detail  in \cite{Foffa:2013vma,Dirian:2014ara,Dirian:2014bma,Dirian:2016puz,Dirian:2017pwp,Belgacem:2017cqo,Belgacem:2018wtb} (see \cite{Maggiore:2016gpx,Belgacem:2017cqo} for review and further references). In both cases
the predictions  for
$d_L^{\,\rm gw}(z)/d_L^{\,\rm em}(z)$ are very accurately reproduced by \eq{eq:fit}, with
$\{\Xi_0\simeq 0.970,n\simeq 2.5\}$ for the RR model~\cite{Belgacem:2017cqo}, and $\{\Xi_0\simeq 0.934, n\simeq 2.6\}$ for the `minimal' RT model~\cite{Belgacem:2019pkk}.\footnote{The `minimal' RT model  is defined as the RT model with vanishing initial conditions set during radiation dominance. An analysis with more general initial conditions set during inflation shows that a much larger result, up to $\Xi_0\simeq 1.6$, can be obtained~\cite{DeltaN64:inprep}.} Actually, recent results~\cite{Belgacem:2018wtb} have shown that the RR model is ruled out by Lunar Laser Ranging, while the RT model is fully viable: it has a background evolution with accelerated expansion in the present epoch, stable cosmological perturbations, passes solar system and Lunar Laser Ranging constraints,
fits the cosmological CMB, BAO and SNa data at a level statistically indistinguishable from $\Lambda$CDM, and predicts $c_{\rm gw}=c$. It therefore gives an example of a phenomenologically viable alternative to  $\Lambda$CDM, which predicts a value of $\Xi_0$ that differs from the GR value $\Xi_0=1$ by about $6.6\%$.
 As we will see below, such a value is well within the reach of ET, and in fact even of the HLVKI network.
Several other modified gravity models have been explored in \cite{Belgacem:2019pkk}, and they all give a modified GW luminosity distance, with an expression that  is very well fitted by \eq{eq:fit}
[with the only exception of bigravity, where the interaction between the two metrics induces an interesting phenomenon of oscillations in $d_L^{\,\rm gw}(z)/d_L^{\,\rm em}(z)$].

In the following we will explore the potential of standard sirens to constrain the DE sector of a modified gravity  theory, using either $(w_0,w_a)$ or $(\Xi_0,w_0)$ as the extra parameters with respect to
$\Lambda$CDM.\footnote{As discussed
 in \cite{Belgacem:2018lbp}, the parameter $n$ plays in general a lesser role, since it just determines the precise way in which the function $d_L^{\,\rm gw}(z)/d_L^{\,\rm em}(z)$  interpolates between its asymptotic value $\Xi_0$ at large redshift, and the value $d_L^{\,\rm gw}/d_L^{\,\rm em}=1$ at $z=0$. Most sources relevant for ET are already in, or close to,  the asymptotic regime. For definiteness, we
will then held $n$ fixed to the value $n=5/2$.} 

As before, we will combine standard sirens with CMB, BAO and SNe. Observe that we use 
$\Lambda$CDM  as our fiducial cosmological model (and GR as our fiducial description of gravity), so in particular our fiducial values for the extra parameters in the DE sector are $w_0=-1$, $w_a=0$ and $\Xi_0=1$,
and we are studying to what accuracy we can get back these values from the data. For this reason, it is correct
to use $\Lambda$CDM (and GR) to compute the cosmological perturbations  when comparing with CMB, BAO and SNe. A generic modified gravity theory might also a priori induce modifications in the GW production mechanism, beside the modification in the GW propagation. The two effects are however largely decoupled, with modified GW propagation affecting the amplitude of the GW signal 
through a modification of the  luminosity distance, in a way that is incremental with the distance to the source and therefore redshift dependent [at least up to some redshift where the effect saturates, for models reproduced by the $(\Xi_0,n)$ parametrization], while modified GW production affects the phase of the signal, at some post-Newtonian order. On top of this, models that provide interesting modifications of GR at cosmological scales must have a form of screening at short distances in order not to spoil the successes of GR at the Solar System and laboratory scales, so they typically do not induce any observable effect on GR production. This is in particular the case for the RT nonlocal model, whose effects at scales small compared to $H_0^{-1}$  are totally negligible~\cite{Maggiore:2013mea,Kehagias:2014sda,Belgacem:2018wtb}. It therefore makes sense to study modified GW propagation in isolation, without worrying about possible corrections on the accuracy on $\Delta d_L/d_L$ that could be induced by modifications in the GW production mechanism.

\subsection{Results for the  HLVKI network}\label{sect:DEwith2G}

We begin by presenting  the constraints that can be obtained on the DE sector by combining standard sirens at the  HLVKI network   with the CMB+BAO+SNe dataset described in Section~\ref{sect:LCDM}.

\begin{figure}[t]
\centering
\includegraphics[width=0.4\textwidth]{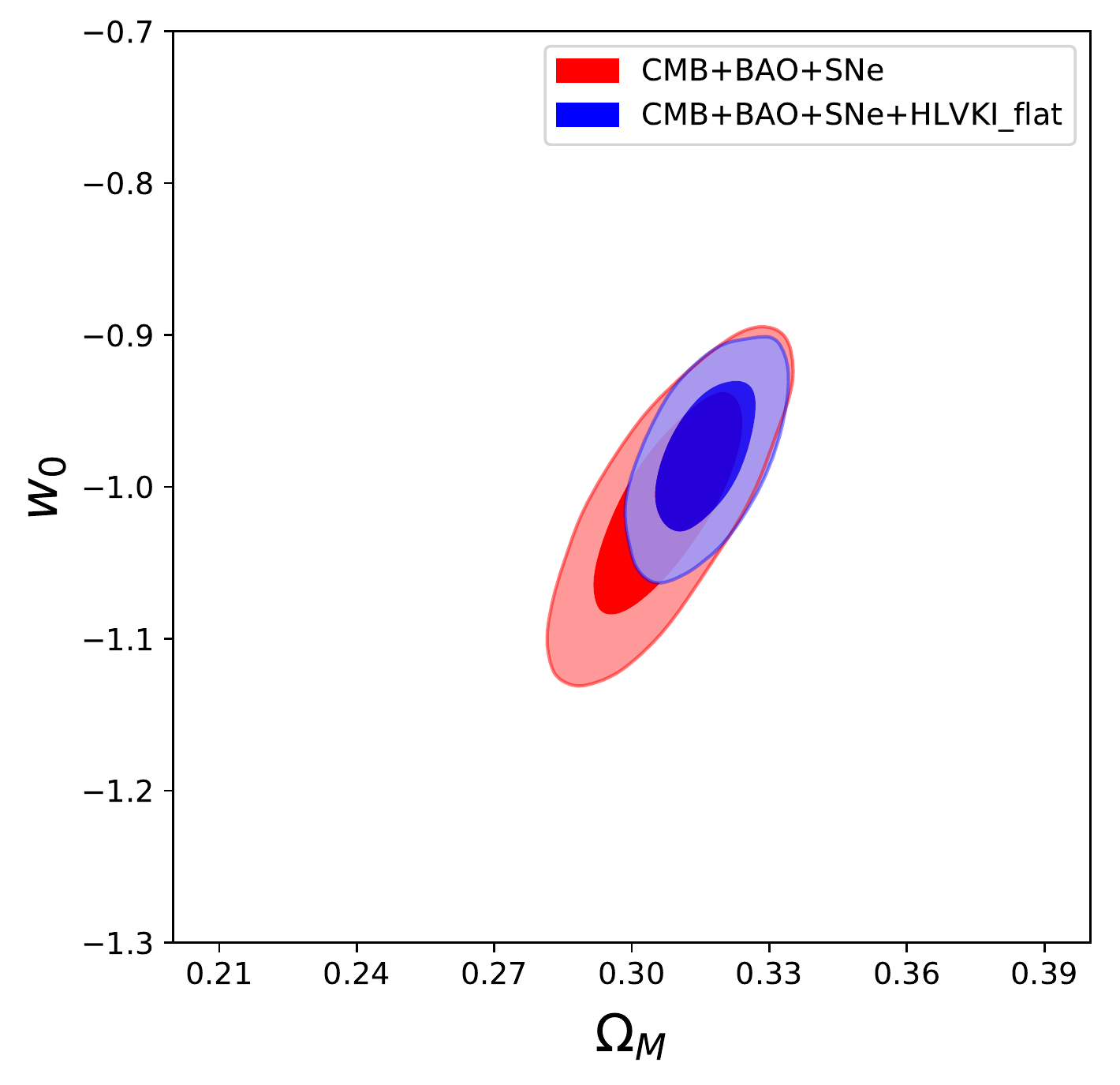}
\includegraphics[width=0.4\textwidth]{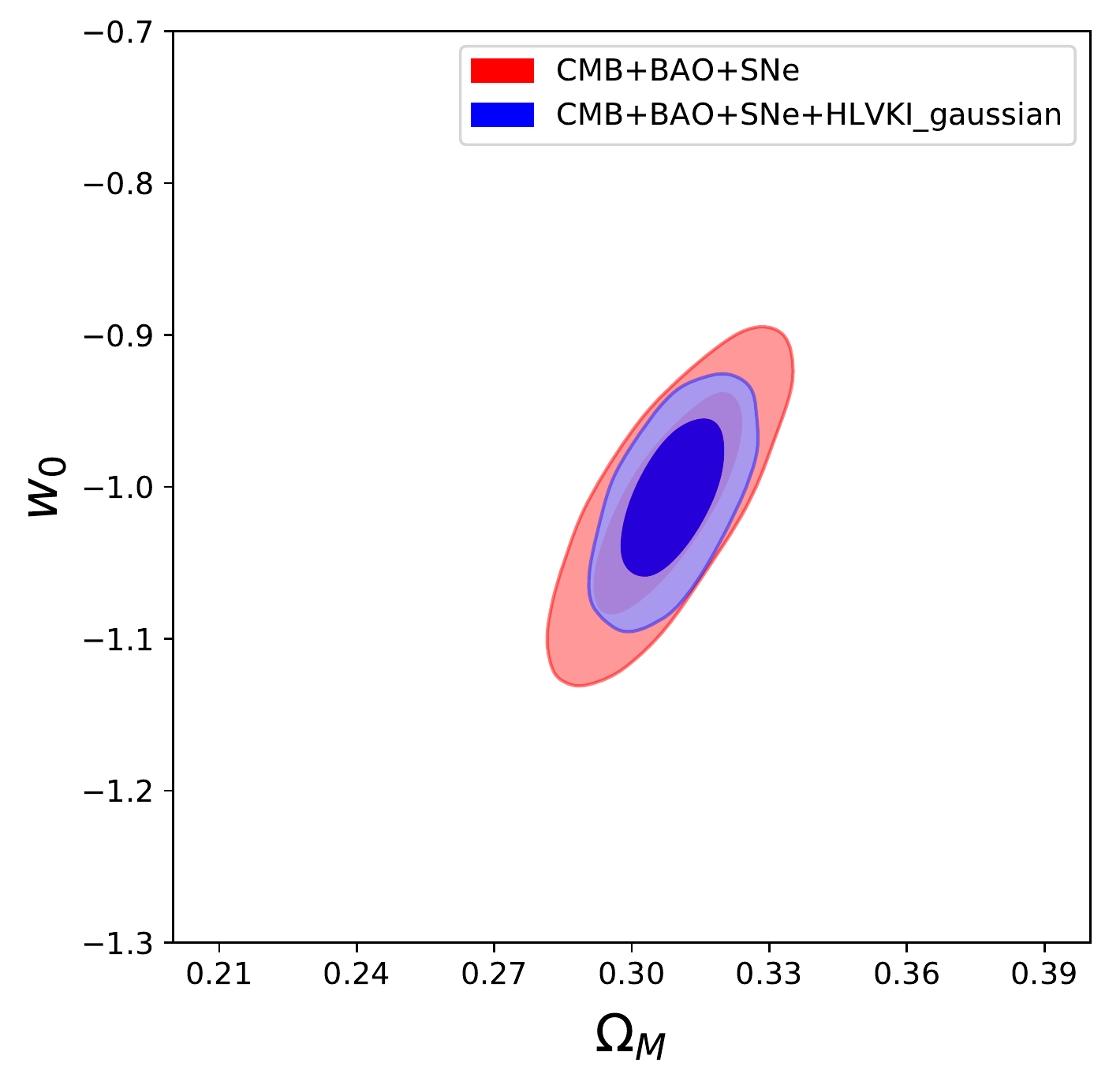}
\caption{The  $1\sigma$ and $2\sigma$
contours  of the two-dimensional likelihood in the $(\oma,w_0)$ plane, in $w$CDM, from CMB+BAO+SNe (red), and the result obtained by combining standard sirens at the HLVKI network with CMB+BAO+SNe (blue). Left: in the case of flat neutron star mass distribution. Right: in the case of gaussian neutron star mass distribution.}
\label{fig:HLVKI_w0}
\end{figure}

Fig.~\ref{fig:HLVKI_w0} shows the likelihood  in the $(\oma,w_0)$ plane in $w$CDM, i.e. when we introduce $w_0$ as the only new parameter that describes the DE sector, while setting $w_a=0$ and excluding also modified GW propagation, i.e setting $\Xi_0=1$, while
Table~\ref{tab:w0with2G} shows
the error on $w_0$  (at $1\sigma$, as in all our tables) from the corresponding one-dimensional likelihood. We give
the result from CMB+BAO+SNe only, and that obtained by  combining CMB+BAO+SNe with standard sirens with flat mass distribution or with the gaussian mass distribution.

\begin{figure}[t]
\centering
\includegraphics[width=0.4\textwidth]{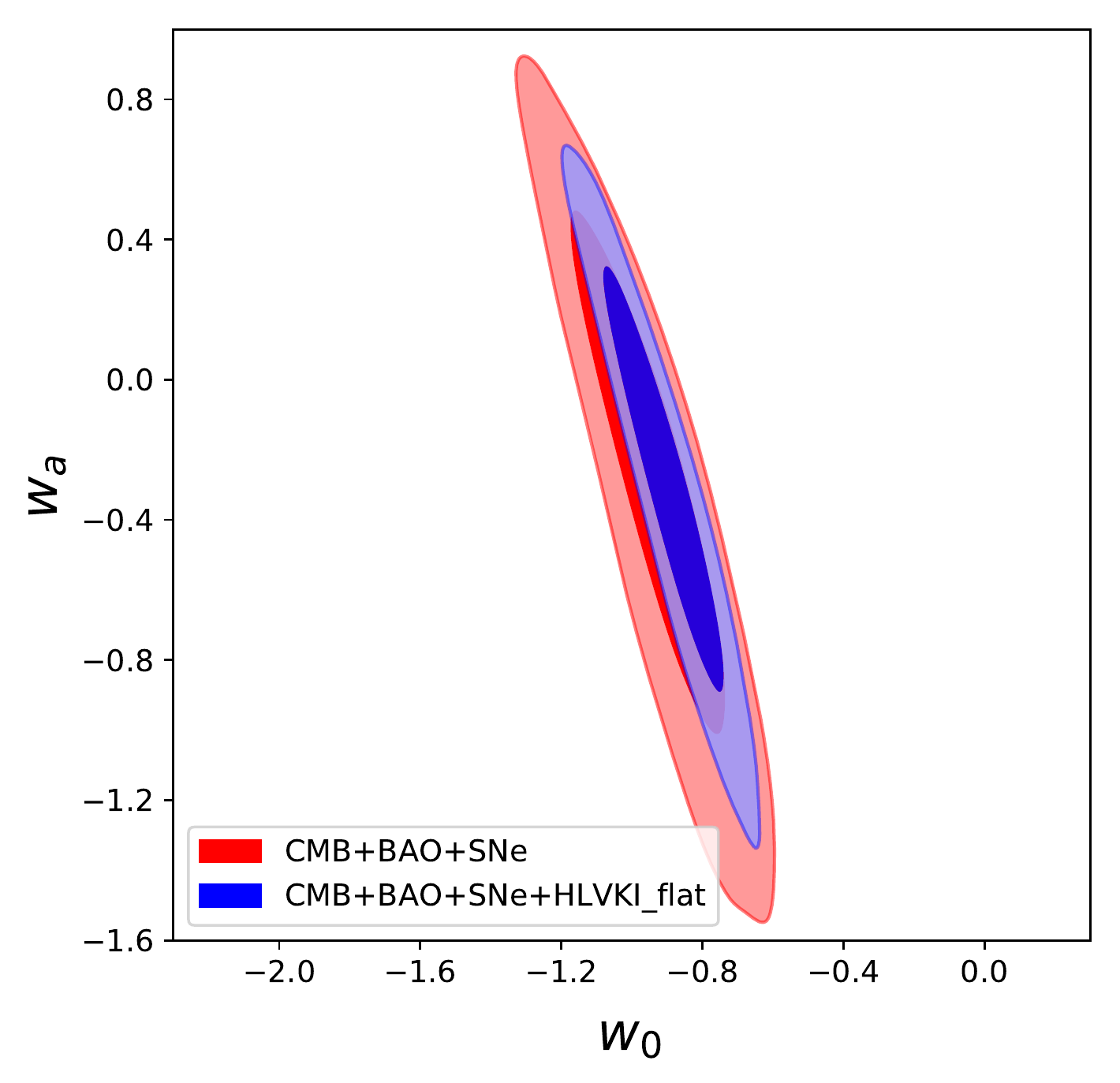}
\includegraphics[width=0.4\textwidth]{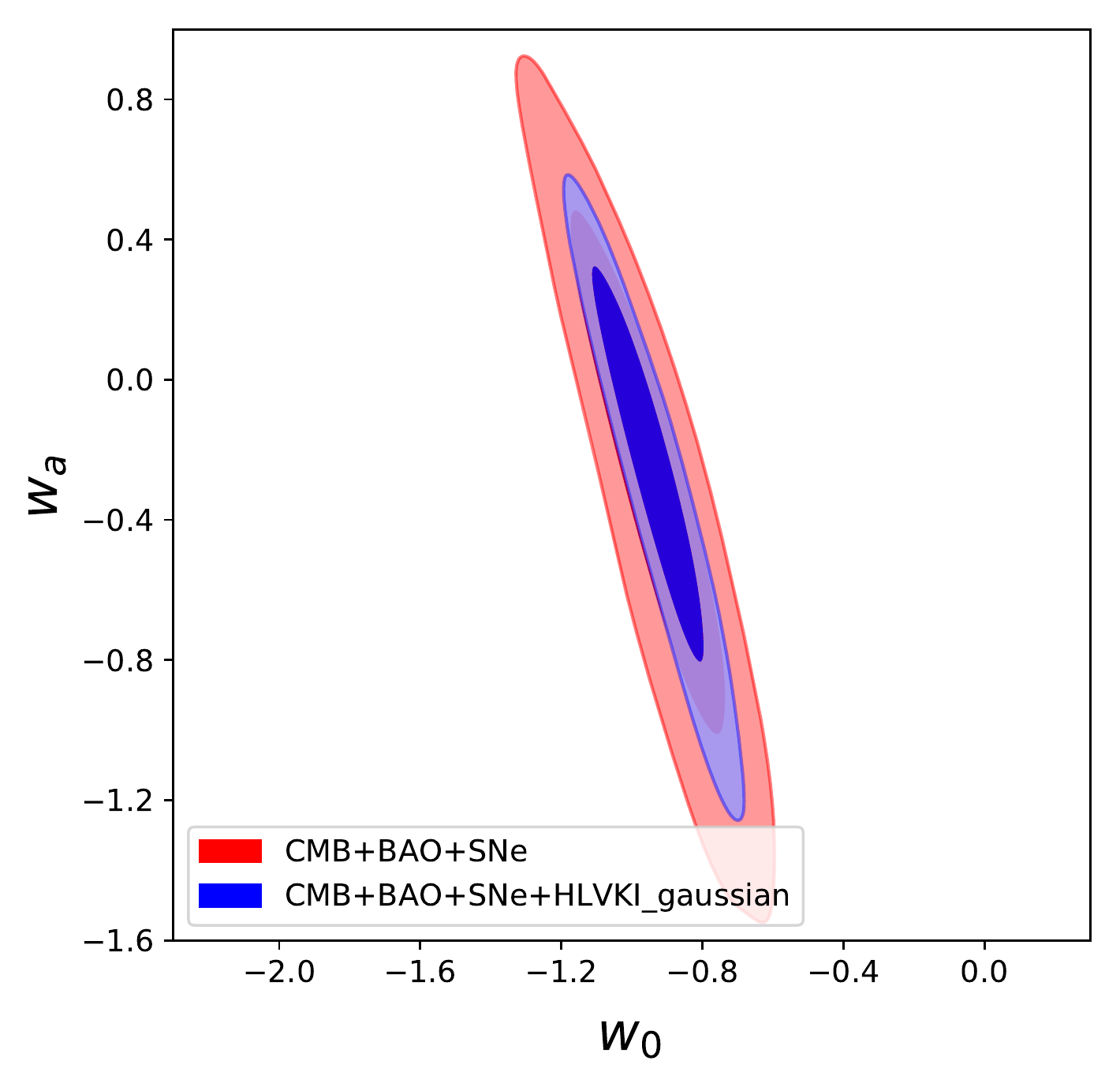}
\caption{The  $1\sigma$ and $2\sigma$
contours  of the two-dimensional likelihood in the $(w_0,w_a)$ plane, in the $(w_0,w_a)$ extension of the DE sector, from CMB+BAO+SNe (red), and the result obtained by combining standard sirens at the HLVKI network with CMB+BAO+SNe (blue). Left: in the case of flat neutron star mass distribution. Right: in the case of gaussian neutron star mass distribution.}
\label{fig:HLVKI_w0wa}
\end{figure}

The results for the  $(w_0,w_a)$ parametrization are shown in Fig.~\ref{fig:HLVKI_w0wa} and Table~\ref{tab:w0wawith2G}. Of course, enlarging the parameter space with one more parameter $w_a$ results in a larger error on $w_0$, compared to the results in Table~\ref{tab:w0with2G}.

\begin{figure}[t]
\centering
\includegraphics[width=0.4\textwidth]{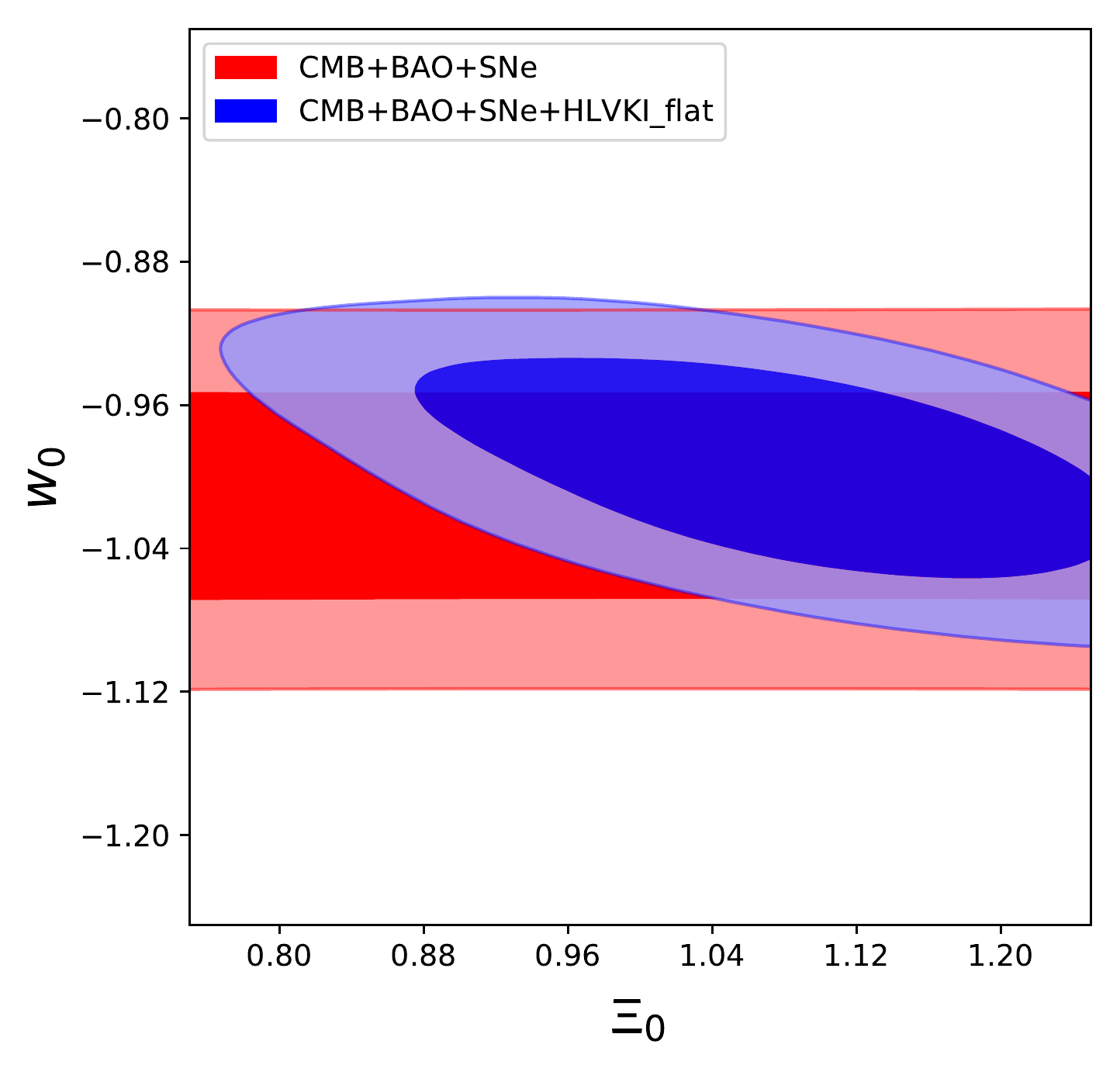}
\includegraphics[width=0.4\textwidth]{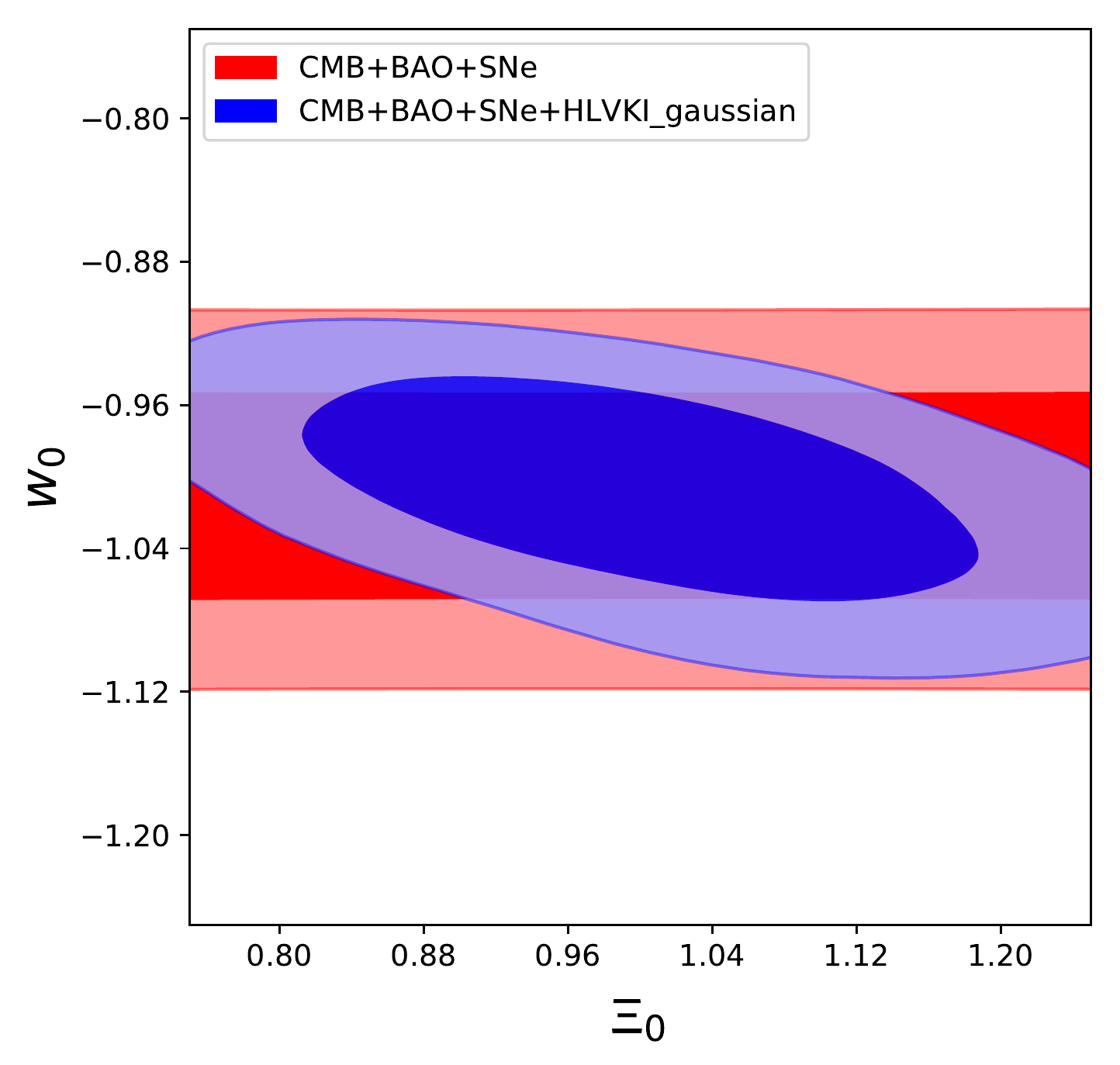}
\caption{The  $1\sigma$ and $2\sigma$
contours  of the two-dimensional likelihood in the $(\Xi_0,w_0)$ plane, in the $(\Xi_0,w_0)$ extension of the DE sector, from CMB+BAO+SNe (red), and the result obtained by combining standard sirens at the HLVKI network with CMB+BAO+SNe (blue). Left: in the case of flat neutron star mass distribution. Right: in the case of gaussian neutron star mass distribution.}
\label{fig:HLVKI_Xi0w0}
\end{figure}

We finally consider the $(\Xi_0,w_0)$ extension of the DE sector. The results are shown in Fig.~\ref{fig:HLVKI_Xi0w0} and Table~\ref{tab:w0Xi0with2G}. The most interesting result is the one for $\Xi_0$, which (with our rather extreme assumption of 10~yr of data taking) can be measured to an accuracy of order $\Delta\Xi_0\simeq 0.1$, i.e. (given that our fiducial value has been taken to be the $\Lambda$CDM value $\Xi_0=1$), a relative accuracy
$\Delta\Xi_0/\Xi_0\simeq 10\%$, which is already in the ballpark of the predictions of interesting modified gravity models~\cite{Belgacem:2019pkk,DeltaN64:inprep}. With a shorter but more realistic time  of data taking, say 3-4~yr, we still expect to get  $\Delta\Xi_0/\Xi_0$ at the level of about $20\%$.

Of course, CMB, BAO and SNe are blind to modified GW propagation, and the corresponding contour is flat in the $\Xi_0$ direction. Standard sirens lift this flat direction. In contrast, we see that the improvement on $w_0$ or $w_a$ from the inclusion of standard sirens is  quite modest.

\begin{table}[t]
\centering
\begin{tabular}{|c|c|c|c|}
 \hline
                               &  CMB+BAO+SNe& combined, flat & combined, gaussian     \\ \hline
$\Delta w_0$         &  0.045                     & 0.033                 & 0.035         \\
\hline
\end{tabular}
\caption{Accuracy ($1\sigma$ level) in the reconstruction of $w_0$ with  CMB+BAO+SNe only,  and  the combined result CMB+BAO+SNe+standard sirens, using the HLVKI detector network and the flat and gaussian mass distributions.
\label{tab:w0with2G}}
\end{table}

\begin{table}[t]
\centering
\begin{tabular}{|c|c|c|c|}
 \hline
                               &  CMB+BAO+SNe& combined, flat & combined, gaussian     \\ \hline
$\Delta w_0$         &  0.140                      & 0.113                 & 0.106         \\
$\Delta w_a$         &  0.483                      & 0.406                  & 0.380            \\
\hline
\end{tabular}
\caption{Accuracy ($1\sigma$ level) in the reconstruction of $w_0$ and $w_a$ with  CMB+BAO+SNe only,  and  the combined result CMB+BAO+SNe+standard sirens, using the HLVKI detector network and the flat and gaussian mass distributions.
\label{tab:w0wawith2G}}
\end{table}

%----------------------

\begin{table}[t]
\centering
\begin{tabular}{|c|c|c|c|}
 \hline
                               &  CMB+BAO+SNe& combined, flat & combined, gaussian     \\ \hline
$\Delta w_0$         &  0.045                     & 0.042                & 0.042         \\
$\Delta \Xi_0$         &  --                          & 0.130               & 0.125            \\
\hline
\end{tabular}
\caption{Accuracy ($1\sigma$ level) in the reconstruction of $w_0$ and $\Xi_0$ with  CMB+BAO+SNe only,  and  the combined result CMB+BAO+SNe+standard sirens, using the HLVKI detector network and the flat and gaussian mass distributions.
\label{tab:w0Xi0with2G}}
\end{table}

\clearpage
\newpage

\subsection{Results for ET}\label{sect:DEwithET}

\begin{figure}[h]
\centering
\includegraphics[width=0.4\textwidth]{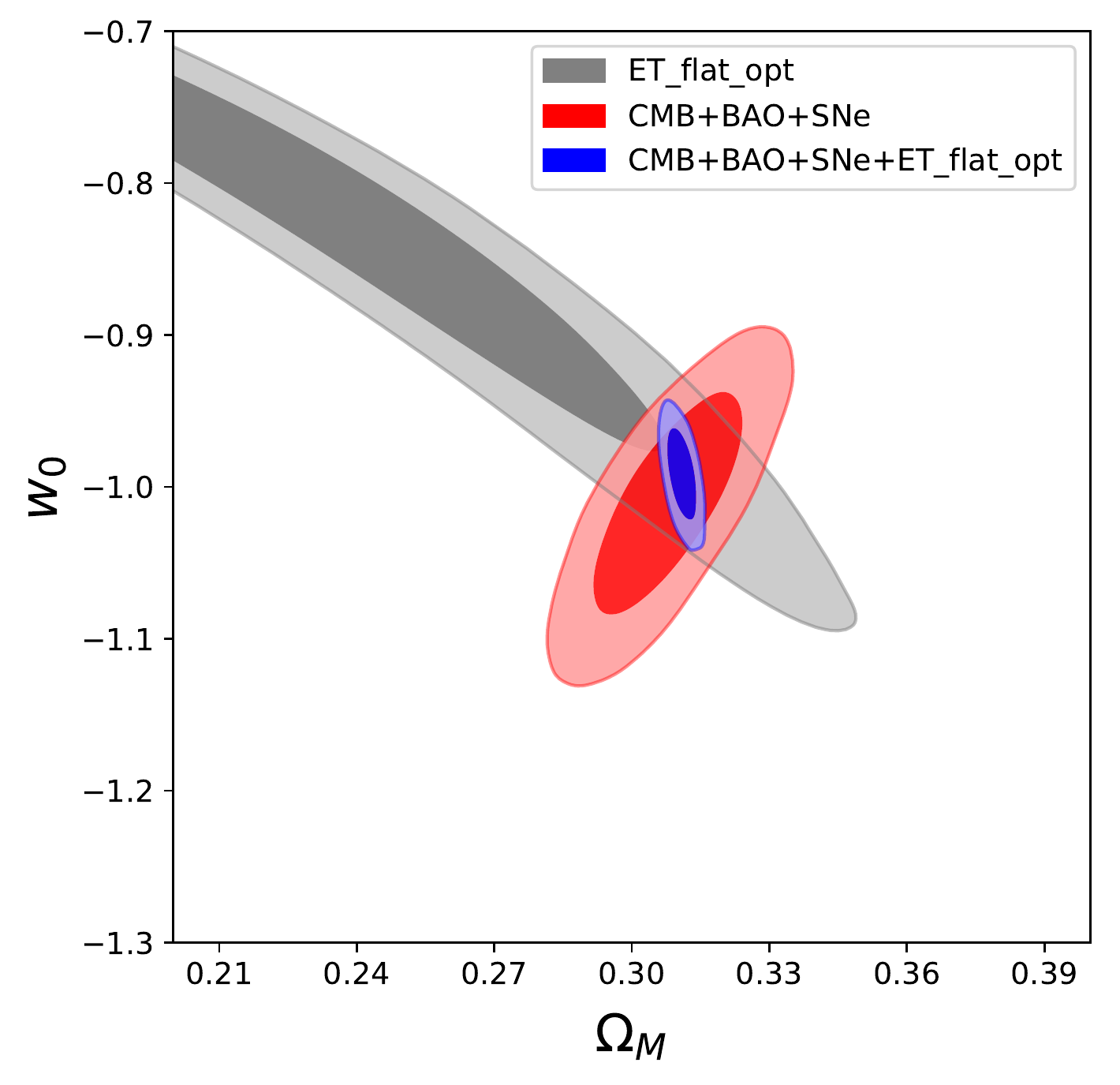}
\includegraphics[width=0.4\textwidth]{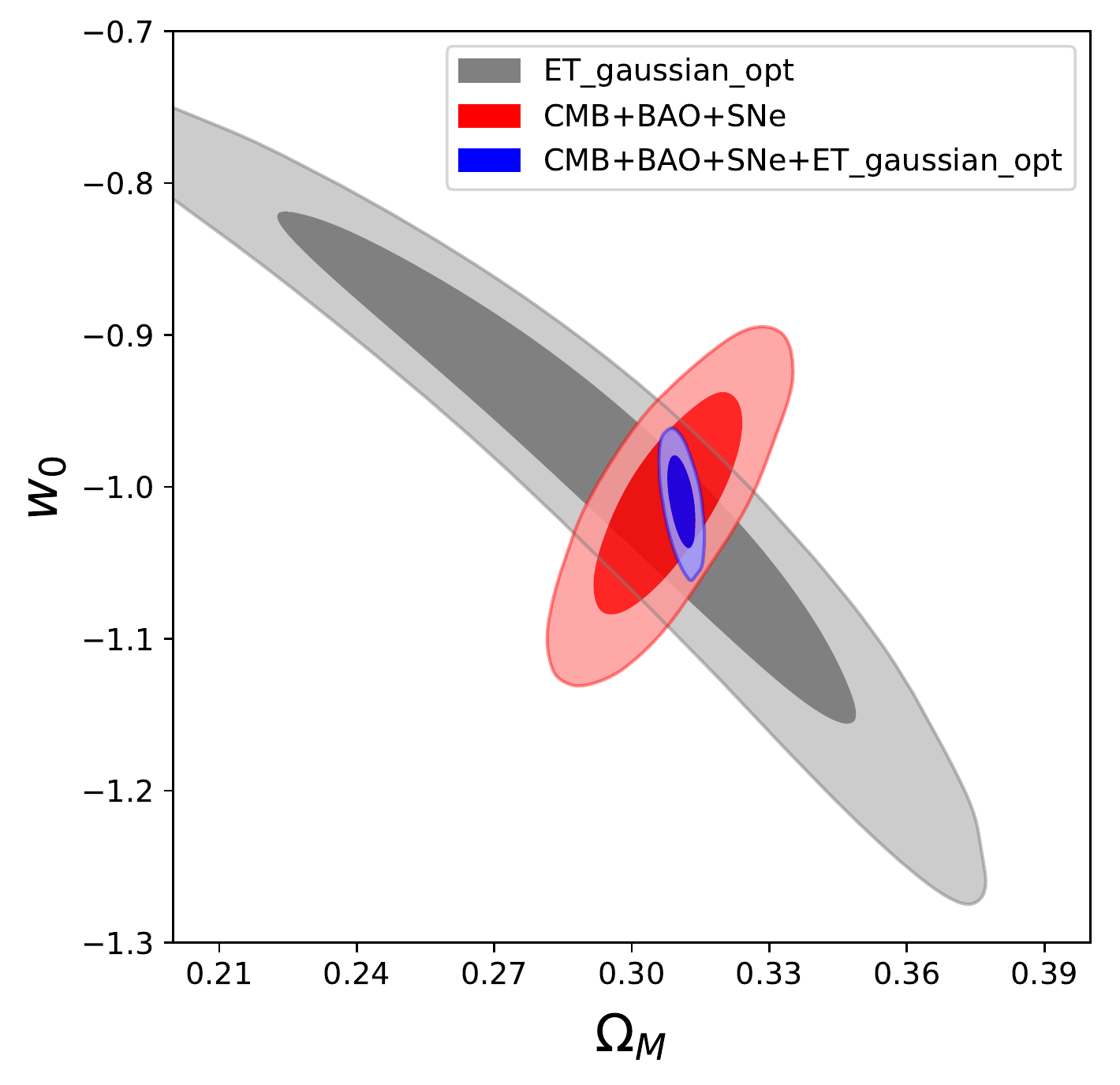}
\caption{The  $1\sigma$ and $2\sigma$
contours  of the two-dimensional likelihood in the $(\oma,w_0)$ plane, in $w$CDM, from CMB+BAO+SNe (red), joint detection of standard sirens at ET and THESEUS (gray) and the result obtained by combining standard sirens   with CMB+BAO+SNe (blue). Left: in the case of flat neutron star mass distribution. Right: in the case of gaussian neutron star mass distribution. We use the optimistic estimate for the FOV of THESEUS.}
\label{fig:ET_w0_opt}
\end{figure}

\begin{figure}[h]
\centering
\includegraphics[width=0.4\textwidth]{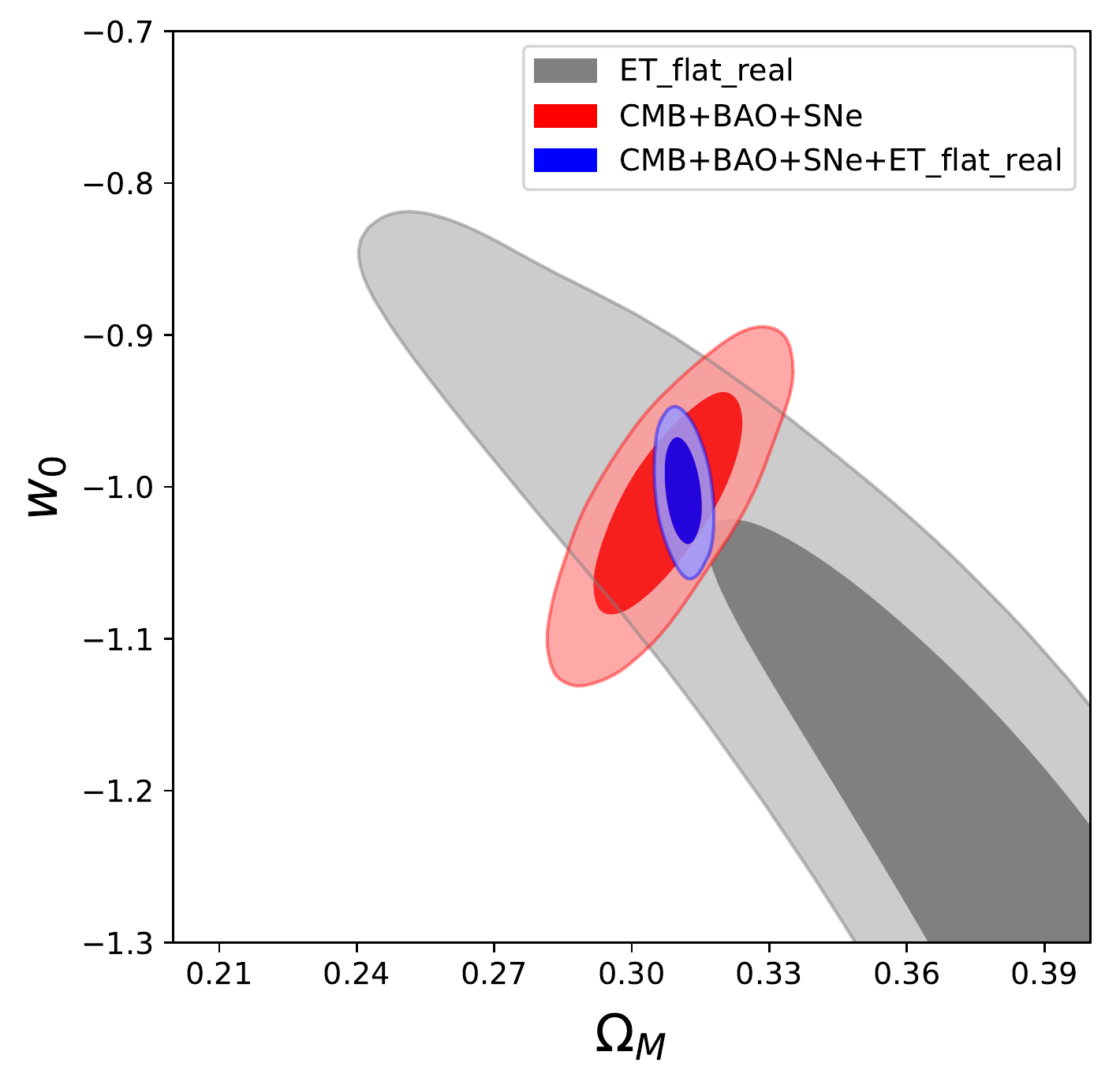}
\includegraphics[width=0.4\textwidth]{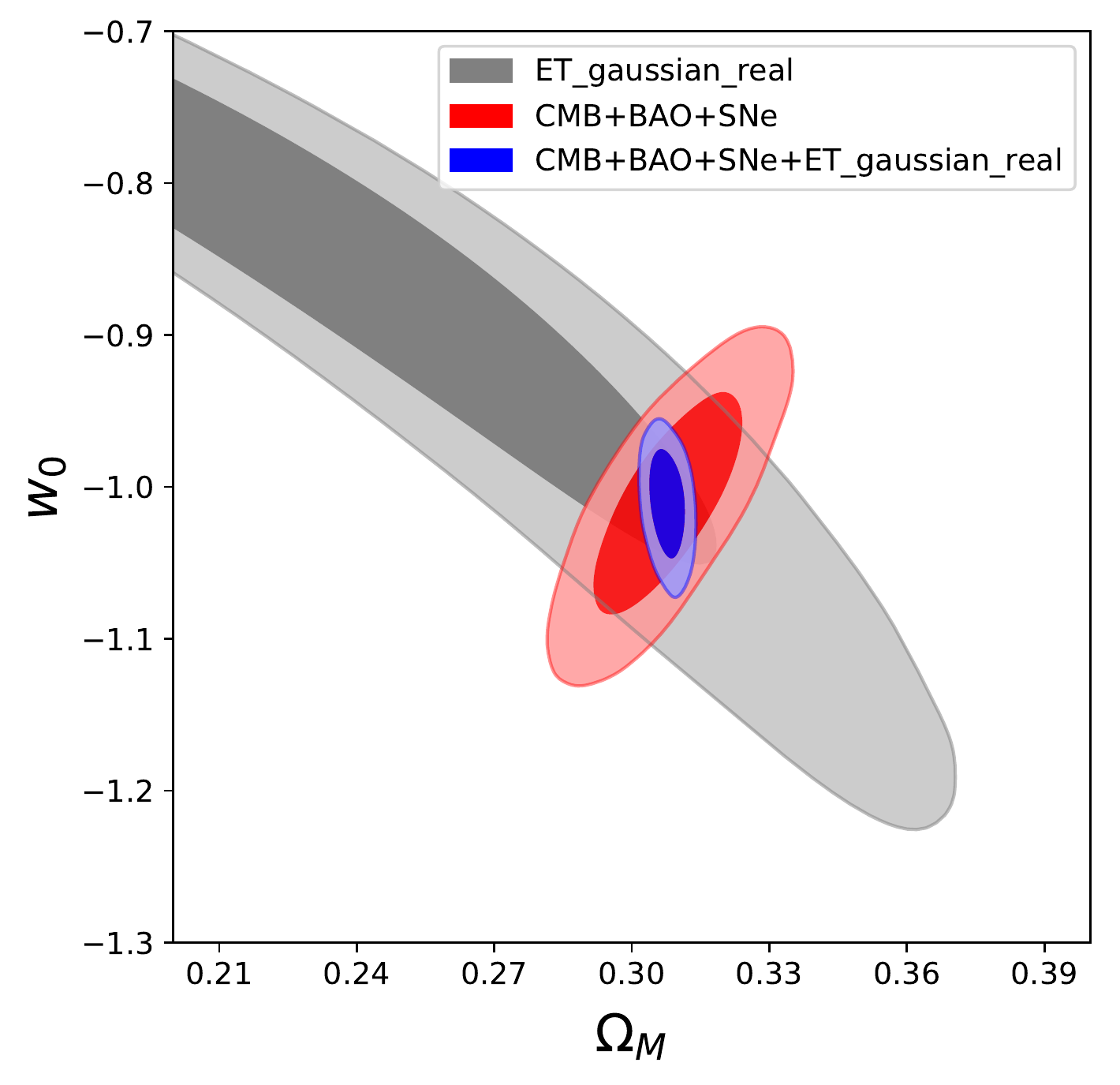}
\caption{As in Fig.~\ref{fig:ET_w0_opt}, with the realistic estimate for the FOV of THESEUS.}
\label{fig:ET_w0_real}
\end{figure}

We next consider a single ET in coincidence with a GRB detector with the characteristics of  THESEUS. Proceeding as before, we show first the results of the $w_0$ extension, in Tables~\ref{tab:w0withETopt} and \ref{tab:w0withETreal} and Figs.~\ref{fig:ET_w0_opt} and \ref{fig:ET_w0_real}, displaying separately the results for the optimistic and realistic FOV of THESEUS. Observe that, contrary to the HLVKI case, now we can obtain some bounds already using standard sirens alone. However, we see from the figures
that the central value of the contour of the standard sirens can happen to be displaced with respect to that from CMB+BAO+SNe. As we already discussed in Section~\ref{sect:LCDMwithET}, this is a statistical effect due to the random scattering of the mock GW data according to the error estimate that, depending on the specific realization, can induce a more or less significant displacement between the two contours.

We see that, adding the joint GW-GRB events to the CMB+BAO+SNe dataset, we can improve the accuracy on $w_0$ by about a factor of 2. This is interesting, although certainly not spectacular. We will see below that the most interesting contribution of a 3G detector such as ET to the exploration of the DE sector rather comes from modified GW propagation.

\vspace{20mm}

\begin{table}[hb]
\centering
\begin{tabular}{|c|c|c|c|c|c|}
 \hline
                               &  CMB+BAO+SNe&ET,      &ET,           & combined,  & combined,      \\
                               &                               &flat      &gaussian   &flat              & gaussian \\ \hline
$\Delta w_0$         &  0.045                    & 0.109  & 0.116      & 0.020          &0.021     \\
\hline
\end{tabular}
\caption{Accuracy ($1\sigma$ level) in the reconstruction of $w_0$ with   only CMB+BAO+SNe,  with only standard sirens (with the flat and gaussian mass distributions, respectively) and  the combined results CMB+BAO+SNe+standard sirens, using ET  and THESEUS and assuming the optimistic FOV of THESEUS.
\label{tab:w0withETopt}}
\end{table}

\vspace{20mm}

\begin{table}[hb]
\centering
\begin{tabular}{|c|c|c|c|c|c|}
 \hline
                               &  CMB+BAO+SNe&ET,      &ET,           & combined,  & combined,      \\
                               &                               &flat      &gaussian   &flat              & gaussian \\ \hline
$\Delta w_0$         &  0.045                    & 0.301  & 0.158      & 0.023          &0.024     \\
\hline
\end{tabular}
\caption{As in Table~\ref{tab:w0withETopt}, assuming the realistic FOV of THESEUS.
\label{tab:w0withETreal}}
\end{table}

\clearpage\newpage

\clearpage\newpage

\begin{figure}[bht]
\centering
\includegraphics[width=0.4\textwidth]{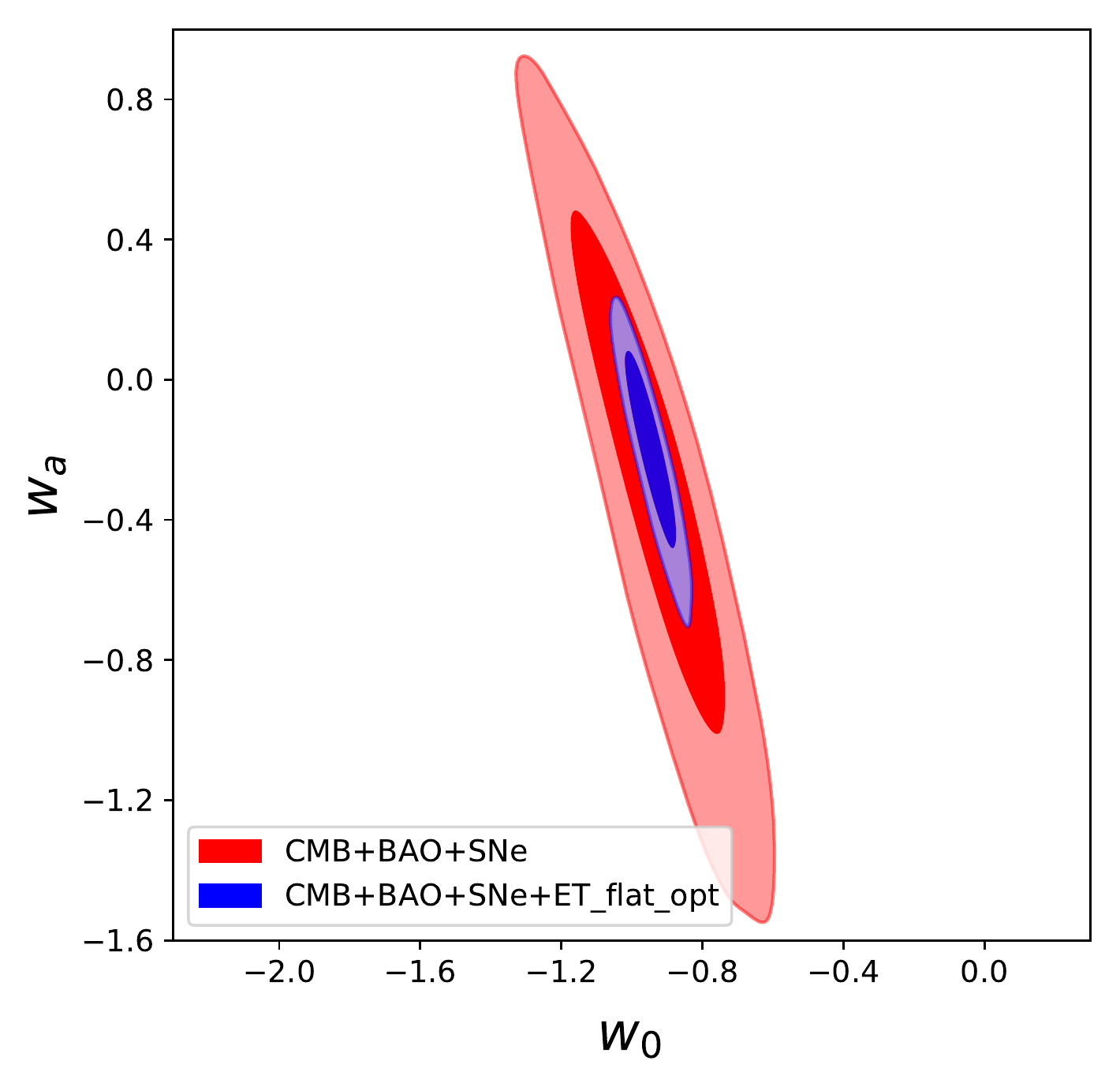}
\includegraphics[width=0.4\textwidth]{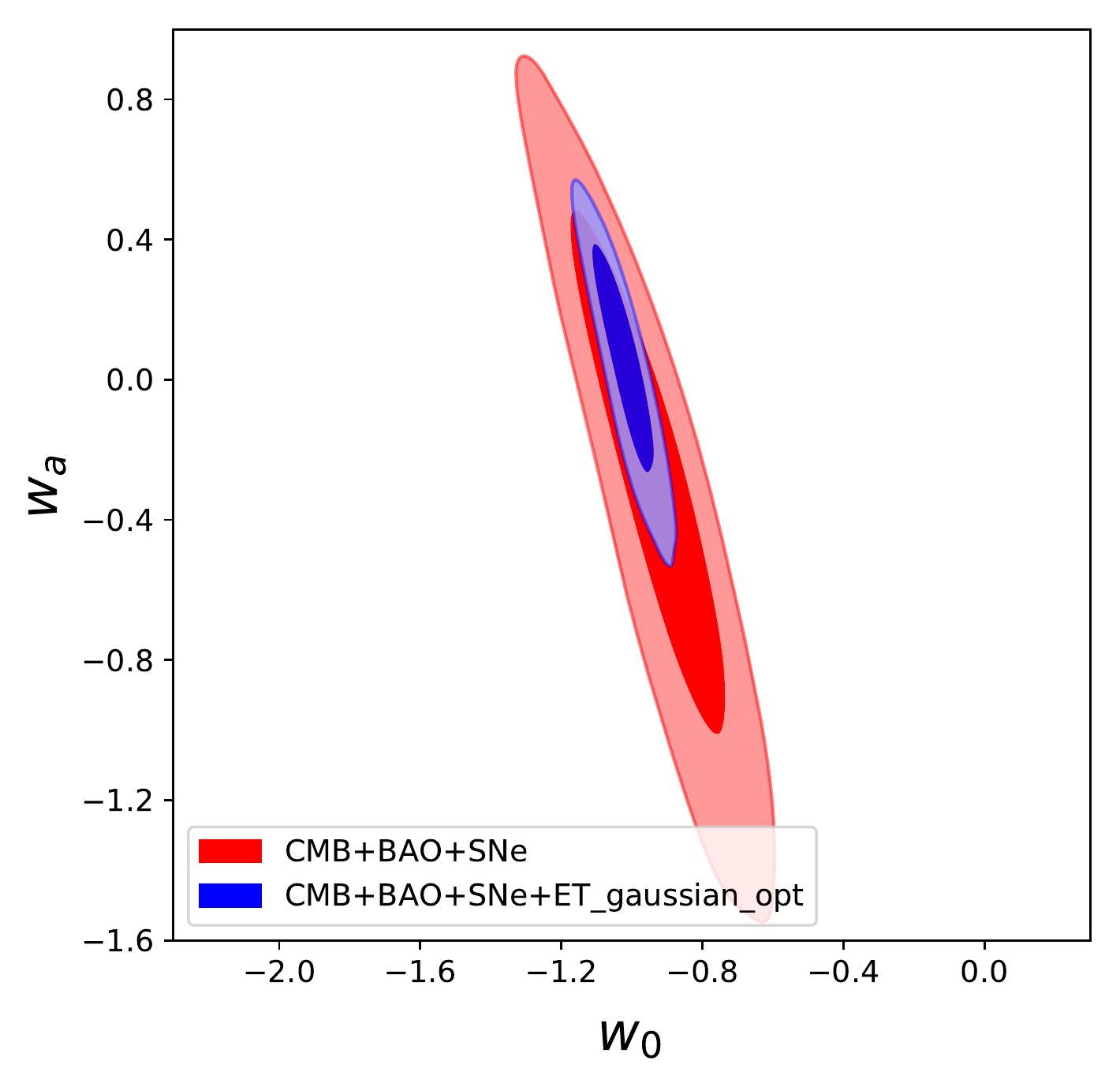}
\caption{The  $1\sigma$ and $2\sigma$
contours  of the two-dimensional likelihood in the $(w_0,w_a)$ plane from CMB+BAO+SNe (red) and the result obtained by combining standard sirens  at ET with CMB+BAO+SNe (blue). Left: in the case of flat neutron star mass distribution. Right: in the case of gaussian neutron star mass distribution. We use the optimistic estimate for the FOV of THESEUS.}
\label{fig:ET_w0wa_opt}
\end{figure}

\begin{figure}[hb]
\centering
\includegraphics[width=0.4\textwidth]{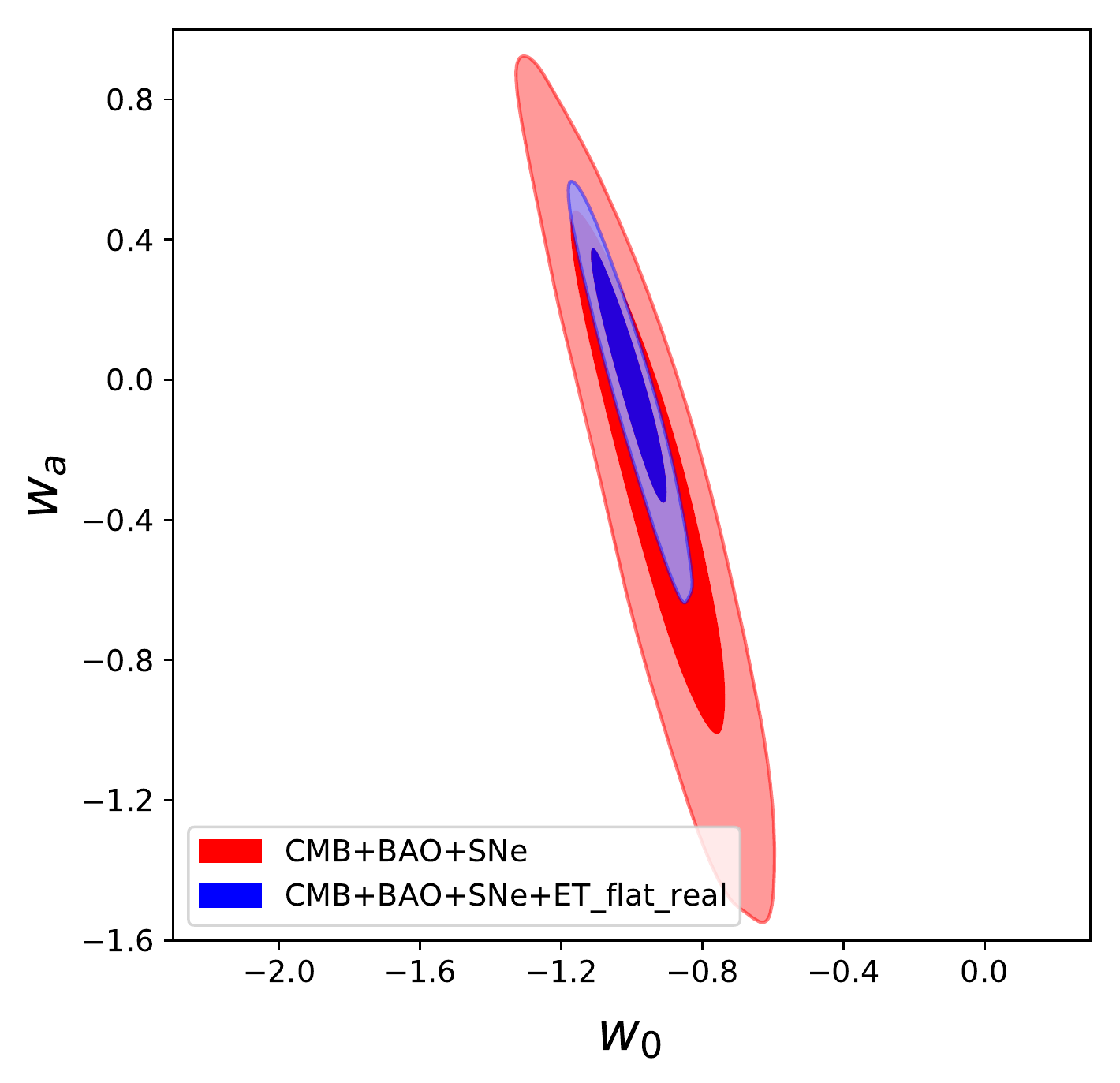}
\includegraphics[width=0.4\textwidth]{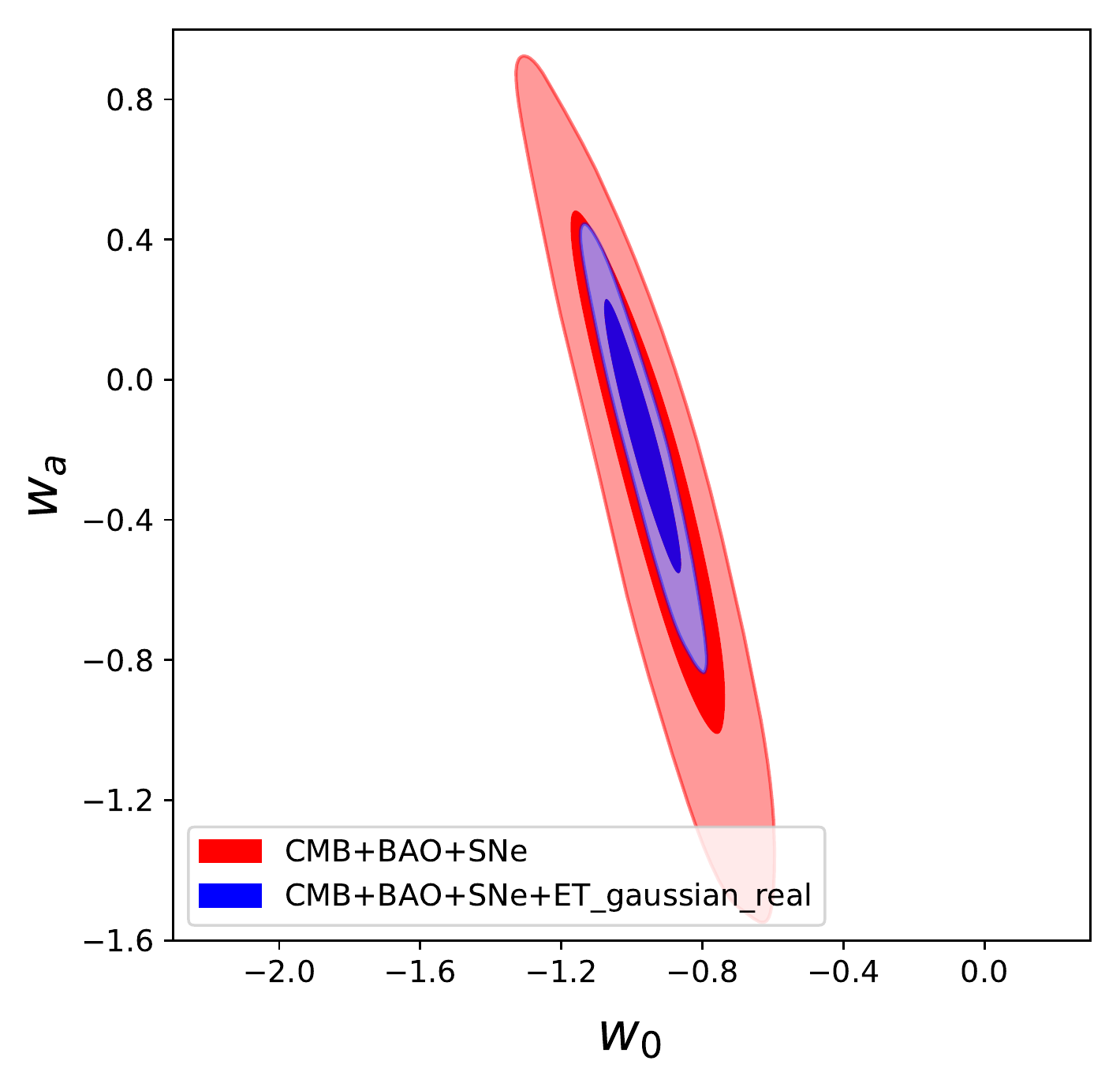}
\caption{As in Fig.~\ref{fig:ET_w0wa_opt}, with the realistic estimate for the FOV of THESEUS.}
\label{fig:ET_w0wa_real}
\end{figure}

We next include $(w_0,w_a)$ as extra parameters. The results are shown in Tables~\ref{tab:w0wawithETopt} and \ref{tab:w0wawithETreal} and in  Figs.~\ref{fig:ET_w0wa_opt} and \ref{fig:ET_w0wa_real}. In this case, with one extra parameter $w_a$, for some scenarios the MCMC does not converge well with standard sirens only, and we just show the combined result for CMB+BAO+SNe+standard sirens, for all scenarios.

\vspace{20mm}

\begin{table}[bh]
\centering
\begin{tabular}{|c|c|c|c|c|c|}
 \hline
                               &  CMB+BAO+SNe & combined,  & combined,      \\
                               &                               &flat               & gaussian \\ \hline
$\Delta w_0$         &  0.140                    & 0.050           & 0.058      \\
$\Delta w_a$         &  0.483                    & 0.193           & 0.224       \\
\hline
\end{tabular}
\caption{Accuracy ($1\sigma$ level) in the reconstruction of $(w_0,w_a)$ with   only CMB+BAO+SNe and  the combined results CMB+BAO+SNe+standard sirens, using ET  and THESEUS and assuming the optimistic FOV of THESEUS.
\label{tab:w0wawithETopt}}
\end{table}

\vspace{20mm}

\begin{table}[bh]
\centering
\begin{tabular}{|c|c|c|c|c|c|}
 \hline
                               &  CMB+BAO+SNe & combined,  & combined,      \\
                               &                               &flat               & gaussian \\ \hline
$\Delta w_0$         &  0.140                    & 0.073           & 0.072      \\
$\Delta w_a$         &  0.483                    & 0.246           & 0.260       \\
\hline
\end{tabular}
\caption{As in Table~\ref{tab:w0wawithETopt}, assuming the realistic FOV of THESEUS.
\label{tab:w0wawithETreal}}
\end{table}

\clearpage\newpage

\vspace{5mm}

\clearpage\newpage
%-----------------------

\begin{figure}[h]
\centering
\includegraphics[width=0.4\textwidth]{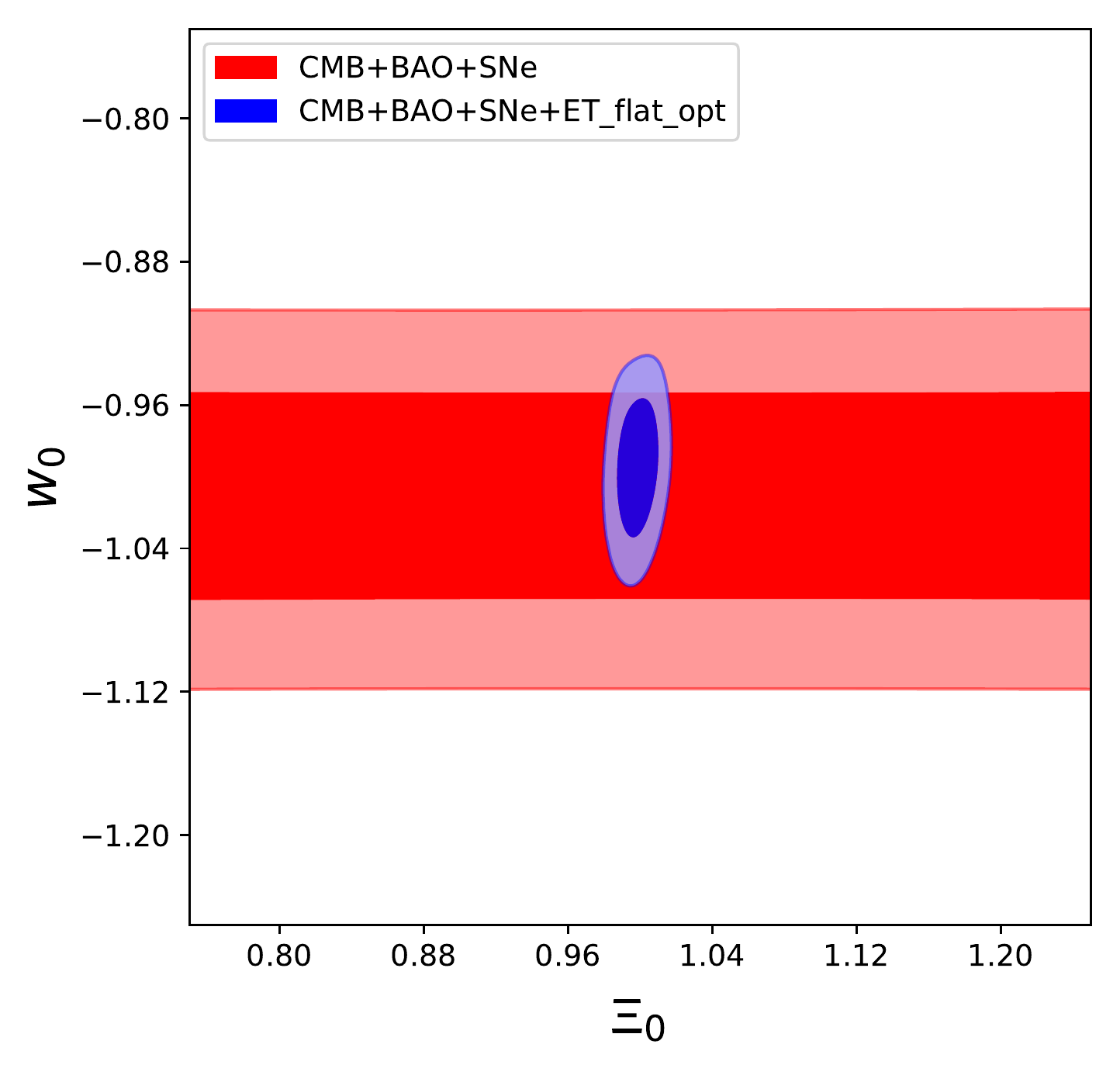}
\includegraphics[width=0.4\textwidth]{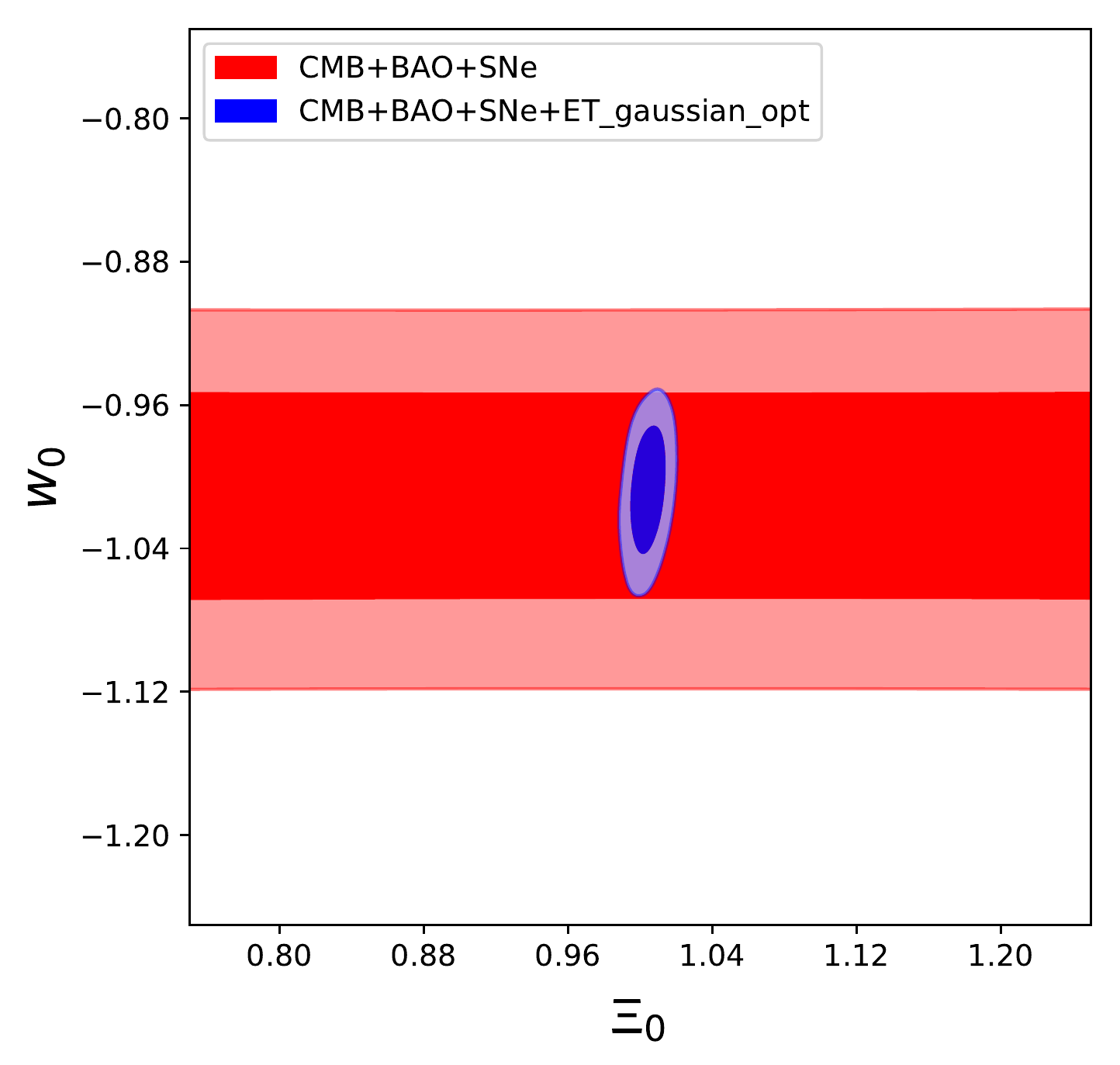}
\caption{The  $1\sigma$ and $2\sigma$
contours  of the two-dimensional likelihood in the $(\Xi_0,w_0)$ plane from CMB+BAO+SNe (red) and the result obtained by combining  joint detections of standard sirens  at ET and THESEUS  with CMB+BAO+SNe (blue). Left: in the case of flat neutron star mass distribution. Right: in the case of gaussian neutron star mass distribution. We use the optimistic estimate for the FOV of THESEUS.}
\label{fig:ET_w0xi0_opt}
\end{figure}

\begin{figure}[hb]
\centering
\includegraphics[width=0.4\textwidth]{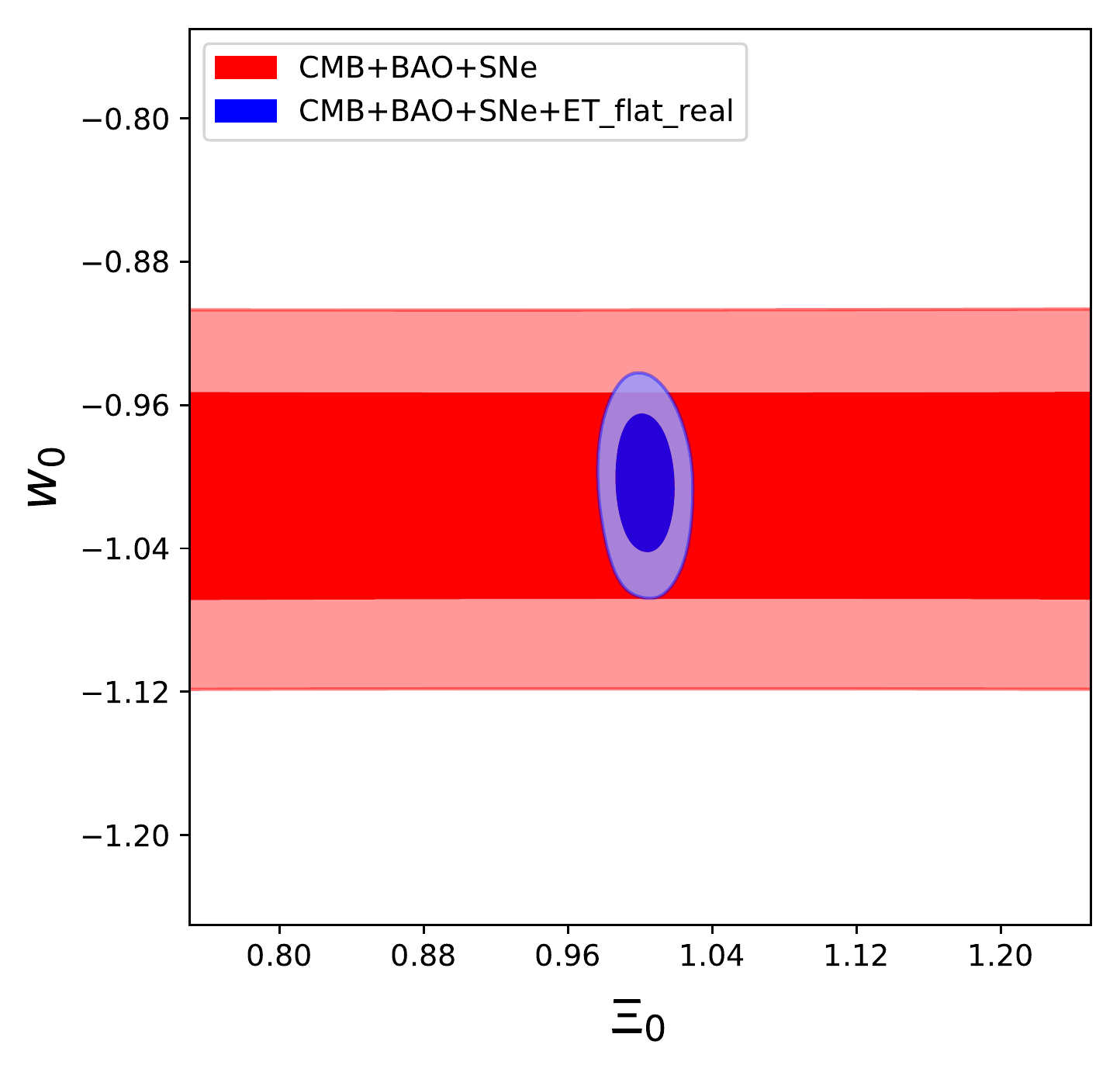}
\includegraphics[width=0.4\textwidth]{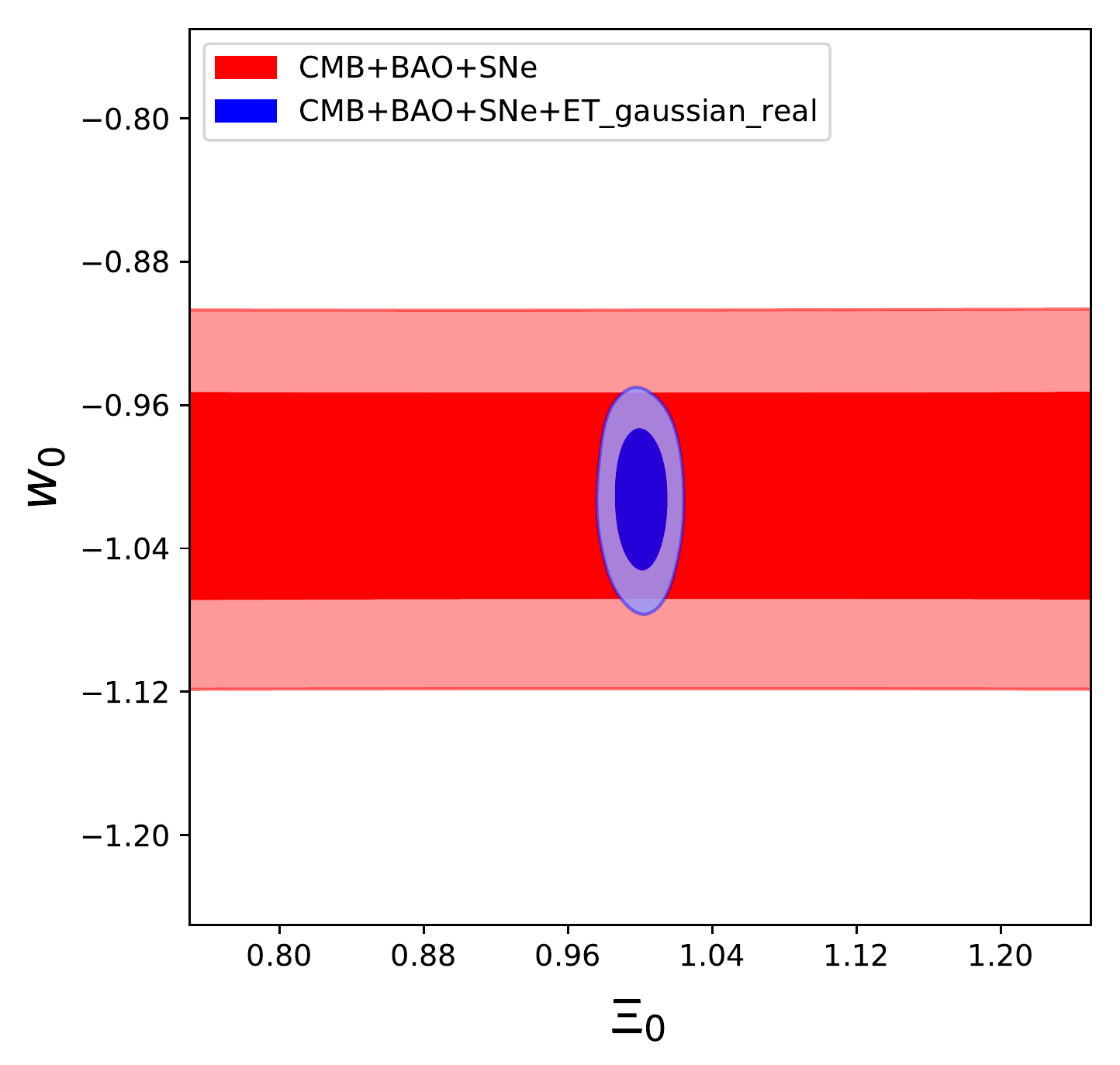}
\caption{As in Fig.~\ref{fig:ET_w0xi0_opt}, with the realistic estimate for the FOV of THESEUS.}
\label{fig:ET_w0xi0_real}
\end{figure}

We finally include modified GW propagation, adding to the baseline  $\Lambda$CDM model the parameters
$(\Xi_0,w_0)$, and writing the GW luminosity distance as in \eq{eq:fit}. We set for simplicity  $n=5/2$, which is of the order of the value predicted by the RT and RR nonlocal gravity model. However, the precise value of $n$ is of limited relevance for the analysis.  Note that, since our catalog of sources has been generated assuming $\Lambda$CDM as our fiducial model,  our fiducial values for these parameters are $\Xi_0=1$ and $w_0=-1$, and we compute the accuracy
$\Delta\Xi_0$ and $\Delta w_0$ to which we can find back these values. The results are shown in Tables~\ref{tab:w0xi0withETopt} and \ref{tab:w0xi0withETreal} and in Figs.~\ref{fig:ET_w0xi0_opt} and \ref{fig:ET_w0xi0_real}. 

It is quite remarkable that, by combining the joint GW-GRB detections from a single ET detector and a GRB detector such as THESEUS, with the current CMB+BAO+SNe dataset, we can reach an accuracy on $\Xi_0$ that, depending on the scenario, is between $0.7\%$ and $1.1\%$. By comparison, the `minimal' RT nonlocal model predicts a deviation from $\Xi_0=1$ at the level of $6.6\%$, almost an order of magnitude larger than this observational sensitivity (and much large values can be obtained with initial conditions set during inflation~\cite{DeltaN64:inprep}). This shows that, while the sensitivity of a single ET detector to $w_0$ will not allow us to obtain a dramatic improvement on the current knowledge of $w_0$, the sensitivity to $\Xi_0$ is extremely interesting and well within the prediction of viable modified gravity models. Furthermore, modified GW propagation, as encoded for instance in the $\Xi_0$ parameter [or, more generally, in the $(\Xi_0,n)$ parameters] is an observable specific to GW detectors, to which electromagnetic observations are blind.

The results of this section can be compared to those  in \cite{Belgacem:2018lbp}, which, following  \cite{Sathyaprakash:2009xt,Zhao:2010sz,Cai:2016sby}, were obtained under the  working hypothesis that ET, over a few years of data taking,  will eventually be able to collect ${\cal O}(10^3)$ BNS with counterpart (without specifying how the counterpart is actually detected), and assuming a redshift distribution proportional to a simple model for the star formation rate (and neglecting the effect of the  delay between binary formation and merger). For the $(w_0,\Xi_0)$ extension of the DE sector, the analysis of ref.~\cite{Belgacem:2018lbp} then led to the forecast
$\Delta\Xi_0=0.008$ and
$\Delta w_0=0.032$, which happens to be very close to the results in Table~\ref{tab:w0xi0withETopt}.

Comparing with our present results, first of all we see from Table~\ref{tab:3G} that the assumption of $10^3$ standard sirens with counterpart, while optimistic, was not unrealistic; in particular, for a gaussian neutron-star mass distribution (and with the optimistic estimate for the FOV of THESEUS), in 10~yr of data we could have  ${\cal O}(500)$ joint GW-GRB events. Furthermore, as we have already mentioned in Section~\ref{sect:count3G},  
in \cite{Zhao:2010sz} was used a threshold for the network SNR obtained combining the three arms of ET given by $\rho_{\rm  threshold}=8$ while we use $\rho_{\rm  threshold}=12$. Lowering the threshold would lead to an increase in the number of GW events.
Still, at first  it could be  surprising that the result for $\Delta\Xi_0$ that we find in this paper happens to be practically identical to that of ref.~\cite{Belgacem:2018lbp}, given  that the number of sources that we are using  here is smaller by a factor $\simeq 2$ compared to the $10^3$ sources used in 
ref.~\cite{Belgacem:2018lbp}. However,  this can be traced to the fact that also the redshift distribution of the sources  is different. Indeed, we see from the right-panel in Fig.~\ref{fig:redshift_dist} that most of the joint GW-GRB detections are at $z<0.5$, while in ref.~\cite{Belgacem:2018lbp} it was assumed that the catalog of sources followed a distribution in redshift determined by the star formation rate; that catalog was peaked at $z\simeq 1$, with long tails at larger $z$, see Fig.~8 of ref.~\cite{Belgacem:2018lbp}. On the other hand, it was also found in  ref.~\cite{Belgacem:2018lbp} that the main contribution to the determination of $\Xi_0$ 
was given by the sources at $z<0.7$, that were about one half of the total, so in the end it is not surprising that our catalog, with about a factor of 2 less sources, but almost all concentrated at $z<0.7$, gives basically the same results as the catalog used in  ref.~\cite{Belgacem:2018lbp}.

%\vspace{5mm}

\begin{table}[t]
\centering
\begin{tabular}{|c|c|c|c|c|c|}
 \hline
                               &  CMB+BAO+SNe & combined,  & combined,      \\
                               &                               &flat               & gaussian \\ \hline
$\Delta w_0$         &  0.045                    & 0.026           & 0.024      \\
$\Delta \Xi_0$       &  --                          & 0.008           & 0.007       \\
\hline
\end{tabular}
\caption{Accuracy ($1\sigma$ level) in the reconstruction of $(w_0,\Xi_0)$ with   only CMB+BAO+SNe and  the combined results CMB+BAO+SNe+standard sirens, using ET  and THESEUS and assuming the optimistic FOV of THESEUS.
\label{tab:w0xi0withETopt}}
\end{table}

%\vspace{20mm}

\begin{table}[h]
\centering
\begin{tabular}{|c|c|c|c|c|c|}
 \hline
                               &  CMB+BAO+SNe & combined,  & combined,      \\
                               &                               &flat               & gaussian \\ \hline
$\Delta w_0$         &  0.045                    & 0.026           & 0.026      \\
$\Delta \Xi_0$       &  --                          & 0.011           & 0.010       \\
\hline
\end{tabular}
\caption{As in Table~\ref{tab:w0xi0withETopt}, assuming the realistic FOV of THESEUS.
\label{tab:w0xi0withETreal}}
\end{table}

%\vspace{20mm}
\clearpage\newpage

\vspace{10mm}

%-----------------------xi0

\clearpage\newpage

\subsection{Results for ET+CE+CE}\label{sect:DEwithET+CE+CE}

\begin{figure}[hb]
\centering
\includegraphics[width=0.4\textwidth]{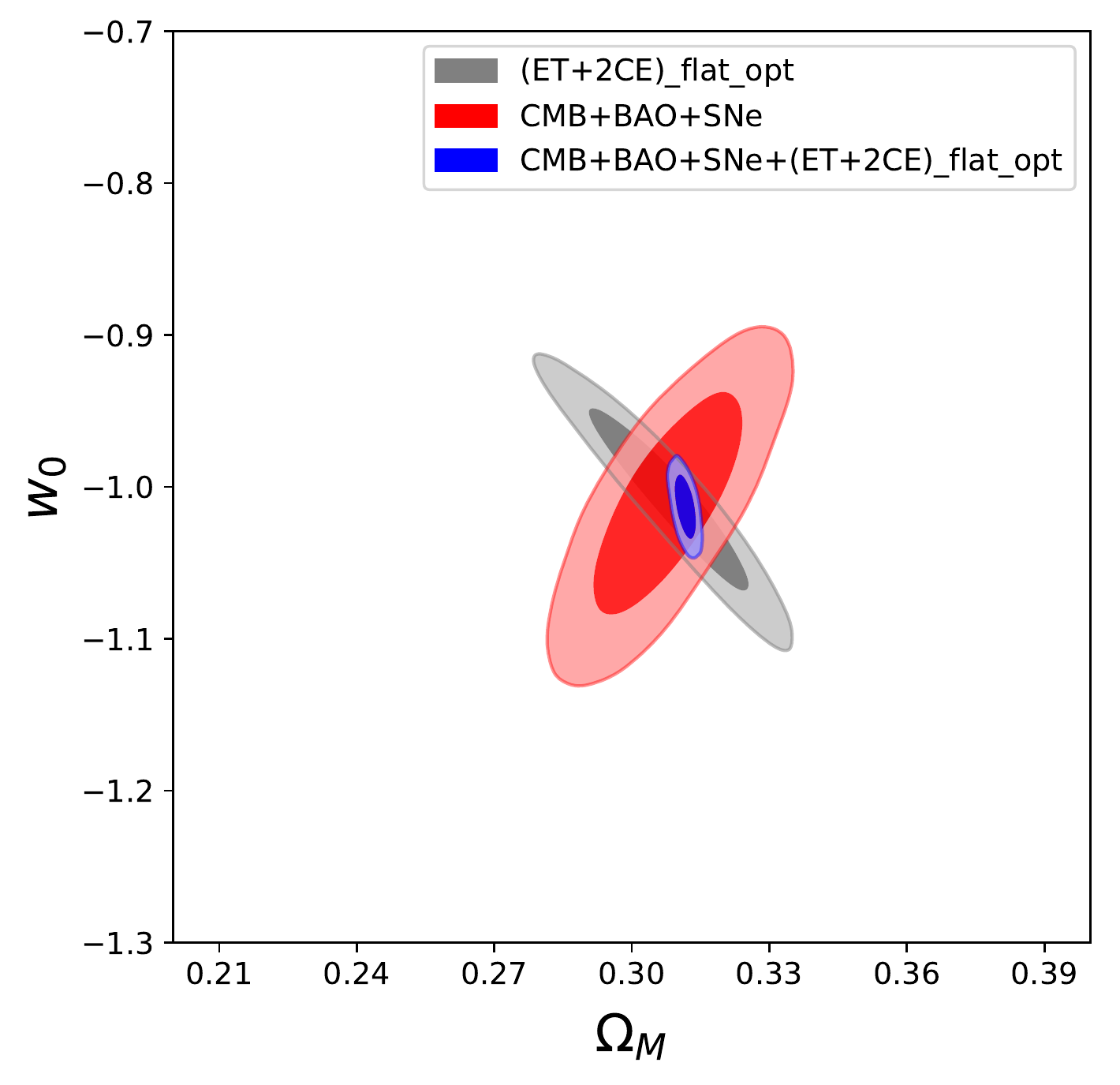}
\includegraphics[width=0.4\textwidth]{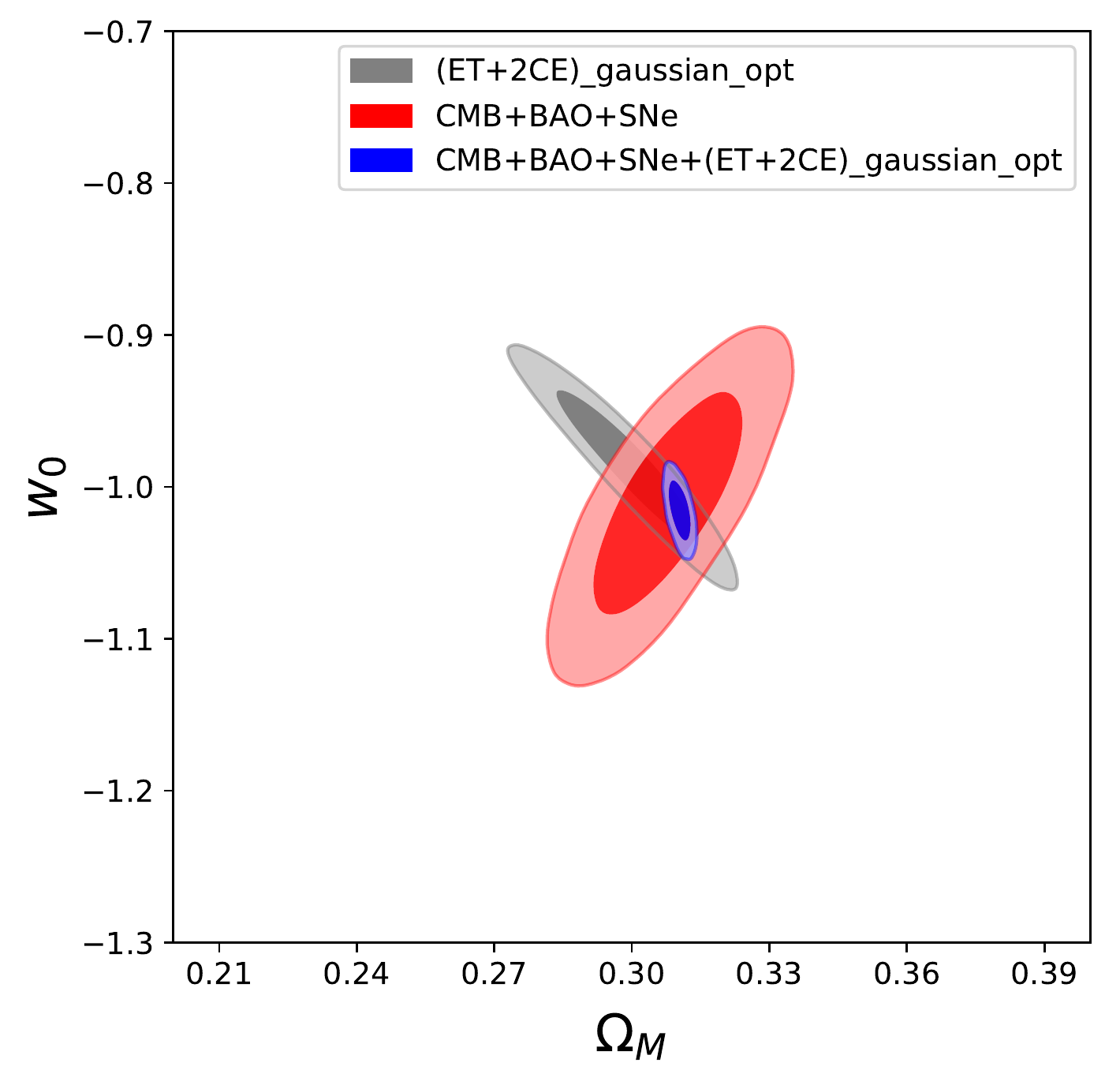}
\caption{The  $1\sigma$ and $2\sigma$
contours  of the two-dimensional likelihood in the $(\oma,w_0)$ plane, in $w$CDM, from CMB+BAO+SNe (red), joint detection of standard sirens at ET+CE+CE and THESEUS (gray) and the result obtained by combining standard sirens   with CMB+BAO+SNe (blue). Left: in the case of flat neutron star mass distribution. Right: in the case of gaussian neutron star mass distribution. We use the optimistic estimate for the FOV of THESEUS.}
\label{fig:ET+2CE_w0_opt}
\end{figure}

\begin{figure}[t]
\centering
\includegraphics[width=0.4\textwidth]{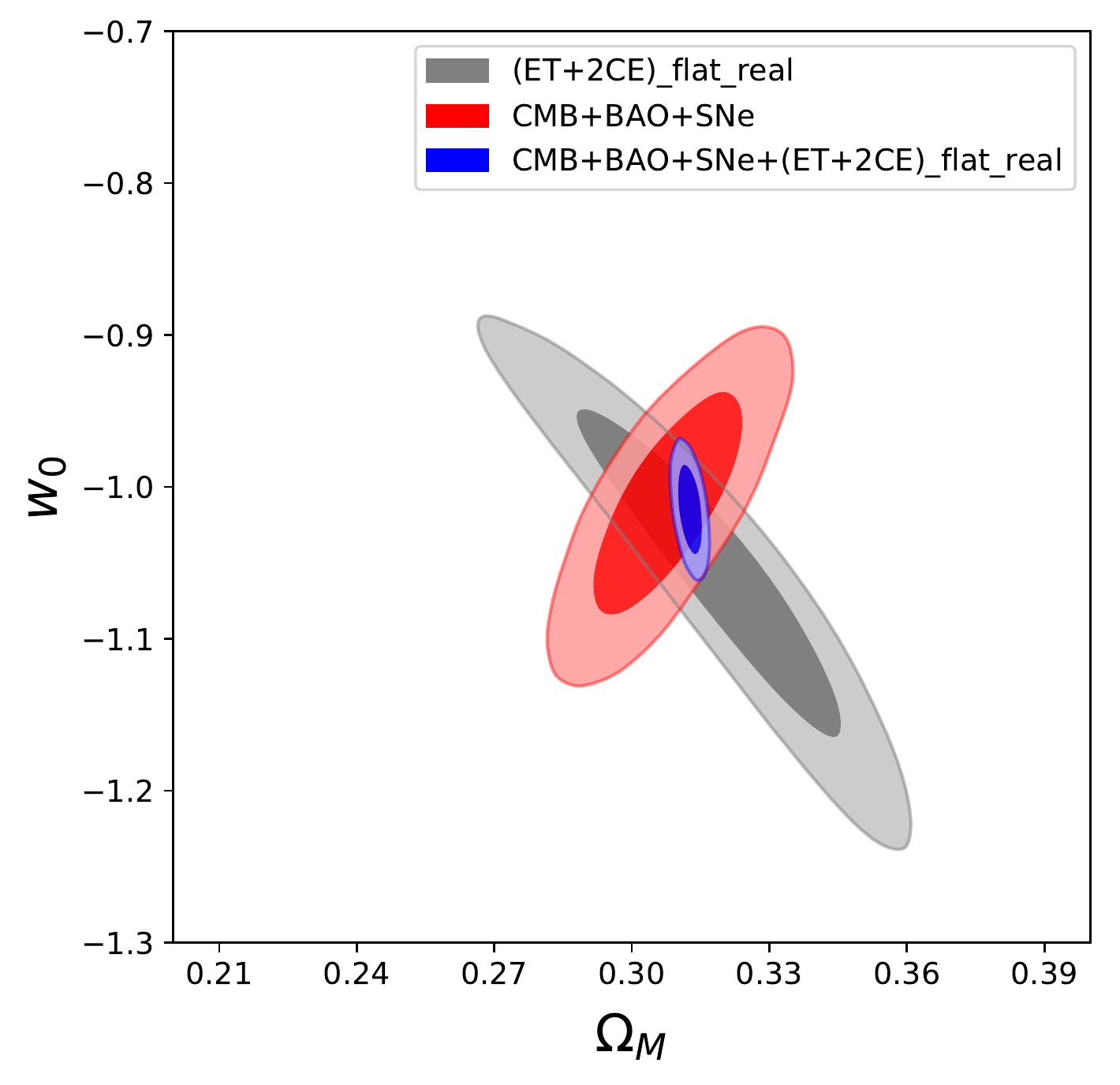}
\includegraphics[width=0.4\textwidth]{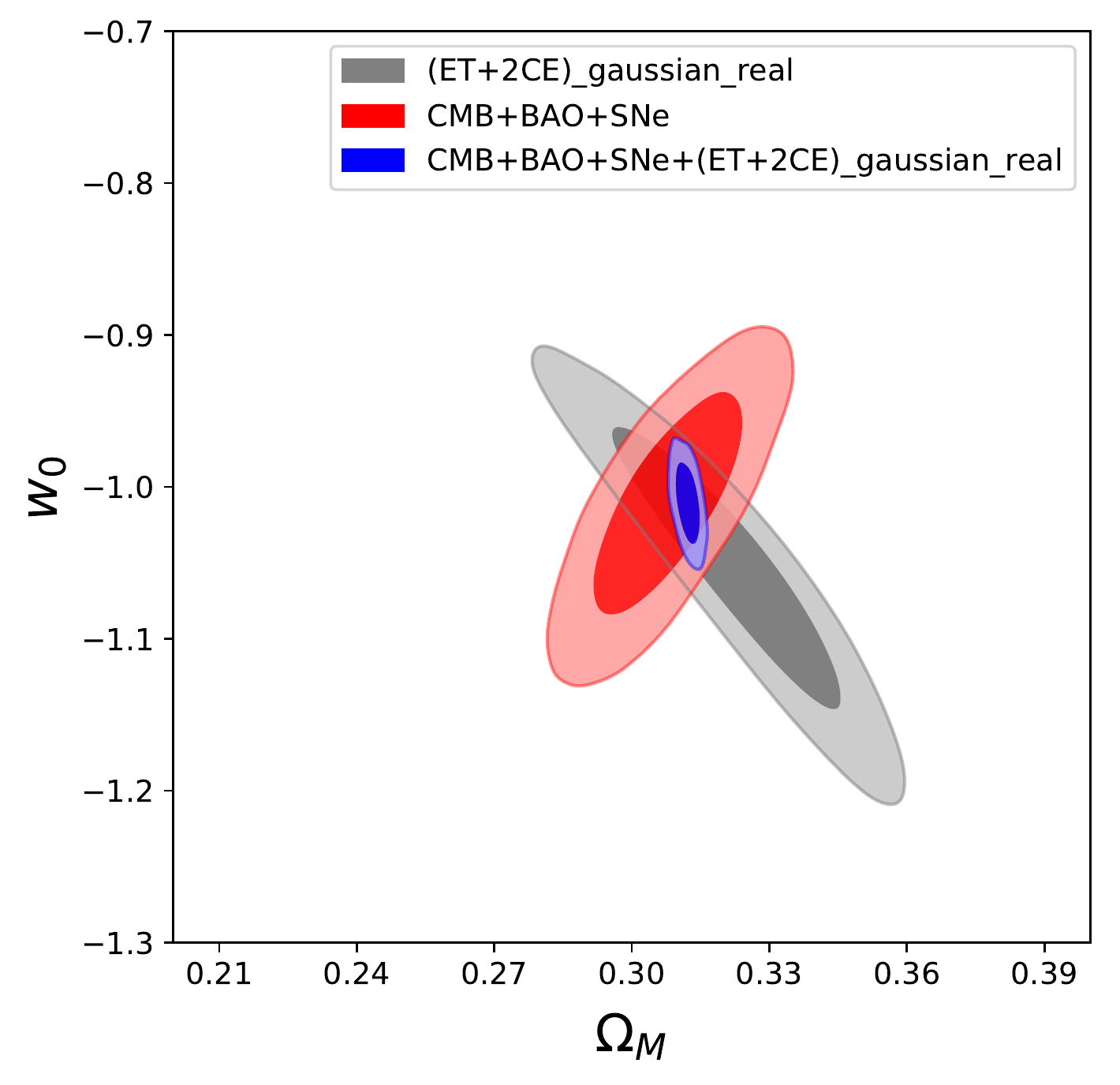}
\caption{As in Fig.~\ref{fig:ET+2CE_w0_opt}, with the realistic estimate for the FOV of THESEUS.}
\label{fig:ET+2CE_w0_real}
\end{figure}

\begin{figure}[t]
\centering
\includegraphics[width=0.4\textwidth]{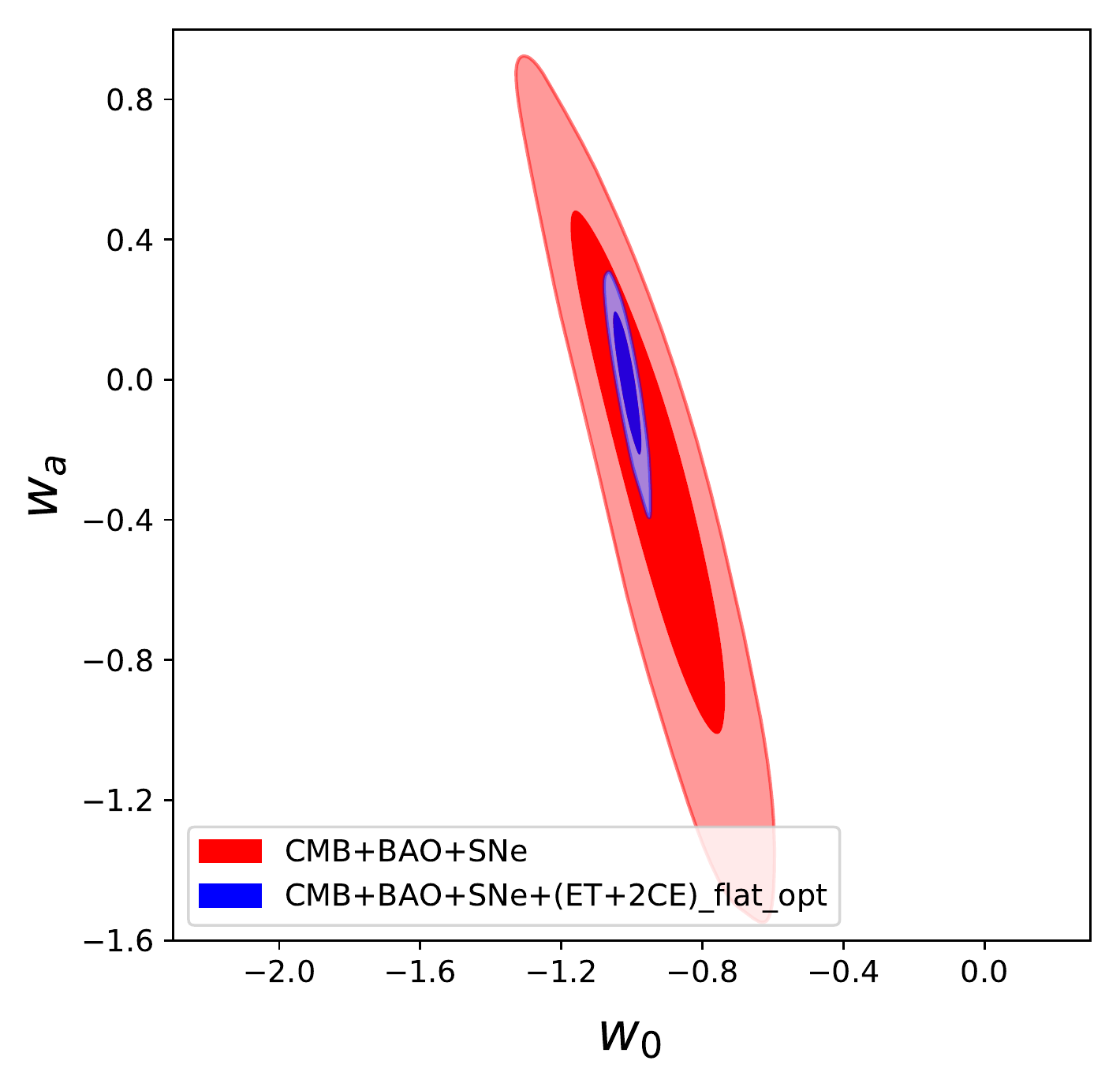}
\includegraphics[width=0.4\textwidth]{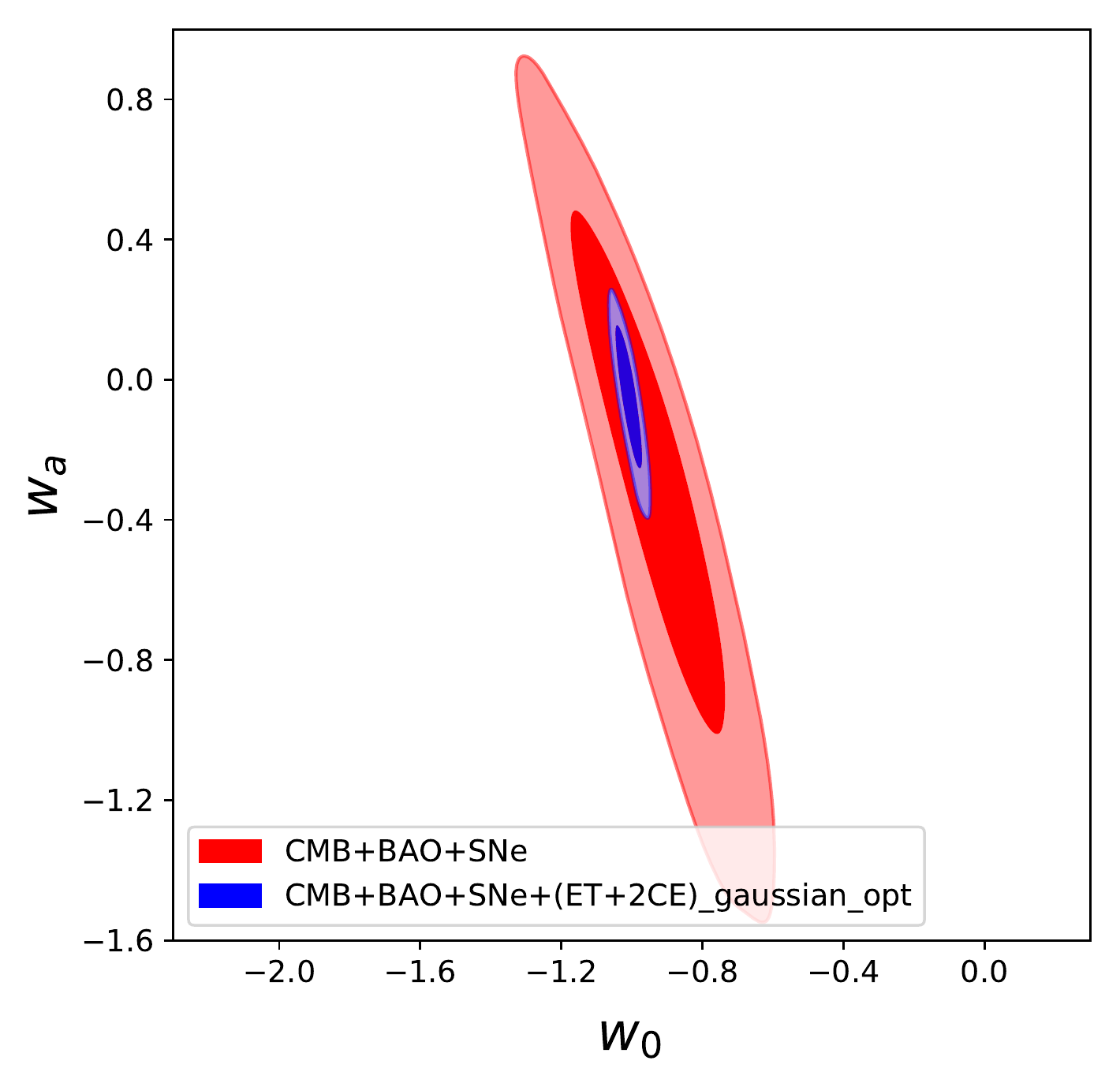}
\caption{The  $1\sigma$ and $2\sigma$
contours  of the two-dimensional likelihood in the $(w_0,w_a)$ plane from CMB+BAO+SNe (red) and the result obtained by combining standard sirens  at ET+CE+CE with CMB+BAO+SNe (blue). Left: in the case of flat neutron star mass distribution. Right: in the case of gaussian neutron star mass distribution. We use the optimistic estimate for the FOV of THESEUS.}
\label{fig:ET+2CE_w0wa_opt}
\end{figure}

\begin{figure}[t]
\centering
\includegraphics[width=0.4\textwidth]{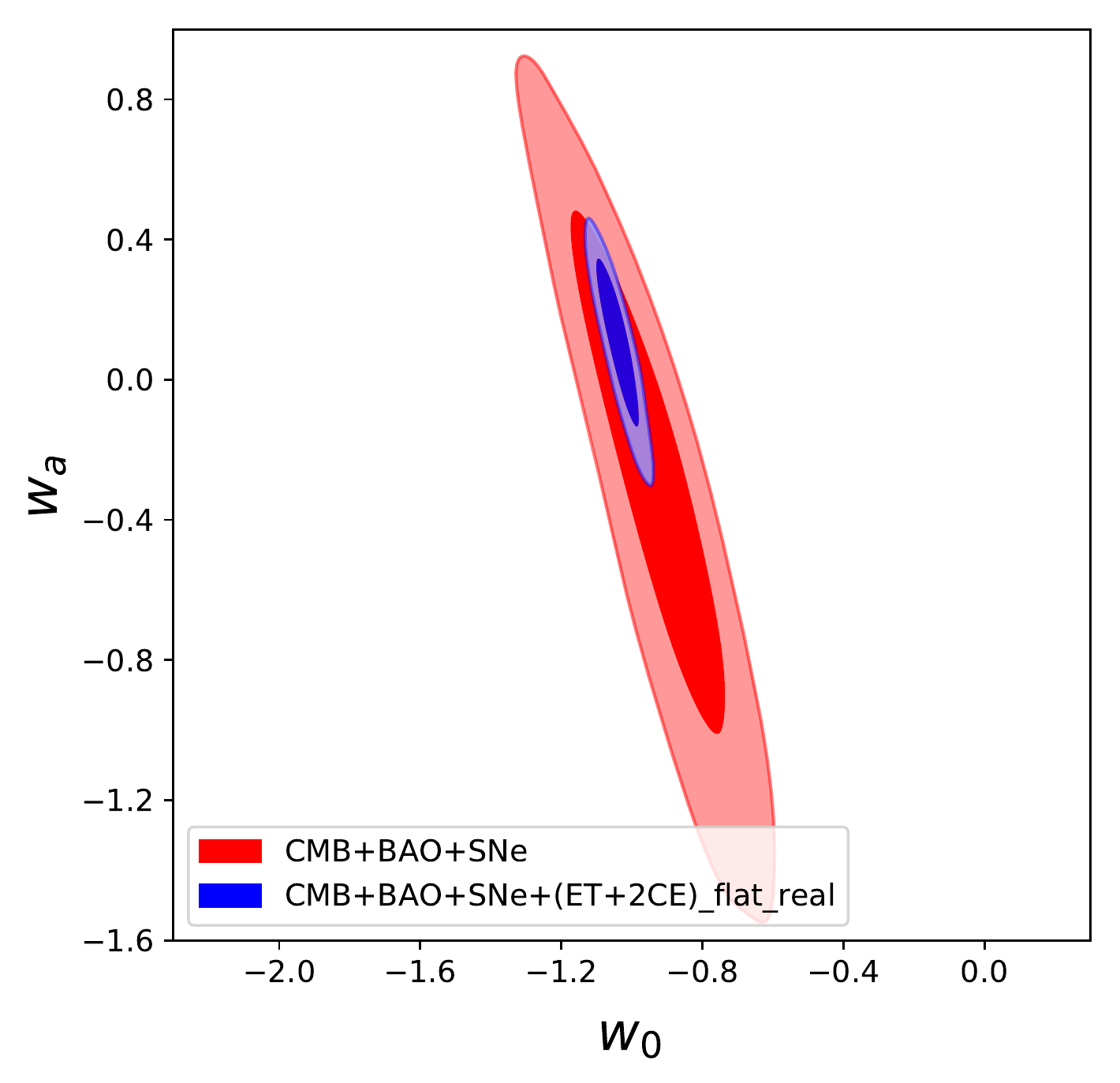}
\includegraphics[width=0.4\textwidth]{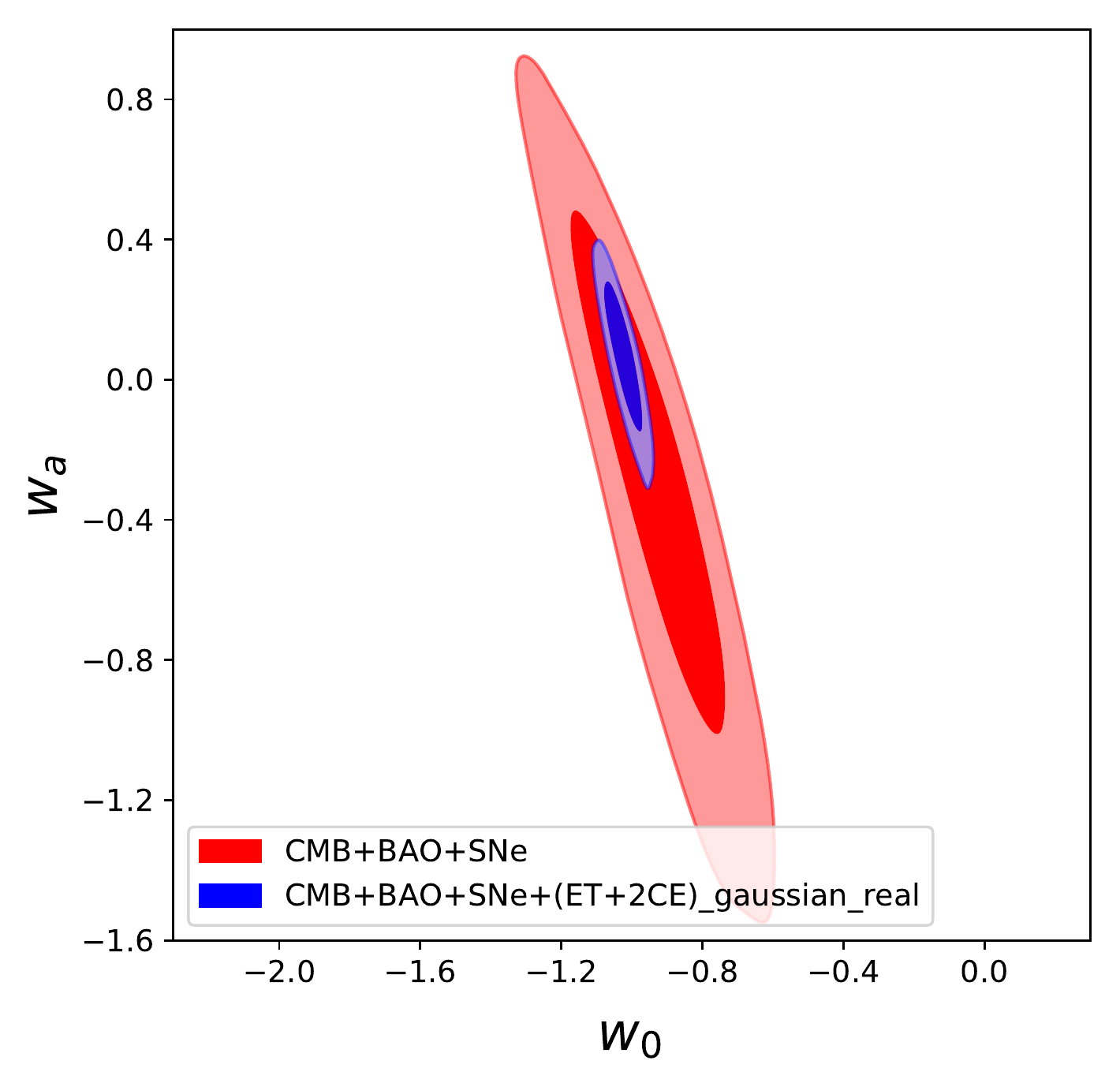}
\caption{As in Fig.~\ref{fig:ET+2CE_w0wa_opt}, with the realistic estimate for the FOV of THESEUS.}
\label{fig:ET+2CE_w0wa_real}
\end{figure}

\begin{figure}[t]
\centering
\includegraphics[width=0.4\textwidth]{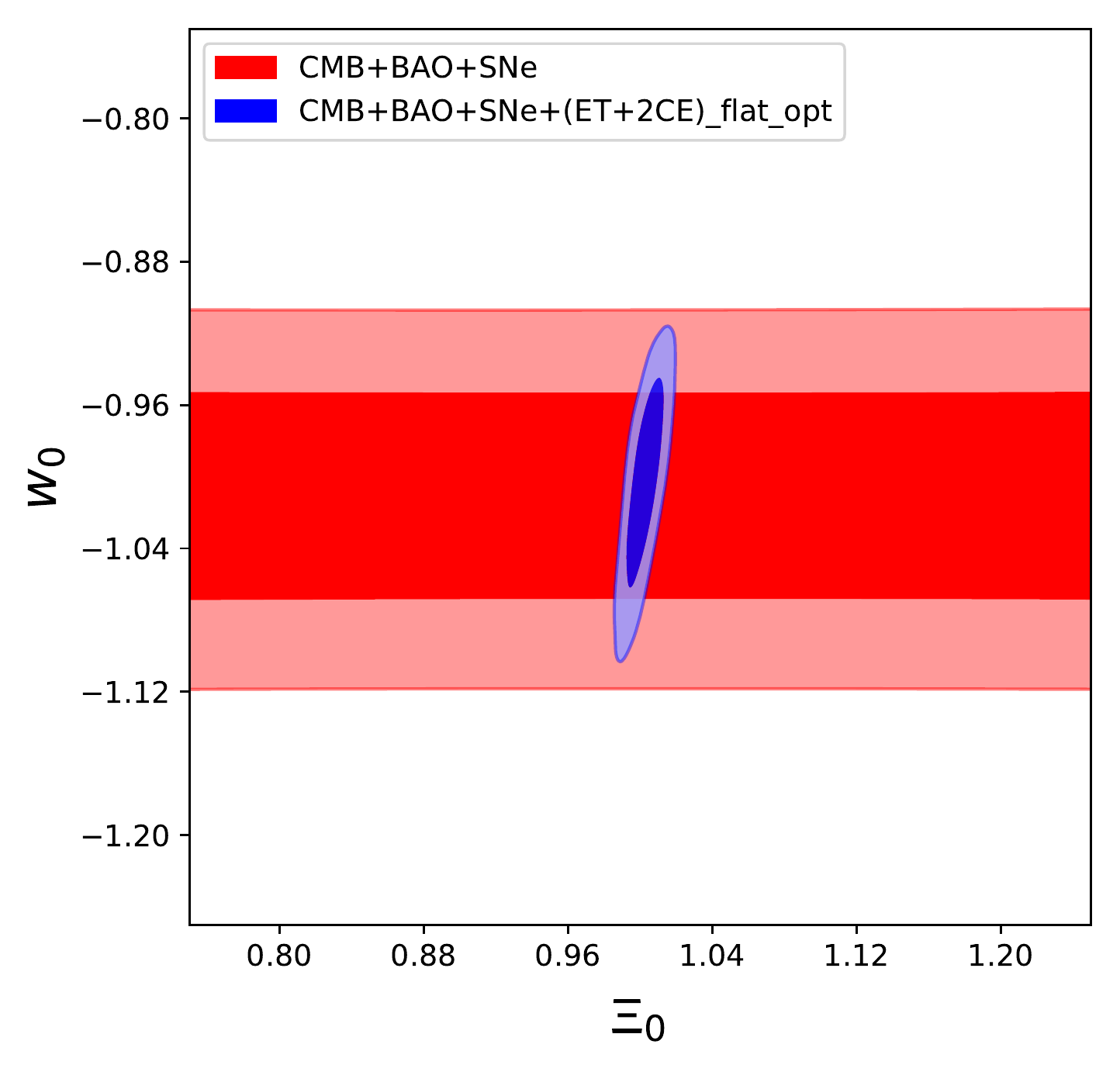}
\includegraphics[width=0.4\textwidth]{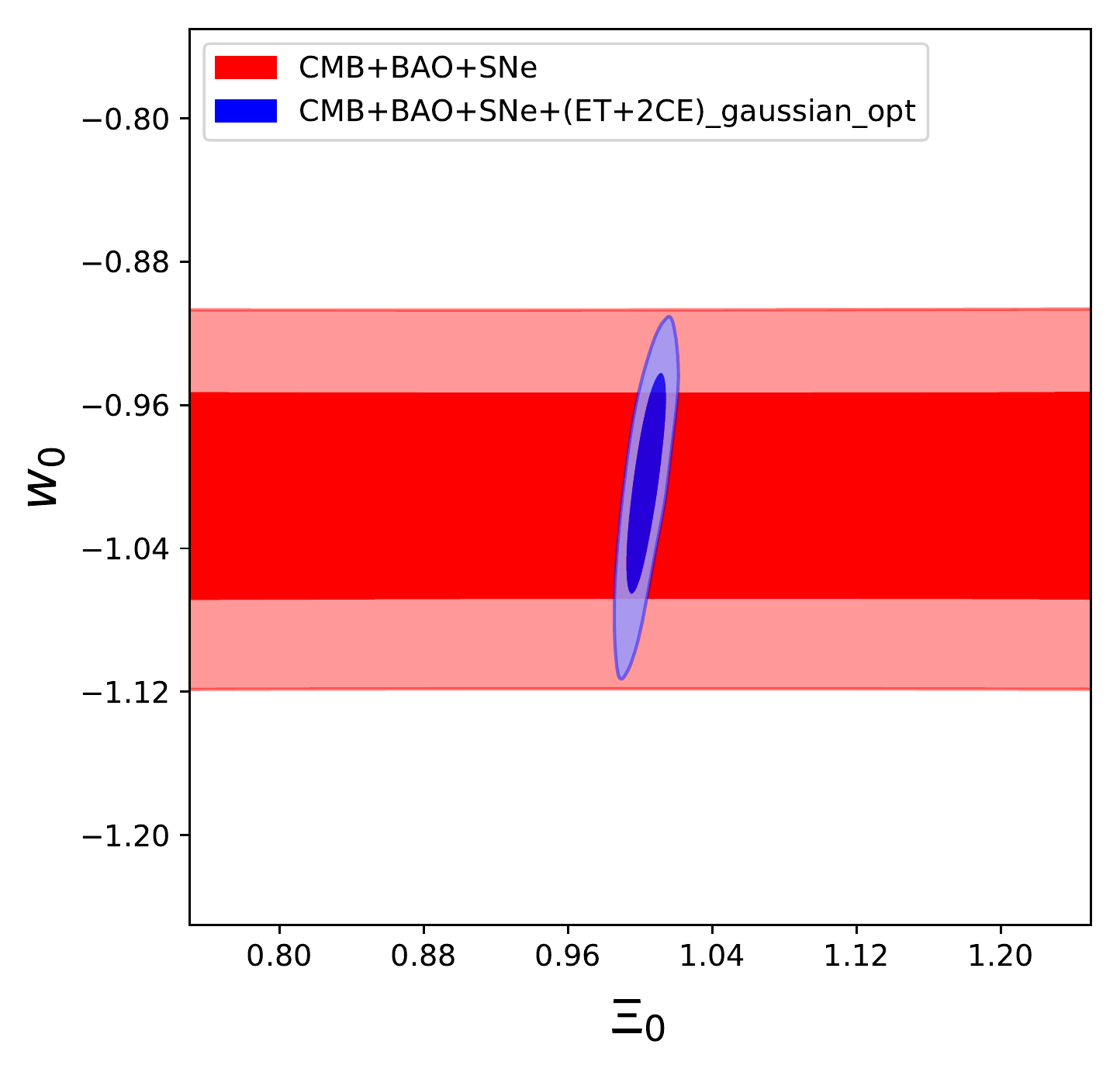}
\caption{The  $1\sigma$ and $2\sigma$
contours  of the two-dimensional likelihood in the $(\Xi_0,w_0)$ plane from CMB+BAO+SNe (red) and the result obtained by combining standard sirens at ET+CE+CE  with CMB+BAO+SNe (blue). Left: in the case of flat neutron star mass distribution. Right: in the case of gaussian neutron star mass distribution. We use the optimistic estimate for the FOV of THESEUS.}
\label{fig:ET+2CE_w0xi0_opt}
\end{figure}

\begin{figure}[t]
\centering
\includegraphics[width=0.4\textwidth]{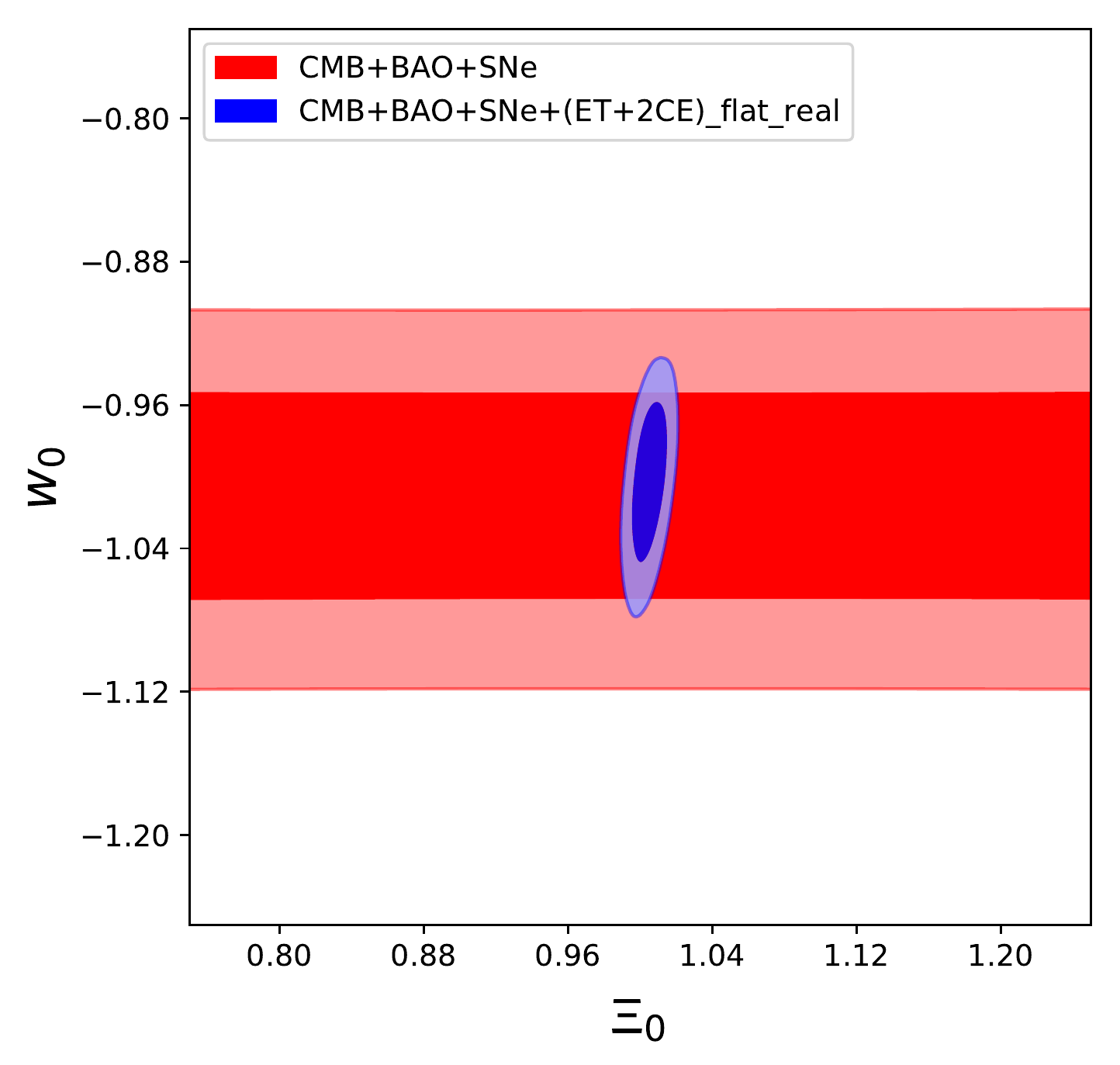}
\includegraphics[width=0.4\textwidth]{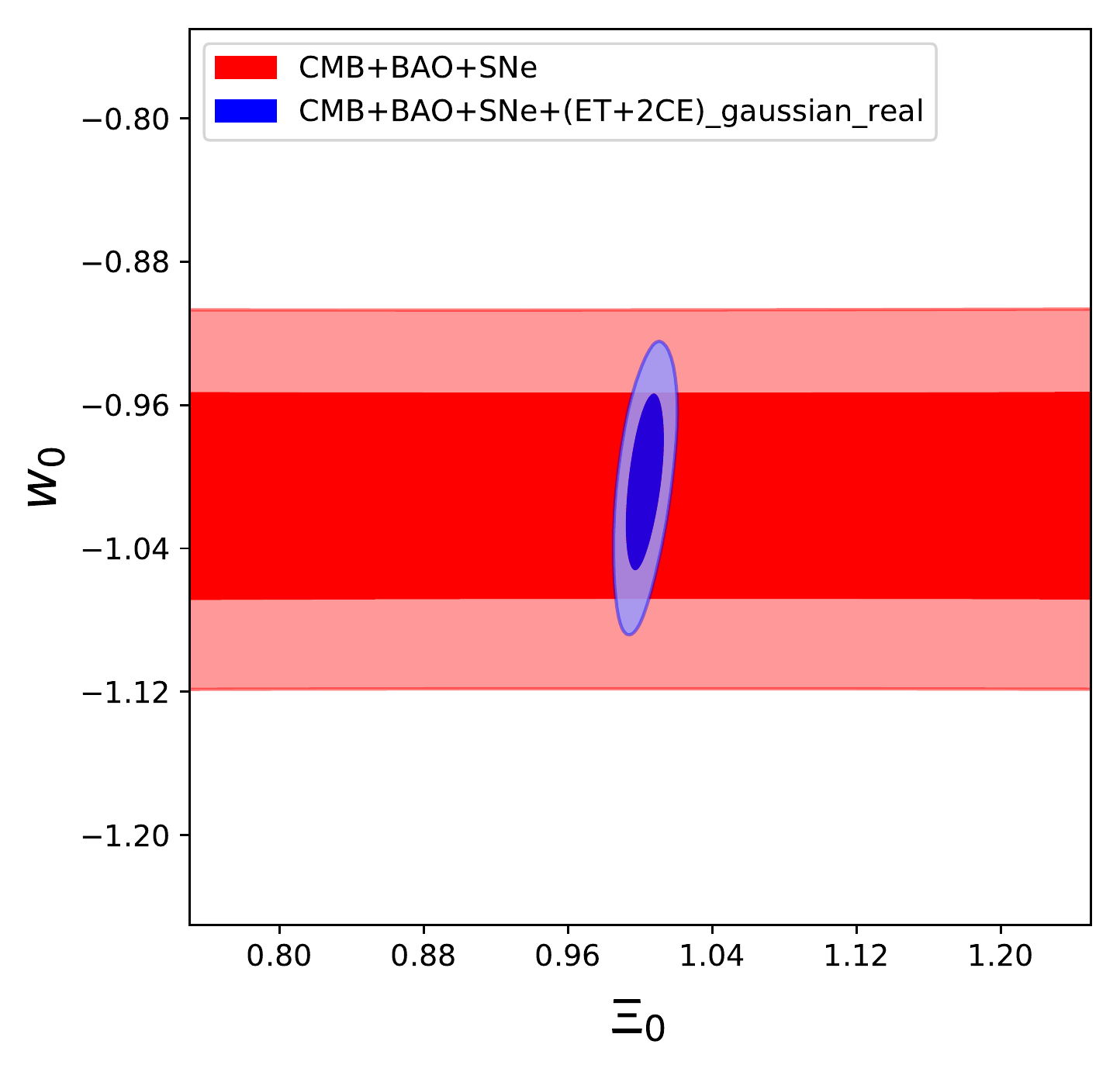}
\caption{As in Fig.~\ref{fig:ET+2CE_w0xi0_opt}, with the realistic estimate for the FOV of THESEUS.}
\label{fig:ET+2CE_w0xi0_real}
\end{figure}

We finally give the corresponding results for the ET+CE+CE network.
For  $w$CDM the results are shown in Figs.~\ref{fig:ET+2CE_w0_opt} and \ref{fig:ET+2CE_w0_real} and Tables~\ref{tab:w0withET+2CEopt} and \ref{tab:w0withET+2CEreal}.
For the $(w_0,w_a)$ extension the results are shown in Figs.~\ref{fig:ET+2CE_w0wa_opt} and \ref{fig:ET+2CE_w0wa_real} and Tables~\ref{tab:w0wawithET+2CEopt} and \ref{tab:w0wawithET+2CEreal}.
For  the
$(\Xi_0,w_0)$ extension the results are shown in Figs.~\ref{fig:ET+2CE_w0xi0_opt} and \ref{fig:ET+2CE_w0xi0_real} and Tables~\ref{tab:w0xi0withET+2CEopt} and \ref{tab:w0xi0withET+2CEreal}.

As we already observed in section~\ref{sect:LCDMwithET+CE+CE} when discussing  parameter estimation in $\Lambda$CDM,  the improvement in the accuracy of the cosmological parameters, compared to the ET-only case, is not very large, because, despite the fact that a network ET+CE+CE detects a number of source larger by an order of magnitude compared to a single ET (and to much larger redshift, see Table~\ref{tab:3G} and Fig.~\ref{fig:redshift_dist}), the corresponding joint GW-GRB detections do not follow the same increase, because of intrinsic limitations in the GRB detections. In particular, as we discussed in section~\ref{sect:count3G},
despite the fact that  the ET+CE+CE network can detect BNS up to $z\simeq 10$ (in our catalog the source with the largest redshift has $z\simeq 9.66$), the joint detections with GRB only reach, in our catalog, $z\simeq 3.38$; all higher redshift sources are lost because their GRB is beyond the flux limit for detection.

To fully exploit the potential of a ET+CE+CE network for standard sirens  it is therefore crucial either to have a more powerful network of multi-messenger observations, for instance  
with IR/optical telescopes (that, guided by the localization capability of the ET+CE+CE network, could provide many more counterparts, at least the region $z<0.5-0.7$ corresponding to their reach),  or else one should resort to statistical methods for the determination of the host galaxy of standard sirens without electromagnetic counterpart. This is also an important message of our analysis.

\begin{table}[hb]
\centering
\begin{tabular}{|c|c|c|c|c|c|}
 \hline
                               &  CMB+BAO+SNe&ET+CE+CE,      &ET+CE+CE,    & combined,  & combined,      \\
                               &                               &flat                     &gaussian           &flat              & gaussian \\ \hline
$\Delta w_0$         &  0.045                    & 0.041                & 0.034               & 0.014          &0.013     \\
\hline
\end{tabular}
\caption{Accuracy ($1\sigma$ level) in the reconstruction of $w_0$ with   only CMB+BAO+SNe,  with only standard sirens (with the flat and gaussian mass distributions, respectively) and  the combined results CMB+BAO+SNe+standard sirens, using ET+CE+CE  and THESEUS and assuming the optimistic FOV of THESEUS.
\label{tab:w0withET+2CEopt}}
\end{table}

\begin{table}[hb]
\centering
\begin{tabular}{|c|c|c|c|c|c|}
 \hline
                               &  CMB+BAO+SNe&ET+CE+CE,      &ET+CE+CE,           & combined,  & combined,      \\
                               &                               &flat                     &gaussian                  &flat              & gaussian \\ \hline
$\Delta w_0$         &  0.045                    & 0.074                & 0.063                      & 0.020          &0.018     \\
\hline
\end{tabular}
\caption{As in Table~\ref{tab:w0withET+2CEopt}, assuming the realistic FOV of THESEUS.
\label{tab:w0withET+2CEreal}}
\end{table}

%------------
\begin{table}[t]
\centering
\begin{tabular}{|c|c|c|c|c|c|}
 \hline
                               &  CMB+BAO+SNe & combined,  & combined,      \\
                               &                               &flat               & gaussian \\ \hline
$\Delta w_0$         &  0.140                    & 0.027           & 0.025      \\
$\Delta w_a$         &  0.483                    & 0.139           & 0.137       \\
\hline
\end{tabular}
\caption{Accuracy ($1\sigma$ level) in the reconstruction of $(w_0,w_a)$ with   only CMB+BAO+SNe and  the combined results CMB+BAO+SNe+standard sirens, using ET+CE+CE  and THESEUS and assuming the optimistic FOV of THESEUS.
\label{tab:w0wawithET+2CEopt}}
\end{table}

\begin{table}[t]
\centering
\begin{tabular}{|c|c|c|c|c|c|}
 \hline
                               &  CMB+BAO+SNe & combined,  & combined,      \\
                               &                               &flat               & gaussian \\ \hline
$\Delta w_0$         &  0.140                    & 0.041           & 0.037      \\
$\Delta w_a$         &  0.483                    & 0.160           & 0.145       \\
\hline
\end{tabular}
\caption{As in Table~\ref{tab:w0wawithET+2CEopt}, assuming the realistic FOV of THESEUS.
\label{tab:w0wawithET+2CEreal}}
\end{table}

%-----------------------

%------------xi0
\begin{table}[t]
\centering
\begin{tabular}{|c|c|c|c|c|c|}
 \hline
                               &  CMB+BAO+SNe & combined,  & combined,      \\
                               &                               &flat               & gaussian \\ \hline
$\Delta w_0$         &  0.045                    & 0.038           & 0.042      \\
$\Delta \Xi_0$       &  --                          & 0.007           & 0.007       \\
\hline
\end{tabular}
\caption{Accuracy ($1\sigma$ level) in the reconstruction of $(w_0,\Xi_0)$ with   only CMB+BAO+SNe and  the combined results CMB+BAO+SNe+standard sirens, using ET+CE+CE  and THESEUS and assuming the optimistic FOV of THESEUS.
\label{tab:w0xi0withET+2CEopt}}
\end{table}

\begin{table}[t]
\centering
\begin{tabular}{|c|c|c|c|c|c|}
 \hline
                               &  CMB+BAO+SNe & combined,  & combined,      \\
                               &                               &flat               & gaussian \\ \hline
$\Delta w_0$         &  0.045                    & 0.030           & 0.033      \\
$\Delta \Xi_0$       &  --                          & 0.006           & 0.007       \\
\hline
\end{tabular}
\caption{As in Table~\ref{tab:w0xi0withET+2CEopt}, assuming the realistic FOV of THESEUS.
\label{tab:w0xi0withET+2CEreal}}
\end{table}

%-----------------------xi0

\clearpage\newpage

\section{Conclusions}\label{sect:conl}

One of  the main  motivations for developing third-generation GW detectors, such as the Einstein Telescope in Europe and Cosmic Explorer in the US, is the possibility of using them for  cosmology studies, in a way that will be complementary to what is done with  electromagnetic probes such as the CMB, type Ia SNe, BAO, or large-scale structures. In particular, the measurement of the luminosity distance to coalescing binaries through the observation of their GWs, combined with an electromagnetic measurement of the redshift (or with statistical methods), gives access to the Hubble parameter $H_0$ and, for sources at sufficiently large redshift, also to the dark-energy equation of state. 

On top of this, a very interesting example of the potential contributions of advanced GW detectors to cosmology
is given by the possibility of studying  modified GW propagation. A specific realization of this  effect was already discussed  more than a decade ago,  in terms of leakage of GWs into extra dimensions in the context of the DGP model~\cite{Deffayet:2007kf}. More recently, several studies of modified GW propagation have been performed~\cite{Saltas:2014dha,Gleyzes:2014rba,Lombriser:2015sxa,Nishizawa:2017nef,Arai:2017hxj,Belgacem:2017ihm,Amendola:2017ovw,Linder:2018jil,Pardo:2018ipy,Belgacem:2018lbp,Lagos:2019kds,Belgacem:2019pkk}, and this effect is now being appreciated  as one of the  most promising tools for testing modified gravity and the dark energy sector of alternatives to $\Lambda$CDM, both with 
third-generation ground-based GW detectors, and with the space interferometer LISA. There are several reasons  for this:

\begin{itemize}

\item Modified GW  propagation emerges very generally in all modified gravity theories, typically as a modification of the friction term in the propagation over a FRW background, and in this form it appears for instance in 
scalar-tensor theories of the Horndeski and DHOST type~\cite{Saltas:2014dha,Lombriser:2015sxa,Arai:2017hxj,Amendola:2017ovw}, in non-local modifications of gravity~\cite{Belgacem:2017cqo,Belgacem:2017ihm,Belgacem:2018lbp,Belgacem:2019pkk}, or in bigravity theories~\cite{Belgacem:2019pkk}, and also emerges naturally in the effective field theory of dark energy~\cite{Gleyzes:2014rba}.
In  all these models (with the exception of bigravity, where there is an interesting phenomenon of oscillations due to the beatings between the two metrics), the effect is very neatly captured by the parametrization~(\ref{eq:fit}), in terms of two parameters $(\Xi_0,n)$, which provides a simple and convenient way of studying it.

\item In a generic modified gravity theory, the effect of modified GW propagation on the luminosity distance of a GW source dominates over that due to the dark energy equation of state. This is due to the fact that, when one changes the equation of state of dark energy with respect to the value $w=-1$ of $\Lambda$CDM, the Bayesian parameter estimation of the other cosmological parameters, such as $H_0$ and $\oma$, as obtained e.g. by comparison with CMB, BAO and SNe data,  of course also gives different values, with respect to their values $\Lambda$CDM, and the change goes in the direction of reducing the change of the luminosity distance at large redshift. Thus, in the luminosity distance, the effect due to the change in the equation of state of DE is largely compensated by that in the cosmological parameters.
In contrast, modified GW propagation is an extra effect on top of this, that is not partially compensated by Bayesian parameter estimation~\cite{Belgacem:2017ihm,Belgacem:2018lbp}. Thus, the sensitivity of GW detectors to the parameter $\Xi_0$ that characterizes modified GW propagation is better than to $w_0$, as  has also been   verified with the explicit MCMC analysis of \cite{Belgacem:2018lbp,Belgacem:2019pkk} and of the present work.

\item Modified GW propagation can only be revealed with GW observations, and is therefore specific to GW detectors. Furthermore, as found in \cite{Belgacem:2018lbp} for ET and in \cite{Belgacem:2019pkk} for LISA, the sensitivity of these experiments to $\Xi_0$ is well within the prediction of some of the most studied alternatives to $\Lambda$CDM.

\end{itemize}

The original motivation of the present paper was to improve on the results of
ref.~\cite{Belgacem:2018lbp} for ET by using a more realistic catalog of sources, in the context of a current effort for building the Science Case for ET and, more generally, for 3G detectors. Indeed, in ref.~\cite{Belgacem:2018lbp} the sensitivity of ET to $\Xi_0$ was computed  assuming that, over a few years of data taking,  ET will be able to collect $10^3$ standard sirens with observed electromagnetic counterpart [out of the ${\cal O}(10^5)$ BNS/yr that ET is expected to  detect]. This is the assumption  that has been usually made in the literature in this context but, in the absence of a concrete study of how to detect the counterpart, it is nothing more than a reasonable working hypothesis. Furthermore, the actual redshift distribution of the joint GW-electromagnetic detections depends not only on the specifications of ET, but also of the detectors used to observe the electromagnetic counterpart.

To go beyond these simple assumptions one needs a concrete scenario for the detection of the electromagnetic counterpart. In this paper we have focused on the possibility of a joint GW-GRB detection using, as  an electromagnetic  partner to 3G detector, the proposed THESEUS mission. As we have repeatedly stressed, this represents only a subset   of the possible electromagnetic counterparts that could be observed, particularly for a network with significant localization capabilities such as HLVKI or ET+CE+CE (in which case it will  probably not even provide the dominant contribution). However,  given the large field of view of GRB satellites,  in this case it is possible to give at least some estimates based uniquely on the characteristics of the detector. In contrast, the estimate for  the number of counterparts detected  at  optical/IR telescope  also strongly depend on issues, such as the fraction of  telescope time that will be devoted to GW follow up, that are currently more difficult to predict.

Our main results are as follows. For the HLVKI network, the number of joint GW-GRB detections  (computed assuming that the Fermi-GBM can make a coincident detection and that \emph{Swift} can  identify an X-ray counterpart) is quite small, of order $1.5$/yr, and is not sufficient to obtains significant cosmological results using standard sirens only. Even when combined with the CMB+BAO+SNe dataset that we have used throughout the paper, it only provides a modest improvement in the accuracy on $H_0$. Indeed, to get interesting accuracy on $H_0$ with standard sirens, say of order $1\%$ or better, as needed to solve the discrepancy between {\em Planck}-$\Lambda$CDM value and the local measurement of $H_0$, a much higher number of events with counterparts are indeed needed, ${\cal O}(50-100)$ \cite{Chen:2017rfc,Feeney:2018mkj}. However, a 2G network such as HLVKI (or, indeed, already HLV) has significant localization accuracy. This number of sources with counterpart could therefore in principle be obtained through optical/IR follow up. Similarly, 
joint GW-GRB detections at the HLVKI network cannot be expected to improve significantly the accuracy on the DE equation of state, compared to what already know from electromagnetic observation, see Tables~\ref{tab:w0with2G} and \ref{tab:w0wawith2G}. The situation is however very different for what concern modified GW propagation, as encoded in the parameter $\Xi_0$. Since electromagnetic observations are blind to it, any  result from 2G detectors will be potentially interesting, and we have found that (after combining with CMB+BAO+SNe to reduce the degeneracy with $H_0$ and $\oma$), the HLVKI network can measure $\Xi_0$ to about $13\%$. This results  assumes 10~yr of data, which is   probably a very optimistic assumption; however, taking e.g. 5~yr of data the result should be approximately worse by about a factor $\sqrt{2}$, so that a measurement at the $20\%$ level should still be possible. This is already in the range of the  predictions from some modified gravity models~\cite{Belgacem:2019pkk,DeltaN64:inprep} and is therefore already a very interesting sensitivity.

For a single ET detector, our results for $H_0$ are given in Tables~\ref{tab:LCDMwithETopt} and \ref{tab:LCDMwithETreal} and show that, already using only standard sirens, and limiting ourselves to the coincidences with GRBs (estimated using a detector with the characteristic of THESEUS), ET can reach an extremely interesting accuracy, with $\Delta H_0/H_0$ between $0.2\%$ and $0.4\%$, depending on the assumptions on the event rate (gaussian vs. flat neutron star mass distribution) and on the FOV of the GRB detector. For the dark energy sector, including both  $w_0$ and $\Xi_0$, the  results are given in 
Table~\ref{tab:w0xi0withETopt}. In particular (again, upon combining with CMB+BAO+SNe to reduce the degeneracies), $\Xi_0$ can be measured to better than $1\%$, more precisely to $(0.7-0.8)\%$. This is an exciting result, since several modified gravity models give predictions significantly higher than this. This result also strengthens the science case for THESEUS. The corresponding result on $w_0$ is still valuable but less exciting, since it just brings the error down from $4.5\%$ to $2.5\%$. 

We have finally studied the configuration ET+CE+CE. In this case, despite the huge increase in GW detections, by a factor ${\cal O}(10)$ compared to a single ET, the final results on the cosmological parameters, limiting ourselves to GW-GRB coincidences, are not significantly better than with a single ET. For instance $\Xi_0$ can now be measured to  $(0.6-0.7)\%$. This is due to the fact that the number of joint GW-GRB detections does not increase correspondingly; in fact, from Table~\ref{tab:3G} we see that it does not even increase by a factor of 2. In other words, there is a bottleneck, due to the fact that a single GRB detector cannot keep the pace, and does not have the reach, of GW detections from a ET+CE+CE network. However, this network would have remarkable localization accuracy, and therefore could benefit from the detection of counterparts from optical/IR/radio telescopes. It is clear from our analysis that a strong effort in the follow up of GW signals would be necessary to really exploit the remarkable potential for cosmology of a network of 3G detectors, whether in the ET+CE+CE configuration that we have studied, or in intermediate configurations involving e.g. the planned Voyager upgrade of the advanced LIGO detectors.

\vspace{5mm}
{\bf Acknowledgments}.
We thank Lorenzo Amati, Matthew Bailes, Carlo Ferrigno, Bruce Gendre, Vuk Mandic, Samaya Nissanke, Bangalore Sathyaprakash,  Volodymyr Savchenko and Giulia Stratta for very useful discussions.
The work  of E.B., S.F. and M.M. is supported by the  Swiss National Science Foundation and  by the SwissMap National Center for Competence in Research. The work of Y.D. is supported by  Swiss National Science Foundation and by  a Consolidator Grant of the European Research Council (ERC-2015-CoG grant 680886).
E.J.H acknowledges support from a Australian Research Council DECRA Fellowship (DE170100891), and parts of his research were supported by the Australian Research Council Centre of Excellence for Gravitational Wave Discovery (OzGrav), through project number CE170100004. 

\appendix

\section{Catalogs of sources}\label{app:catalog}

In this appendix we provide some explicit tables with  informations on the realizations of the catalogs that we have used, in order to facilitate comparisons with our work. For the 2G HLVKI network, the explict realization of the catalog of joint GW-GRB detections is given in Table~\ref{tab:cat2G}. 

For 3G detectors, given the large number of events, we rather give their distribution in bins of frequency
as well as the  mean value and the variance of the ET instrumental contribution to   $\Delta d_L/d_L$, averaged over the events   in the given redshift bin,  for  the specific realization of the catalog of joint GW-GRB detections shown in
Fig.~\ref{fig:redshift_dist}. The four case that we consider, corresponding to flat or gaussian mass distribution and optimistic or realistic FOV of THESEUS, are given in Tables~\ref{tab:relerr_ET_flat_opt}-\ref{tab:relerr_ET_gaussian_real}.

\begin{table}[t]
\centering
\begin{tabular}{|c|c|c|}
\hline
$z$ & $d_L$ (Mpc) & $\Delta d_L$ (Mpc) \\ \hline
0.030550 & 142.195 & 4.321 \\
0.036110 & 178.107 & 7.666 \\
0.053624 & 250.543 & 9.899 \\
0.055359 & 256.893 & 9.889 \\
0.064093 & 310.823 & 11.951 \\
0.069841 & 337.359 & 17.002 \\
0.082936 & 372.050 & 29.253 \\
0.087941 & 413.191 & 29.292 \\
0.089227 & 418.742 & 35.200 \\
0.092182 & 422.677 & 27.394 \\
0.104137 & 510.429 & 35.753 \\
0.108121 & 487.845 & 36.081 \\
0.108558 & 555.629 & 38.370 \\
0.119288 & 587.063 & 46.405 \\
\hline
\end{tabular}
\hspace{8mm}
\begin{tabular}{|c|c|c|}
\hline
$z$ & $d_L$ (Mpc) & $\Delta d_L$ (Mpc) \\ \hline
0.029271 & 134.815 & 4.000 \\
0.035195 & 157.475 & 5.636 \\
0.060585 & 283.567 & 18.706 \\
0.066283 & 316.373 & 14.509 \\
0.071053 & 327.381 & 20.085 \\
0.071730 & 342.952 & 16.957 \\
0.076180 & 341.595 & 22.360 \\
0.081819 & 418.469 & 30.238 \\
0.088698 & 396.734 & 25.757 \\
0.091869 & 402.590 & 34.170 \\
0.094237 & 406.423 & 31.472 \\
0.095288 & 432.996 & 36.423 \\
0.099956 & 491.071 & 31.721 \\
0.102531 & 461.627 & 36.858 \\
0.114869 & 626.939 & 43.010 \\
\hline
\end{tabular}
\caption{The events in the specific realization of the catalog of joint GW-GRB detections  for the HLVKI network, shown in
Fig.~\ref{fig:HLVKI_histogram}, for the flat mass distribution (left table) and the Gaussian mass distribution (right table).
\label{tab:cat2G}}
\end{table}

\begin{table}[]
\centering
\begin{tabular}{|c|c|c|c|c|}
\hline
redshift  &       number of joint         & mean & mean  & standard deviation \\
    bin        &  GW-GRB events &  redshift                             &                  $\Delta d_L/d_L$                  & of $\Delta d_L/d_L$ \\ \hline
(0 , 0.1) & 14 & 0.07539 & 0.00746&0.00306 \\
(0.1 , 0.2) & 47 &0.15233 &0.01590&0.00667 \\
(0.2 , 0.3) & 58 & 0.24660 &0.02372&0.00818 \\
(0.3 , 0.4) & 66 & 0.35216 &0.03337&0.01247 \\
(0.4 , 0.5) & 51 &0.44213 &0.04290&0.01439 \\
(0.5 , 0.6) & 35 &0.54520 & 0.05069&0.01422 \\
(0.6 , 0.7) & 35 & 0.64652 &0.05424&0.01391 \\
(0.7 , 0.8) & 23 &0.74763 & 0.05267&0.01265 \\
(0.8 , 0.9) & 19 & 0.84150 &0.05823&0.01161 \\
(0.9 , 1.0) & 13 &0.95992 & 0.05893&0.01139 \\
(1.0 , 1.1) & 13 &1.03980 &0.06516&0.01056 \\
(1.1 , 1.2) & 7 &1.15798 &0.06457&0.00814 \\
(1.2 , 1.3) & 1 &1.27448 &0.07567&0 \\
(1.3 , 1.4)& 2 &1.35285& 0.07305&0.00350 \\
(1.4 , 1.5) & 2 & 1.45724 & 0.07000&0.00482 \\
(1.5 , 1.6) & 2 & 1.57550 &0.06693&0.00042 \\
(1.6 , 1.7) & 1 & 1.62090 & 0.06890&0 \\
\hline
\end{tabular}
\caption{The mean value and the variance of the ET instrumental contribution to   $\Delta d_L/d_L$, averaging over the events   in the given redshift bin,  for  the specific realization of the catalog of joint GW-GRB detections shown in
Fig.~\ref{fig:redshift_dist}, for the flat distribution of neutron star masses and the `optimistic' scenario for the FOV of THESEUS.
\label{tab:relerr_ET_flat_opt}}
\end{table}

\begin{table}[]
\centering
\begin{tabular}{|c|c|c|c|c|}
\hline
redshift  &       number of joint         & mean & mean  & standard deviation \\
    bin        &  GW-GRB events &  redshift                             &                  $\Delta d_L/d_L$                  & of $\Delta d_L/d_L$ \\ \hline
(0 , 0.1) & 18 &0.07076 &0.00853&0.00342 \\
(0.1 , 0.2) & 62 &0.15528 &0.01791&0.00659 \\
(0.2 , 0.3) & 73 &0.24902 &0.02826&0.00877 \\
(0.3 , 0.4) & 74 & 0.35218 &0.03670&0.01168 \\
(0.4 , 0.5) & 65 &0.44866 &0.04737&0.01509 \\
(0.5 , 0.6) & 55 &0.54196 &0.05672&0.01726 \\
(0.6 , 0.7) & 47 & 0.64370 &0.05495&0.01398 \\
(0.7 , 0.8) & 33 &0.74970 & 0.06027&0.01339 \\
(0.8 , 0.9) & 26 & 0.84755 &0.06457&0.01007 \\
(0.9 , 1.0) & 22 &0.94282 & 0.06533&0.01079 \\
(1.0 , 1.1) & 12 &1.05950 &0.06478&0.00645 \\
(1.1 , 1.2) & 10 &1.14518 &0.06993&0.00508 \\
(1.2 , 1.3) & 3 &1.27270 &0.07187&0.00153 \\
(1.3 , 1.4)& 3 &1.34741& 0.07152&0.00292 \\
(1.4 , 1.5) & 6 &1.46689 & 0.07687&0.00418 \\
(1.5 , 1.6) & 1 & 1.58407 &0.07577&0 \\
(1.6 , 1.7) & 1 &1.62843 &0.07947&0 \\
\hline
\end{tabular}
\caption{As in
Table~\ref{tab:relerr_ET_flat_opt}, for the Gaussian distribution of neutron star masses and the `optimistic' scenario for the FOV of THESEUS.
\label{tab:relerr_ET_gaussian_opt}}
\end{table}

\begin{table}[]
\centering
\begin{tabular}{|c|c|c|c|c|}
\hline
redshift  &       number of joint         & mean & mean  & standard deviation \\
    bin        &  GW-GRB events &  redshift                             &                  $\Delta d_L/d_L$                  & of $\Delta d_L/d_L$ \\ \hline
(0 , 0.1) & 2 & 0.09183 & 0.00582&0.00023 \\
(0.1 , 0.2) & 19 &0.14579 & 0.01675&0.00731 \\
(0.2 , 0.3) & 19 & 0.24404 &0.02630&0.00789 \\
(0.3 , 0.4) & 25 & 0.35527 & 0.03324&0.01065 \\
(0.4 , 0.5) & 23 & 0.44039 & 0.04143&0.01561 \\
(0.5 , 0.6) & 10 & 0.54527 & 0.05305&0.00887 \\
(0.6 , 0.7) & 12 & 0.64379 &0.05181&0.01299 \\
(0.7 , 0.8) & 3 & 0.74185 & 0.05141&0.00496 \\
(0.8 , 0.9) & 4 & 0.85262 &0.06450&0.00922 \\
(0.9 , 1.0) & 4 & 0.95465 & 0.05803&0.00518 \\
(1.0 , 1.1) & 1 & 1.03492 & 0.06103&0 \\
(1.1 , 1.2) & 5 &1.16308 &0.06442&0.00667 \\
(1.2 , 1.3) & -- &-- & --&-- \\
(1.3 , 1.4)& -- &-- & --&-- \\
(1.4 , 1.5) & 1 & 1.41882 & 0.07482&0 \\
\hline
\end{tabular}
\caption{As in
Table~\ref{tab:relerr_ET_flat_opt}, for the flat distribution of neutron star masses and the `realistic' scenario for the FOV of THESEUS.
\label{tab:relerr_ET_flat_real}}
\end{table}

\begin{table}[]
\centering
\begin{tabular}{|c|c|c|c|c|}
\hline
redshift  &       number of joint         & mean & mean  & standard deviation \\
    bin        &  GW-GRB events &  redshift                             &                  $\Delta d_L/d_L$                  & of $\Delta d_L/d_L$ \\ \hline
(0 , 0.1) & 4 &0.07108 & 0.00868&0.00244 \\
(0.1 , 0.2) & 24 &0.15001 &0.01784&0.00692 \\
(0.2 , 0.3) & 24 &0.24043 &0.02558&0.00680 \\
(0.3 , 0.4) & 27 &0.35355 & 0.03529&0.01004 \\
(0.4 , 0.5) & 28 &0.44966 & 0.04843&0.01528 \\
(0.5 , 0.6) & 9 & 0.53785 & 0.05646&0.01807 \\
(0.6 , 0.7) & 14 & 0.64540 &0.05329&0.01318 \\
(0.7 , 0.8) & 13 &0.73793 &0.05493&0.01368 \\
(0.8 , 0.9) & 8 &0.85497 &0.06413&0.00746 \\
(0.9 , 1.0) & 4 & 0.93702 & 0.06257&0.01228 \\
(1.0 , 1.1) & 6 & 1.05334 & 0.06494&0.00651 \\
(1.1 , 1.2) & 3 &1.15162 &0.06749&0.00246 \\
(1.2 , 1.3) & 1 &1.25943 & 0.07373&0 \\
(1.3 , 1.4)& -- &-- & --&-- \\
(1.4 , 1.5) &2 & 1.45375 &0.07851&0.00398 \\
(1.5 , 1.6) & 1 & 1.58407 &0.07577&0 \\
(1.6 , 1.7) & 1 & 1.62843 & 0.07947&0 \\
\hline
\end{tabular}
\caption{As in
Table~\ref{tab:relerr_ET_flat_opt}, for the Gaussian distribution of neutron star masses and the `realistic' scenario for the FOV of THESEUS.
\label{tab:relerr_ET_gaussian_real}}
\end{table}

\clearpage

\bibliographystyle{utphys}
\bibliography{myrefs}

\providecommand{\href}[2]{#2}\begingroup\raggedright\begin{thebibliography}{10}

\bibitem{Abbott:2016blz}
B.~P. Abbott {\em et~al.}, ``{Observation of Gravitational Waves from a Binary
  Black Hole Merger},''
  \href{http://dx.doi.org/10.1103/PhysRevLett.116.061102}{{\em Phys. Rev.
  Lett.} {\bfseries 116} no.~6, (2016) 061102},
\href{http://arxiv.org/abs/1602.03837}{{\ttfamily arXiv:1602.03837 [gr-qc]}}.
%%CITATION = ARXIV:1602.03837;%%.

\bibitem{Abbott:2016nmj}
B.~P. Abbott {\em et~al.}, ``{GW151226: Observation of Gravitational Waves from
  a 22-Solar-Mass Binary Black Hole Coalescence},''
  \href{http://dx.doi.org/10.1103/PhysRevLett.116.241103}{{\em Phys. Rev.
  Lett.} {\bfseries 116} no.~24, (2016) 241103},
\href{http://arxiv.org/abs/1606.04855}{{\ttfamily arXiv:1606.04855 [gr-qc]}}.
%%CITATION = ARXIV:1606.04855;%%.

\bibitem{Abbott:2017vtc}
B.~P. Abbott {\em et~al.}, ``{GW170104: Observation of a 50-Solar-Mass Binary
  Black Hole Coalescence at Redshift 0.2}''
  \href{http://dx.doi.org/10.1103/PhysRevLett.118.221101}{{\em Phys. Rev.
  Lett.} {\bfseries 118} no.~22, (2017) 221101},
\href{http://arxiv.org/abs/1706.01812}{{\ttfamily arXiv:1706.01812 [gr-qc]}}.
%%CITATION = ARXIV:1706.01812;%%.

\bibitem{Abbott:2017gyy}
B.~P. Abbott {\em et~al.}, ``{GW170608: Observation of a 19-solar-mass Binary
  Black Hole Coalescence},''
  \href{http://dx.doi.org/10.3847/2041-8213/aa9f0c}{{\em Astrophys. J.}
  {\bfseries 851} no.~2, (2017) L35},
\href{http://arxiv.org/abs/1711.05578}{{\ttfamily arXiv:1711.05578
  [astro-ph.HE]}}.
%%CITATION = ARXIV:1711.05578;%%.

\bibitem{Abbott:2017oio}
B.~P. Abbott {\em et~al.}, ``{GW170814: A Three-Detector Observation of
  Gravitational Waves from a Binary Black Hole Coalescence},''
  \href{http://dx.doi.org/10.1103/PhysRevLett.119.141101}{{\em Phys. Rev.
  Lett.} {\bfseries 119} no.~14, (2017) 141101},
\href{http://arxiv.org/abs/1709.09660}{{\ttfamily arXiv:1709.09660 [gr-qc]}}.
%%CITATION = ARXIV:1709.09660;%%.

\bibitem{LIGOScientific:2018mvr}
B.~P. Abbott {\em et~al.}, ``{GWTC-1: A Gravitational-Wave Transient Catalog of
  Compact Binary Mergers Observed by LIGO and Virgo during the First and Second
  Observing Runs},''
\href{http://arxiv.org/abs/1811.12907}{{\ttfamily arXiv:1811.12907
  [astro-ph.HE]}}.
%%CITATION = ARXIV:1811.12907;%%.

\bibitem{TheLIGOScientific:2017qsa}
B.~Abbott {\em et~al.}, ``{GW170817: Observation of Gravitational Waves from a
  Binary Neutron Star Inspiral},''
  \href{http://dx.doi.org/10.1103/PhysRevLett.119.161101}{{\em Phys. Rev.
  Lett.} {\bfseries 119} (2017) 161101},
\href{http://arxiv.org/abs/1710.05832}{{\ttfamily arXiv:1710.05832 [gr-qc]}}.
%%CITATION = ARXIV:1710.05832;%%.

\bibitem{Goldstein:2017mmi}
A.~Goldstein {\em et~al.}, ``{An Ordinary Short Gamma-Ray Burst with
  Extraordinary Implications: Fermi-GBM Detection of GRB 170817A},''
  \href{http://dx.doi.org/10.3847/2041-8213/aa8f41}{{\em Astrophys. J.}
  {\bfseries 848} (2017) L14},
\href{http://arxiv.org/abs/1710.05446}{{\ttfamily arXiv:1710.05446
  [astro-ph.HE]}}.
%%CITATION = ARXIV:1710.05446;%%.

\bibitem{Savchenko:2017ffs}
V.~Savchenko {\em et~al.}, ``{INTEGRAL Detection of the First Prompt Gamma-Ray
  Signal Coincident with the Gravitational-wave Event GW170817},''
  \href{http://dx.doi.org/10.3847/2041-8213/aa8f94}{{\em Astrophys. J.}
  {\bfseries 848} (2017) L15},
\href{http://arxiv.org/abs/1710.05449}{{\ttfamily arXiv:1710.05449
  [astro-ph.HE]}}.
%%CITATION = ARXIV:1710.05449;%%.

\bibitem{Monitor:2017mdv}
B.~P. Abbott {\em et~al.}, ``{Gravitational Waves and Gamma-rays from a Binary
  Neutron Star Merger: GW170817 and GRB 170817A},''
  \href{http://dx.doi.org/10.3847/2041-8213/aa920c}{{\em Astrophys. J.}
  {\bfseries 848} (2017) L13},
\href{http://arxiv.org/abs/1710.05834}{{\ttfamily arXiv:1710.05834
  [astro-ph.HE]}}.
%%CITATION = ARXIV:1710.05834;%%.

\bibitem{GBM:2017lvd}
B.~P. Abbott {\em et~al.}, ``{Multi-messenger Observations of a Binary Neutron
  Star Merger},'' \href{http://dx.doi.org/10.3847/2041-8213/aa91c9}{{\em
  Astrophys. J.} {\bfseries 848} no.~2, (2017) L12},
\href{http://arxiv.org/abs/1710.05833}{{\ttfamily arXiv:1710.05833
  [astro-ph.HE]}}.
%%CITATION = ARXIV:1710.05833;%%.

\bibitem{Audley:2017drz}
P.~Amaro-Seoane {\em et~al.}, ``{Laser Interferometer Space Antenna},''
\href{http://arxiv.org/abs/1702.00786}{{\ttfamily arXiv:1702.00786
  [astro-ph.IM]}}.
%%CITATION = ARXIV:1702.00786;%%.

\bibitem{Punturo:2010zz}
M.~Punturo {\em et~al.}, ``{The Einstein Telescope: A third-generation
  gravitational wave observatory},''
\href{http://dx.doi.org/10.1088/0264-9381/27/19/194002}{{\em Class. Quant.
  Grav.} {\bfseries 27} (2010) 194002}.
%%CITATION = CQGRD,27,194002;%%.

\bibitem{Sathyaprakash:2012jk}
B.~Sathyaprakash {\em et~al.}, ``{Scientific Objectives of Einstein
  Telescope},'' \href{http://dx.doi.org/10.1088/0264-9381/29/12/124013,
  10.1088/0264-9381/30/7/079501}{{\em Class. Quant. Grav.} {\bfseries 29}
  (2012) 124013}, \href{http://arxiv.org/abs/1206.0331}{{\ttfamily
  arXiv:1206.0331 [gr-qc]}}.
[Erratum: Class. Quant. Grav.30,079501(2013)].
%%CITATION = ARXIV:1206.0331;%%.

\bibitem{Dwyer:2014fpa}
S.~Dwyer, D.~Sigg, S.~W. Ballmer, L.~Barsotti, N.~Mavalvala, and M.~Evans,
  ``{Gravitational wave detector with cosmological reach},''
  \href{http://dx.doi.org/10.1103/PhysRevD.91.082001}{{\em Phys. Rev.}
  {\bfseries D91} no.~8, (2015) 082001},
\href{http://arxiv.org/abs/1410.0612}{{\ttfamily arXiv:1410.0612
  [astro-ph.IM]}}.
%%CITATION = ARXIV:1410.0612;%%.

\bibitem{Schutz:1986gp}
B.~F. Schutz, ``{Determining the Hubble Constant from Gravitational Wave
  Observations},''
\href{http://dx.doi.org/10.1038/323310a0}{{\em Nature} {\bfseries 323} (1986)
  310--311}.
%%CITATION = NATUA,323,310;%%.

\bibitem{Holz:2005df}
D.~E. Holz and S.~A. Hughes, ``{Using gravitational-wave standard sirens},''
  \href{http://dx.doi.org/10.1086/431341}{{\em Astrophys. J.} {\bfseries 629}
  (2005) 15--22},
\href{http://arxiv.org/abs/astro-ph/0504616}{{\ttfamily arXiv:astro-ph/0504616
  [astro-ph]}}.
%%CITATION = ASTRO-PH/0504616;%%.

\bibitem{Dalal:2006qt}
N.~Dalal, D.~E. Holz, S.~A. Hughes, and B.~Jain, ``{Short GRB and binary black
  hole standard sirens as a probe of dark energy},''
  \href{http://dx.doi.org/10.1103/PhysRevD.74.063006}{{\em Phys. Rev.}
  {\bfseries D74} (2006) 063006},
\href{http://arxiv.org/abs/astro-ph/0601275}{{\ttfamily arXiv:astro-ph/0601275
  [astro-ph]}}.
%%CITATION = ASTRO-PH/0601275;%%.

\bibitem{MacLeod:2007jd}
C.~L. MacLeod and C.~J. Hogan, ``{Precision of Hubble constant derived using
  black hole binary absolute distances and statistical redshift information},''
  \href{http://dx.doi.org/10.1103/PhysRevD.77.043512}{{\em Phys. Rev.}
  {\bfseries D77} (2008) 043512},
\href{http://arxiv.org/abs/0712.0618}{{\ttfamily arXiv:0712.0618 [astro-ph]}}.
%%CITATION = ARXIV:0712.0618;%%.

\bibitem{Nissanke:2009kt}
S.~Nissanke, D.~E. Holz, S.~Hughes, N.~Dalal, and J.~L. Sievers, ``{Exploring
  short gamma-ray bursts as gravitational-wave standard sirens},''
  \href{http://dx.doi.org/10.1088/0004-637X/725/1/496}{{\em Astrophys. J.}
  {\bfseries 725} (2010) 496--514},
\href{http://arxiv.org/abs/0904.1017}{{\ttfamily arXiv:0904.1017
  [astro-ph.CO]}}.
%%CITATION = ARXIV:0904.1017;%%.

\bibitem{Cutler:2009qv}
C.~Cutler and D.~E. Holz, ``{Ultra-high precision cosmology from gravitational
  waves},'' \href{http://dx.doi.org/10.1103/PhysRevD.80.104009}{{\em Phys.
  Rev.} {\bfseries D80} (2009) 104009},
\href{http://arxiv.org/abs/0906.3752}{{\ttfamily arXiv:0906.3752
  [astro-ph.CO]}}.
%%CITATION = ARXIV:0906.3752;%%.

\bibitem{Sathyaprakash:2009xt}
B.~S. Sathyaprakash, B.~F. Schutz, and C.~Van Den~Broeck, ``{Cosmography with
  the Einstein Telescope},''
  \href{http://dx.doi.org/10.1088/0264-9381/27/21/215006}{{\em Class. Quant.
  Grav.} {\bfseries 27} (2010) 215006},
\href{http://arxiv.org/abs/0906.4151}{{\ttfamily arXiv:0906.4151
  [astro-ph.CO]}}.
%%CITATION = ARXIV:0906.4151;%%.

\bibitem{Zhao:2010sz}
W.~Zhao, C.~Van Den~Broeck, D.~Baskaran, and T.~G.~F. Li, ``{Determination of
  Dark Energy by the Einstein Telescope: Comparing with CMB, BAO and SNIa
  Observations},'' \href{http://dx.doi.org/10.1103/PhysRevD.83.023005}{{\em
  Phys. Rev.} {\bfseries D83} (2011) 023005},
\href{http://arxiv.org/abs/1009.0206}{{\ttfamily arXiv:1009.0206
  [astro-ph.CO]}}.
%%CITATION = ARXIV:1009.0206;%%.

\bibitem{DelPozzo:2011yh}
W.~Del~Pozzo, ``{Inference of the cosmological parameters from gravitational
  waves: application to second generation interferometers},''
  \href{http://dx.doi.org/10.1103/PhysRevD.86.043011}{{\em Phys. Rev.}
  {\bfseries D86} (2012) 043011},
\href{http://arxiv.org/abs/1108.1317}{{\ttfamily arXiv:1108.1317
  [astro-ph.CO]}}.
%%CITATION = ARXIV:1108.1317;%%.

\bibitem{Nishizawa:2011eq}
A.~Nishizawa, K.~Yagi, A.~Taruya, and T.~Tanaka, ``{Cosmology with space-based
  gravitational-wave detectors --- dark energy and primordial gravitational
  waves ---},'' \href{http://dx.doi.org/10.1103/PhysRevD.85.044047}{{\em Phys.
  Rev.} {\bfseries D85} (2012) 044047},
\href{http://arxiv.org/abs/1110.2865}{{\ttfamily arXiv:1110.2865
  [astro-ph.CO]}}.
%%CITATION = ARXIV:1110.2865;%%.

\bibitem{Taylor:2012db}
S.~R. Taylor and J.~R. Gair, ``{Cosmology with the lights off: standard sirens
  in the Einstein Telescope era},''
  \href{http://dx.doi.org/10.1103/PhysRevD.86.023502}{{\em Phys. Rev.}
  {\bfseries D86} (2012) 023502},
\href{http://arxiv.org/abs/1204.6739}{{\ttfamily arXiv:1204.6739
  [astro-ph.CO]}}.
%%CITATION = ARXIV:1204.6739;%%.

\bibitem{Camera:2013xfa}
S.~Camera and A.~Nishizawa, ``{Beyond Concordance Cosmology with Magnification
  of Gravitational-Wave Standard Sirens},''
  \href{http://dx.doi.org/10.1103/PhysRevLett.110.151103}{{\em Phys. Rev.
  Lett.} {\bfseries 110} (2013) 151103},
\href{http://arxiv.org/abs/1303.5446}{{\ttfamily arXiv:1303.5446
  [astro-ph.CO]}}.
%%CITATION = ARXIV:1303.5446;%%.

\bibitem{Tamanini:2016zlh}
N.~Tamanini, C.~Caprini, E.~Barausse, A.~Sesana, A.~Klein, and A.~Petiteau,
  ``{Science with the space-based interferometer eLISA. III: Probing the
  expansion of the Universe using gravitational wave standard sirens},''
  \href{http://dx.doi.org/10.1088/1475-7516/2016/04/002}{{\em JCAP} {\bfseries
  1604} (2016) 002},
\href{http://arxiv.org/abs/1601.07112}{{\ttfamily arXiv:1601.07112
  [astro-ph.CO]}}.
%%CITATION = ARXIV:1601.07112;%%.

\bibitem{Caprini:2016qxs}
C.~Caprini and N.~Tamanini, ``{Constraining early and interacting dark energy
  with gravitational wave standard sirens: the potential of the eLISA
  mission},'' \href{http://dx.doi.org/10.1088/1475-7516/2016/10/006}{{\em JCAP}
  {\bfseries 1610} (2016) 006},
\href{http://arxiv.org/abs/1607.08755}{{\ttfamily arXiv:1607.08755
  [astro-ph.CO]}}.
%%CITATION = ARXIV:1607.08755;%%.

\bibitem{Cai:2016sby}
R.-G. Cai and T.~Yang, ``{Estimating cosmological parameters by the simulated
  data of gravitational waves from the Einstein Telescope},''
  \href{http://dx.doi.org/10.1103/PhysRevD.95.044024}{{\em Phys. Rev.}
  {\bfseries D95} (2017) 044024},
\href{http://arxiv.org/abs/1608.08008}{{\ttfamily arXiv:1608.08008
  [astro-ph.CO]}}.
%%CITATION = ARXIV:1608.08008;%%.

\bibitem{DelPozzo:2017kme}
W.~Del~Pozzo, A.~Sesana, and A.~Klein, ``{Stellar binary black holes in the
  LISA band: a new class of standard sirens},''
  \href{http://dx.doi.org/10.1093/mnras/sty057}{{\em Mon. Not. Roy. Astron.
  Soc.} {\bfseries 475} (2018) 3485--3492},
\href{http://arxiv.org/abs/1703.01300}{{\ttfamily arXiv:1703.01300
  [astro-ph.CO]}}.
%%CITATION = ARXIV:1703.01300;%%.

\bibitem{Belgacem:2017ihm}
E.~Belgacem, Y.~Dirian, S.~Foffa, and M.~Maggiore, ``{The gravitational-wave
  luminosity distance in modified gravity theories},'' {\em Phys. Rev.}
  {\bfseries D97} (2018) 104066,
\href{http://arxiv.org/abs/1712.08108}{{\ttfamily arXiv:1712.08108
  [astro-ph.CO]}}.
%%CITATION = ARXIV:1712.08108;%%.

\bibitem{Belgacem:2018lbp}
E.~Belgacem, Y.~Dirian, S.~Foffa, and M.~Maggiore, ``{Modified
  gravitational-wave propagation and standard sirens},''
  \href{http://dx.doi.org/10.1103/PhysRevD.98.023510}{{\em Phys. Rev.}
  {\bfseries D98} (2018) 023510},
\href{http://arxiv.org/abs/1805.08731}{{\ttfamily arXiv:1805.08731 [gr-qc]}}.
%%CITATION = ARXIV:1805.08731;%%.

\bibitem{Mendonca:2019yfo}
J.~Mendon\c{c}a and R.~Sturani, ``{Cosmological model selection from standard
  siren detections by third generation gravitational wave obervatories},''
\href{http://arxiv.org/abs/1905.03848}{{\ttfamily arXiv:1905.03848 [gr-qc]}}.
%%CITATION = ARXIV:1905.03848;%%.

\bibitem{Sathyaprakash:2019nnu}
B.~S. Sathyaprakash {\em et~al.}, ``{Cosmology and the Early Universe},''
\href{http://arxiv.org/abs/1903.09260}{{\ttfamily arXiv:1903.09260
  [astro-ph.HE]}}.
%%CITATION = ARXIV:1903.09260;%%.

\bibitem{Sathyaprakash:2019rom}
B.~S. Sathyaprakash {\em et~al.}, ``{Multimessenger Universe with Gravitational
  Waves from Binaries},''
\href{http://arxiv.org/abs/1903.09277}{{\ttfamily arXiv:1903.09277
  [astro-ph.HE]}}.
%%CITATION = ARXIV:1903.09277;%%.

\bibitem{Sathyaprakash:2019yqt}
B.~S. Sathyaprakash {\em et~al.}, ``{Extreme Gravity and Fundamental
  Physics},''
\href{http://arxiv.org/abs/1903.09221}{{\ttfamily arXiv:1903.09221
  [astro-ph.HE]}}.
%%CITATION = ARXIV:1903.09221;%%.

\bibitem{Belgacem:2019pkk}
E.~Belgacem and others (LISA Cosmology Working~Group), ``{Testing modified
  gravity at cosmological distances with LISA standard sirens},''
\href{http://arxiv.org/abs/1906.01593}{{\ttfamily arXiv:1906.01593
  [astro-ph.CO]}}.
%%CITATION = ARXIV:1906.01593;%%.

\bibitem{2012PhRvD..86l2001R}
T.~{Regimbau}, T.~{Dent}, W.~{Del Pozzo}, S.~{Giampanis}, T.~G.~F. {Li},
  C.~{Robinson}, C.~{Van Den Broeck}, D.~{Meacher}, C.~{Rodriguez}, B.~S.
  {Sathyaprakash}, and K.~{W{\'o}jcik}, ``{Mock data challenge for the Einstein
  Gravitational-Wave Telescope},''
  \href{http://dx.doi.org/10.1103/PhysRevD.86.122001}{{\em Phys. Rev. D}
  {\bfseries 86} (2012) 122001},
  \href{http://arxiv.org/abs/1201.3563}{{\ttfamily arXiv:1201.3563 [gr-qc]}}.

\bibitem{2014PhRvD..89h4046R}
T.~{Regimbau}, D.~{Meacher}, and M.~{Coughlin}, ``{Second Einstein Telescope
  mock science challenge: Detection of the gravitational-wave stochastic
  background from compact binary coalescences},''
  \href{http://dx.doi.org/10.1103/PhysRevD.89.084046}{{\em Phys. Rev. D}
  {\bfseries 89} (2014) 084046},
  \href{http://arxiv.org/abs/1404.1134}{{\ttfamily arXiv:1404.1134}}.

\bibitem{Regimbau:2014nxa}
T.~Regimbau, K.~Siellez, D.~Meacher, B.~Gendre, and M.~Bo{\"e}r, ``{Revisiting
  coincidence rate between Gravitational Wave detection and short Gamma-Ray
  Burst for the Advanced and third generation},''
  \href{http://dx.doi.org/10.1088/0004-637X/799/1/69}{{\em Astrophys. J.}
  {\bfseries 799} no.~1, (2015) 69},
\href{http://arxiv.org/abs/1410.2739}{{\ttfamily arXiv:1410.2739
  [astro-ph.HE]}}.
%%CITATION = ARXIV:1410.2739;%%.

\bibitem{2015PhRvD..92f3002M}
D.~{Meacher}, M.~{Coughlin}, S.~{Morris}, T.~{Regimbau}, N.~{Christensen},
  S.~{Kandhasamy}, V.~{Mandic}, J.~D. {Romano}, and E.~{Thrane}, ``{Mock data
  and science challenge for detecting an astrophysical stochastic
  gravitational-wave background with Advanced LIGO and Advanced Virgo},''
  \href{http://dx.doi.org/10.1103/PhysRevD.92.063002}{{\em Phys. Rev. D}
  {\bfseries 92} (2015) 063002},
  \href{http://arxiv.org/abs/1506.06744}{{\ttfamily arXiv:1506.06744
  [astro-ph.HE]}}.

\bibitem{2016PhRvD..93b4018M}
D.~{Meacher}, K.~{Cannon}, C.~{Hanna}, T.~{Regimbau}, and B.~S.
  {Sathyaprakash}, ``{Second Einstein Telescope mock data and science
  challenge: Low frequency binary neutron star data analysis},''
  \href{http://dx.doi.org/10.1103/PhysRevD.93.024018}{{\em Phys. Rev. D}
  {\bfseries 93} no.~2, (Jan, 2016) 024018},
  \href{http://arxiv.org/abs/1511.01592}{{\ttfamily arXiv:1511.01592 [gr-qc]}}.

\bibitem{2017PhRvL.118o1105R}
T.~{Regimbau}, M.~{Evans}, N.~{Christensen}, E.~{Katsavounidis},
  B.~{Sathyaprakash}, and S.~{Vitale}, ``{Digging Deeper: Observing Primordial
  Gravitational Waves below the Binary-Black-Hole-Produced Stochastic
  Background},'' \href{http://dx.doi.org/10.1103/PhysRevLett.118.151105}{{\em
  Phys. Rev. Lett.} {\bfseries 118} (2017) 151105},
  \href{http://arxiv.org/abs/1611.08943}{{\ttfamily arXiv:1611.08943}}.

\bibitem{Amati:2017npy}
{\bfseries THESEUS} Collaboration, L.~Amati {\em et~al.}, ``{The THESEUS space
  mission concept: science case, design and expected performances},''
  \href{http://dx.doi.org/10.1016/j.asr.2018.03.010}{{\em Adv. Space Res.}
  {\bfseries 62} (2018) 191--244},
\href{http://arxiv.org/abs/1710.04638}{{\ttfamily arXiv:1710.04638
  [astro-ph.IM]}}.
%%CITATION = ARXIV:1710.04638;%%.

\bibitem{Stratta:2017bwq}
{\bfseries THESEUS} Collaboration, G.~Stratta {\em et~al.}, ``{THESEUS: a key
  space mission concept for Multi-Messenger Astrophysics},''
  \href{http://dx.doi.org/10.1016/j.asr.2018.04.013}{{\em Adv. Space Res.}
  {\bfseries 62} (2018) 662--682},
\href{http://arxiv.org/abs/1712.08153}{{\ttfamily arXiv:1712.08153
  [astro-ph.HE]}}.
%%CITATION = ARXIV:1712.08153;%%.

\bibitem{Stratta:2018ldl}
G.~Stratta, L.~Amati, R.~Ciolfi, and S.~Vinciguerra, ``{THESEUS in the era of
  Multi-Messenger Astronomy},'' {\em Mem. Soc. Ast. It.} {\bfseries 89} (2018)
  205,
\href{http://arxiv.org/abs/1802.01677}{{\ttfamily arXiv:1802.01677
  [astro-ph.IM]}}.
%%CITATION = ARXIV:1802.01677;%%.

\bibitem{Deffayet:2007kf}
C.~Deffayet and K.~Menou, ``{Probing Gravity with Spacetime Sirens},''
  \href{http://dx.doi.org/10.1086/522931}{{\em Astrophys. J.} {\bfseries 668}
  (2007) L143--L146},
\href{http://arxiv.org/abs/0709.0003}{{\ttfamily arXiv:0709.0003 [astro-ph]}}.
%%CITATION = ARXIV:0709.0003;%%.

\bibitem{Yunes:2010yf}
N.~Yunes, R.~O'Shaughnessy, B.~J. Owen, and S.~Alexander, ``{Testing
  gravitational parity violation with coincident gravitational waves and short
  gamma-ray bursts},'' \href{http://dx.doi.org/10.1103/PhysRevD.82.064017}{{\em
  Phys. Rev.} {\bfseries D82} (2010) 064017},
\href{http://arxiv.org/abs/1005.3310}{{\ttfamily arXiv:1005.3310 [gr-qc]}}.
%%CITATION = ARXIV:1005.3310;%%.

\bibitem{Saltas:2014dha}
I.~D. Saltas, I.~Sawicki, L.~Amendola, and M.~Kunz, ``{Anisotropic Stress as a
  Signature of Nonstandard Propagation of Gravitational Waves},''
  \href{http://dx.doi.org/10.1103/PhysRevLett.113.191101}{{\em Phys. Rev.
  Lett.} {\bfseries 113} (2014) 191101},
\href{http://arxiv.org/abs/1406.7139}{{\ttfamily arXiv:1406.7139
  [astro-ph.CO]}}.
%%CITATION = ARXIV:1406.7139;%%.

\bibitem{Gleyzes:2014rba}
J.~Gleyzes, D.~Langlois, and F.~Vernizzi, ``{A unifying description of dark
  energy},'' \href{http://dx.doi.org/10.1142/S021827181443010X}{{\em Int. J.
  Mod. Phys.} {\bfseries D23} (2014) 1443010},
\href{http://arxiv.org/abs/1411.3712}{{\ttfamily arXiv:1411.3712 [hep-th]}}.
%%CITATION = ARXIV:1411.3712;%%.

\bibitem{Lombriser:2015sxa}
L.~Lombriser and A.~Taylor, ``{Breaking a Dark Degeneracy with Gravitational
  Waves},'' \href{http://dx.doi.org/10.1088/1475-7516/2016/03/031}{{\em JCAP}
  {\bfseries 1603} (2016) 031},
\href{http://arxiv.org/abs/1509.08458}{{\ttfamily arXiv:1509.08458
  [astro-ph.CO]}}.
%%CITATION = ARXIV:1509.08458;%%.

\bibitem{Nishizawa:2017nef}
A.~Nishizawa, ``{Generalized framework for testing gravity with
  gravitational-wave propagation. I. Formulation},''
  \href{http://dx.doi.org/10.1103/PhysRevD.97.104037}{{\em Phys. Rev.}
  {\bfseries D97} (2018) 104037},
\href{http://arxiv.org/abs/1710.04825}{{\ttfamily arXiv:1710.04825 [gr-qc]}}.
%%CITATION = ARXIV:1710.04825;%%.

\bibitem{Arai:2017hxj}
S.~Arai and A.~Nishizawa, ``{Generalized framework for testing gravity with
  gravitational-wave propagation. II. Constraints on Horndeski theory},''
  \href{http://dx.doi.org/10.1103/PhysRevD.97.104038}{{\em Phys. Rev.}
  {\bfseries D97} (2018) 104038},
\href{http://arxiv.org/abs/1711.03776}{{\ttfamily arXiv:1711.03776 [gr-qc]}}.
%%CITATION = ARXIV:1711.03776;%%.

\bibitem{Amendola:2017ovw}
L.~Amendola, I.~Sawicki, M.~Kunz, and I.~D. Saltas, ``{Direct detection of
  gravitational waves can measure the time variation of the Planck mass},''
  \href{http://dx.doi.org/10.1088/1475-7516/2018/08/030}{{\em JCAP} {\bfseries
  1808} (2018) 030},
\href{http://arxiv.org/abs/1712.08623}{{\ttfamily arXiv:1712.08623
  [astro-ph.CO]}}.
%%CITATION = ARXIV:1712.08623;%%.

\bibitem{Linder:2018jil}
E.~V. Linder, ``{No Slip Gravity},''
  \href{http://dx.doi.org/10.1088/1475-7516/2018/03/005}{{\em JCAP} {\bfseries
  1803} (2018) 005},
\href{http://arxiv.org/abs/1801.01503}{{\ttfamily arXiv:1801.01503
  [astro-ph.CO]}}.
%%CITATION = ARXIV:1801.01503;%%.

\bibitem{Pardo:2018ipy}
K.~Pardo, M.~Fishbach, D.~E. Holz, and D.~N. Spergel, ``{Limits on the number
  of spacetime dimensions from GW170817},''
  \href{http://dx.doi.org/10.1088/1475-7516/2018/07/048}{{\em JCAP} {\bfseries
  1807} (2018) 048},
\href{http://arxiv.org/abs/1801.08160}{{\ttfamily arXiv:1801.08160 [gr-qc]}}.
%%CITATION = ARXIV:1801.08160;%%.

\bibitem{Lagos:2019kds}
M.~Lagos, M.~Fishbach, P.~Landry, and D.~E. Holz, ``{Standard sirens with a
  running Planck mass},''
  \href{http://dx.doi.org/10.1103/PhysRevD.99.083504}{{\em Phys. Rev.}
  {\bfseries D99} (2019) 083504},
\href{http://arxiv.org/abs/1901.03321}{{\ttfamily arXiv:1901.03321
  [astro-ph.CO]}}.
%%CITATION = ARXIV:1901.03321;%%.

\bibitem{2018PhRvL.120i1101A}
B.~P. {Abbott}, R.~{Abbott}, T.~D. {Abbott}, F.~{Acernese}, K.~{Ackley},
  C.~{Adams}, T.~{Adams}, P.~{Addesso}, R.~X. {Adhikari}, V.~B. {Adya}, and
  et~al., ``{GW170817: Implications for the Stochastic Gravitational-Wave
  Background from Compact Binary Coalescences},''
  \href{http://dx.doi.org/10.1103/PhysRevLett.120.091101}{{\em Phys. Rev.
  Lett.} {\bfseries 120} (2018) 091101},
  \href{http://arxiv.org/abs/1710.05837}{{\ttfamily arXiv:1710.05837 [gr-qc]}}.

\bibitem{2015MNRAS.447.2575V}
E.~{Vangioni}, K.~A. {Olive}, T.~{Prestegard}, J.~{Silk}, P.~{Petitjean}, and
  V.~{Mandic}, ``{The impact of star formation and gamma-ray burst rates at
  high redshift on cosmic chemical evolution and reionization},''
  \href{http://dx.doi.org/10.1093/mnras/stu2600}{{\em Mon. Not. Roy. Astron.
  Soc.} {\bfseries 447} (2015) 2575--2587},
  \href{http://arxiv.org/abs/1409.2462}{{\ttfamily arXiv:1409.2462}}.

\bibitem{Madau:2014bja}
P.~Madau and M.~Dickinson, ``{Cosmic Star Formation History},''
  \href{http://dx.doi.org/10.1146/annurev-astro-081811-125615}{{\em Ann. Rev.
  Astron. Astrophys.} {\bfseries 52} (2014) 415--486},
\href{http://arxiv.org/abs/1403.0007}{{\ttfamily arXiv:1403.0007
  [astro-ph.CO]}}.
%%CITATION = ARXIV:1403.0007;%%.

\bibitem{Vitale:2018yhm}
S.~Vitale and W.~M. Farr, ``{Measuring the star formation rate with
  gravitational waves from binary black holes},''
\href{http://arxiv.org/abs/1808.00901}{{\ttfamily arXiv:1808.00901
  [astro-ph.HE]}}.
%%CITATION = ARXIV:1808.00901;%%.

\bibitem{Chen:2017rfc}
H.-Y. Chen, M.~Fishbach, and D.~E. Holz, ``{Precision standard siren
  cosmology},''
\href{http://arxiv.org/abs/1712.06531}{{\ttfamily arXiv:1712.06531
  [astro-ph.CO]}}.
%%CITATION = ARXIV:1712.06531;%%.

\bibitem{Berger:2013jza}
E.~Berger, ``{Short-Duration Gamma-Ray Bursts},''
  \href{http://dx.doi.org/10.1146/annurev-astro-081913-035926}{{\em Ann. Rev.
  Astron. Astrophys.} {\bfseries 52} (2014) 43--105},
\href{http://arxiv.org/abs/1311.2603}{{\ttfamily arXiv:1311.2603
  [astro-ph.HE]}}.
%%CITATION = ARXIV:1311.2603;%%.

\bibitem{evans_swift_2017}
P.~A. Evans {\em et~al.}, ``Swift and {NuSTAR} observations of {GW}170817:
  {Detection} of a blue kilonova,''
  \href{http://dx.doi.org/10.1126/science.aap9580}{{\em Science} {\bfseries
  358} no.~6370, (Dec., 2017) 1565--1570}.
  \url{http://science.sciencemag.org/content/358/6370/1565}.

\bibitem{Howell_rates_2018}
E.~J. {Howell}, K.~{Ackley}, A.~{Rowlinson}, and D.~{Coward}, ``{Joint
  gravitational wave - gamma-ray burst detection rates in the aftermath of
  GW170817},'' \href{http://dx.doi.org/10.1093/mnras/stz455}{{\em Mon. Not.
  Roy. Astron. Soc.} {\bfseries 485} no.~1, (May, 2019) 1435--1447},
  \href{http://arxiv.org/abs/1811.09168}{{\ttfamily arXiv:1811.09168
  [astro-ph.HE]}}.

\bibitem{WandermanPiran2015MNRAS}
D.~{Wanderman} and T.~{Piran}, ``{The rate, luminosity function and time delay
  of non-Collapsar short GRBs},''
  \href{http://dx.doi.org/10.1093/mnras/stv123}{{\em Mon. Not. Roy. Astron.
  Soc.} {\bfseries 448} (Apr., 2015) 3026--3037},
  \href{http://arxiv.org/abs/1405.5878}{{\ttfamily arXiv:1405.5878
  [astro-ph.HE]}}.

\bibitem{Burns2016ApJ}
E.~{Burns}, V.~{Connaughton}, B.-B. {Zhang}, A.~{Lien}, M.~S. {Briggs},
  A.~{Goldstein}, V.~{Pelassa}, and E.~{Troja}, ``{Do the Fermi Gamma-Ray Burst
  Monitor and Swift Burst Alert Telescope see the Same Short Gamma-Ray
  Bursts?},'' \href{http://dx.doi.org/10.3847/0004-637X/818/2/110}{{\em
  Astrophys. J.} {\bfseries 818} (Feb., 2016) 110},
  \href{http://arxiv.org/abs/1512.00923}{{\ttfamily arXiv:1512.00923
  [astro-ph.HE]}}.

\bibitem{Planck_2015_1}
{\bfseries Planck} Collaboration, R.~Adam {\em et~al.}, ``{Planck 2015 results.
  I. Overview of products and scientific results},''
\href{http://arxiv.org/abs/1502.01582}{{\ttfamily arXiv:1502.01582
  [astro-ph.CO]}}.
%%CITATION = ARXIV:1502.01582;%%.

\bibitem{Ade:2015rim}
{\bfseries Planck} Collaboration, P.~A.~R. Ade {\em et~al.}, ``{Planck 2015
  results. XIV. Dark energy and modified gravity},''
  \href{http://dx.doi.org/10.1051/0004-6361/201525814}{{\em Astron. Astrophys.}
  {\bfseries 594} (2016) A14},
\href{http://arxiv.org/abs/1502.01590}{{\ttfamily arXiv:1502.01590
  [astro-ph.CO]}}.
%%CITATION = ARXIV:1502.01590;%%.

\bibitem{Planck_2015_Lkl}
{\bfseries Planck} Collaboration, N.~Aghanim {\em et~al.}, ``{Planck 2015
  results. XI. CMB power spectra, likelihoods, and robustness of parameters},''
  \href{http://dx.doi.org/10.1051/0004-6361/201526926}{{\em Astron. Astrophys.}
  {\bfseries 594} (2016) A11},
\href{http://arxiv.org/abs/1507.02704}{{\ttfamily arXiv:1507.02704
  [astro-ph.CO]}}.
%%CITATION = ARXIV:1507.02704;%%.

\bibitem{Planck_2015_lens}
{\bfseries Planck} Collaboration, P.~A.~R. Ade {\em et~al.}, ``{Planck 2015
  results. XV. Gravitational lensing},''
\href{http://arxiv.org/abs/1502.01591}{{\ttfamily arXiv:1502.01591
  [astro-ph.CO]}}.
%%CITATION = ARXIV:1502.01591;%%.

\bibitem{Betoule:2014frx}
{\bfseries SDSS} Collaboration, M.~Betoule {\em et~al.}, ``{Improved
  cosmological constraints from a joint analysis of the SDSS-II and SNLS
  supernova samples},''
  \href{http://dx.doi.org/10.1051/0004-6361/201423413}{{\em Astron. Astrophys.}
  {\bfseries 568} (2014) A22},
\href{http://arxiv.org/abs/1401.4064}{{\ttfamily arXiv:1401.4064
  [astro-ph.CO]}}.
%%CITATION = ARXIV:1401.4064;%%.

\bibitem{Beutler:2011hx}
F.~Beutler, C.~Blake, M.~Colless, D.~H. Jones, L.~Staveley-Smith, {\em et~al.},
  ``{The 6dF Galaxy Survey: Baryon Acoustic Oscillations and the Local Hubble
  Constant},'' \href{http://dx.doi.org/10.1111/j.1365-2966.2011.19250.x}{{\em
  Mon.Not.Roy.Astron.Soc.} {\bfseries 416} (2011) 3017--3032},
\href{http://arxiv.org/abs/1106.3366}{{\ttfamily arXiv:1106.3366
  [astro-ph.CO]}}.
%%CITATION = ARXIV:1106.3366;%%.

\bibitem{Ross_SDSS_2014}
A.~J. Ross, L.~Samushia, C.~Howlett, W.~J. Percival, A.~Burden, and M.~Manera,
  ``{The clustering of the SDSS DR7 main Galaxy sample - I. A 4 per cent
  distance measure at $z = 0.15$},''
  \href{http://dx.doi.org/10.1093/mnras/stv154}{{\em Mon. Not. Roy. Astron.
  Soc.} {\bfseries 449} no.~1, (2015) 835--847},
\href{http://arxiv.org/abs/1409.3242}{{\ttfamily arXiv:1409.3242
  [astro-ph.CO]}}.
%%CITATION = ARXIV:1409.3242;%%.

\bibitem{Anderson:2013zyy}
{\bfseries BOSS} Collaboration, L.~Anderson {\em et~al.}, ``{The clustering of
  galaxies in the SDSS-III Baryon Oscillation Spectroscopic Survey: Baryon
  Acoustic Oscillations in the Data Release 10 and 11 galaxy samples},''
\href{http://arxiv.org/abs/1312.4877}{{\ttfamily arXiv:1312.4877
  [astro-ph.CO]}}.
%%CITATION = ARXIV:1312.4877;%%.

\bibitem{Class}
D.~{Blas}, J.~{Lesgourgues}, and T.~{Tram}, ``{The Cosmic Linear Anisotropy
  Solving System (CLASS). Part II: Approximation schemes},''
  \href{http://dx.doi.org/10.1088/1475-7516/2011/07/034}{{\em JCAP} {\bfseries
  7} (July, 2011) 34}, \href{http://arxiv.org/abs/1104.2933}{{\ttfamily
  arXiv:1104.2933 [astro-ph.CO]}}.

\bibitem{Planck_2015_CP}
{\bfseries Planck} Collaboration, P.~A.~R. Ade {\em et~al.}, ``{Planck 2015
  results. XIII. Cosmological parameters},''
  \href{http://dx.doi.org/10.1051/0004-6361/201525830}{{\em Astron. Astrophys.}
  {\bfseries 594} (2016) A13},
\href{http://arxiv.org/abs/1502.01589}{{\ttfamily arXiv:1502.01589
  [astro-ph.CO]}}.
%%CITATION = ARXIV:1502.01589;%%.

\bibitem{Feeney:2018mkj}
S.~M. Feeney, H.~V. Peiris, A.~R. Williamson, S.~M. Nissanke, D.~J. Mortlock,
  J.~Alsing, and D.~Scolnic, ``{Prospects for resolving the Hubble constant
  tension with standard sirens},''
\href{http://arxiv.org/abs/1802.03404}{{\ttfamily arXiv:1802.03404
  [astro-ph.CO]}}.
%%CITATION = ARXIV:1802.03404;%%.

\bibitem{Riess:2019cxk}
A.~G. Riess, S.~Casertano, W.~Yuan, L.~M. Macri, and D.~Scolnic, ``{Large
  Magellanic Cloud Cepheid Standards Provide a 1\% Foundation for the
  Determination of the Hubble Constant and Stronger Evidence for Physics Beyond
  LambdaCDM},'' \href{http://dx.doi.org/10.3847/1538-4357/ab1422}{{\em
  Astrophys. J.} {\bfseries 876} no.~1, (2019) 85},
\href{http://arxiv.org/abs/1903.07603}{{\ttfamily arXiv:1903.07603
  [astro-ph.CO]}}.
%%CITATION = ARXIV:1903.07603;%%.

\bibitem{Chevallier:2000qy}
M.~Chevallier and D.~Polarski, ``{Accelerating universes with scaling dark
  matter},'' \href{http://dx.doi.org/10.1142/S0218271801000822}{{\em
  Int.J.Mod.Phys.} {\bfseries D10} (2001) 213--224},
\href{http://arxiv.org/abs/gr-qc/0009008}{{\ttfamily arXiv:gr-qc/0009008
  [gr-qc]}}.
%%CITATION = GR-QC/0009008;%%.

\bibitem{Linder:2002et}
E.~V. Linder, ``{Exploring the expansion history of the universe},''
  \href{http://dx.doi.org/10.1103/PhysRevLett.90.091301}{{\em Phys.Rev.Lett.}
  {\bfseries 90} (2003) 091301},
\href{http://arxiv.org/abs/astro-ph/0208512}{{\ttfamily arXiv:astro-ph/0208512
  [astro-ph]}}.
%%CITATION = ASTRO-PH/0208512;%%.

\bibitem{Belgacem:2017cqo}
E.~Belgacem, Y.~Dirian, S.~Foffa, and M.~Maggiore, ``{Nonlocal gravity.
  Conceptual aspects and cosmological predictions},''
  \href{http://dx.doi.org/10.1088/1475-7516/2018/03/002}{{\em JCAP} {\bfseries
  1803} (2018) 002},
\href{http://arxiv.org/abs/1712.07066}{{\ttfamily arXiv:1712.07066 [hep-th]}}.
%%CITATION = ARXIV:1712.07066;%%.

\bibitem{Maggiore:2018zz}
M.~Maggiore, {\em {Gravitational Waves. Vol. 2. Astrophysics and Cosmology}}.
\newblock Oxford University Press, 848 p, 2018.

\bibitem{Maggiore:2013mea}
M.~Maggiore, ``{Phantom dark energy from nonlocal infrared modifications of
  general relativity},'' {\em Phys.Rev.} {\bfseries D89} (2014) 043008,
\href{http://arxiv.org/abs/1307.3898}{{\ttfamily arXiv:1307.3898 [hep-th]}}.
%%CITATION = ARXIV:1307.3898;%%.

\bibitem{Maggiore:2014sia}
M.~Maggiore and M.~Mancarella, ``{Non-local gravity and dark energy},''
  \href{http://dx.doi.org/10.1103/PhysRevD.90.023005}{{\em Phys.Rev.}
  {\bfseries D90} (2014) 023005},
\href{http://arxiv.org/abs/1402.0448}{{\ttfamily arXiv:1402.0448 [hep-th]}}.
%%CITATION = ARXIV:1402.0448;%%.

\bibitem{Foffa:2013vma}
S.~Foffa, M.~Maggiore, and E.~Mitsou, ``{Cosmological dynamics and dark energy
  from non-local infrared modifications of gravity},'' {\em Int.J.Mod.Phys.}
  {\bfseries A29} (2014) 1450116,
\href{http://arxiv.org/abs/1311.3435}{{\ttfamily arXiv:1311.3435 [hep-th]}}.
%%CITATION = ARXIV:1311.3435;%%.

\bibitem{Dirian:2014ara}
Y.~Dirian, S.~Foffa, N.~Khosravi, M.~Kunz, and M.~Maggiore, ``{Cosmological
  perturbations and structure formation in nonlocal infrared modifications of
  general relativity},''
  \href{http://dx.doi.org/10.1088/1475-7516/2014/06/033}{{\em JCAP} {\bfseries
  1406} (2014) 033},
\href{http://arxiv.org/abs/1403.6068}{{\ttfamily arXiv:1403.6068
  [astro-ph.CO]}}.
%%CITATION = ARXIV:1403.6068;%%.

\bibitem{Dirian:2014bma}
Y.~Dirian, S.~Foffa, M.~Kunz, M.~Maggiore, and V.~Pettorino, ``{Non-local
  gravity and comparison with observational datasets},''
  \href{http://dx.doi.org/10.1088/1475-7516/2015/04/044}{{\em JCAP} {\bfseries
  1504} (2015) 044},
\href{http://arxiv.org/abs/1411.7692}{{\ttfamily arXiv:1411.7692
  [astro-ph.CO]}}.
%%CITATION = ARXIV:1411.7692;%%.

\bibitem{Dirian:2016puz}
Y.~Dirian, S.~Foffa, M.~Kunz, M.~Maggiore, and V.~Pettorino, ``{Non-local
  gravity and comparison with observational datasets. II. Updated results and
  Bayesian model comparison with $\Lambda$CDM},''
  \href{http://dx.doi.org/10.1088/1475-7516/2016/05/068}{{\em JCAP} {\bfseries
  1605} (2016) 068},
\href{http://arxiv.org/abs/1602.03558}{{\ttfamily arXiv:1602.03558
  [astro-ph.CO]}}.
%%CITATION = ARXIV:1602.03558;%%.

\bibitem{Dirian:2017pwp}
Y.~Dirian, ``{Changing the Bayesian prior: Absolute neutrino mass constraints
  in nonlocal gravity},''
  \href{http://dx.doi.org/10.1103/PhysRevD.96.083513}{{\em Phys. Rev.}
  {\bfseries D96} (2017) 083513},
\href{http://arxiv.org/abs/1704.04075}{{\ttfamily arXiv:1704.04075
  [astro-ph.CO]}}.
%%CITATION = ARXIV:1704.04075;%%.

\bibitem{Belgacem:2018wtb}
E.~Belgacem, A.~Finke, A.~Frassino, and M.~Maggiore, ``{Testing nonlocal
  gravity with Lunar Laser Ranging},''
  \href{http://dx.doi.org/10.1088/1475-7516/2019/02/035}{{\em JCAP} {\bfseries
  1902} (2019) 035},
\href{http://arxiv.org/abs/1812.11181}{{\ttfamily arXiv:1812.11181 [gr-qc]}}.
%%CITATION = ARXIV:1812.11181;%%.

\bibitem{Maggiore:2016gpx}
M.~Maggiore, ``{Nonlocal Infrared Modifications of Gravity. A Review},''
  \href{http://dx.doi.org/10.1007/978-3-319-51700-1_16}{{\em Fundam. Theor.
  Phys.} {\bfseries 187} (2017) 221--281},
\href{http://arxiv.org/abs/1606.08784}{{\ttfamily arXiv:1606.08784 [hep-th]}}.
%%CITATION = ARXIV:1606.08784;%%.

\bibitem{DeltaN64:inprep}
E.~Belgacem, Y.~Dirian, A.~Finke, S.~Foffa, and M.~Maggiore, ``{Nonlocal
  gravity and gravitational-wave observations},'' {\em to appear} .

\bibitem{Kehagias:2014sda}
A.~Kehagias and M.~Maggiore, ``{Spherically symmetric static solutions in a
  non-local infrared modification of General Relativity},''
  \href{http://dx.doi.org/10.1007/JHEP08(2014)029}{{\em JHEP} {\bfseries 1408}
  (2014) 029},
\href{http://arxiv.org/abs/1401.8289}{{\ttfamily arXiv:1401.8289 [hep-th]}}.
%%CITATION = ARXIV:1401.8289;%%.

\end{thebibliography}\endgroup

\end{document}